\documentclass{article}

\usepackage{microtype}
\usepackage{graphicx}
\usepackage{subcaption}
\usepackage{booktabs} % for professional tables

\usepackage{hyperref}

\usepackage[preprint]{icml2026}

\usepackage{amsmath}
\usepackage{amssymb}
\usepackage{mathtools}
\usepackage{amsthm}

\usepackage[capitalize,noabbrev]{cleveref}

\theoremstyle{plain}

\theoremstyle{definition}

\theoremstyle{remark}

\usepackage[textsize=tiny]{todonotes}

\usepackage{lipsum}
\usepackage{threeparttable}

\newcommand{\blue}[1]{{\color{black}#1}}

\usepackage{enumitem}

\usepackage{subcaption} %for subfigure

\usepackage{tablefootnote} %for footnote in the table
\usepackage{colortbl} %for color in the table

\newcommand{\ignore}[1]{}

\usepackage[capitalize,noabbrev]{cleveref}
\usepackage{times}
\usepackage{wrapfig}
\usepackage{xspace}
\usepackage{makecell}
 \usepackage{multirow}

\newcommand{\Suite}{DIYHealth Suite\xspace}

\newcommand{\Name}{DIYHealthGPT\xspace}
\newcommand{\Engine}{DIYHealth Data Engine\xspace}

\newcommand{\Dataset}{DIYHealth-900K\xspace}

\newcommand{\Benchmark}{DIYHealthBench\xspace}

\begin{document}

\icmltitlerunning{\Suite: Dataset, Model, and Benchmark for Health Management at Home}

\twocolumn[
  \icmltitle{\Suite: Dataset, Model, and Benchmark \\ for Health Management at Home}

      \icmlsetsymbol{equal}{*}

  \begin{icmlauthorlist}
    \icmlauthor{Changshuo Liu}{nus}
    \icmlauthor{Junran Wu}{nus}
    \icmlauthor{Zhongle Xie}{zju2}
    \icmlauthor{Wenqiao Zhang}{zju2}
    \icmlauthor{Kaiping Zheng}{nus}
    \icmlauthor{Jiaqi Zhu}{nus,equal}
    \icmlauthor{Qingpeng Cai}{nus}
    \icmlauthor{Gene Anne Ooi}{sgh}
    \icmlauthor{Marcus Chun Jin Tan}{nuh}
    \icmlauthor{Jianwei Yin}{zju,zju2}
    \icmlauthor{James Wei Luen Yip}{nuh}
    \icmlauthor{Beng Chin Ooi}{zju,zju2}
  \end{icmlauthorlist}

  \icmlaffiliation{nus}{School of Computing, National University of Singapore, Singapore}
  \icmlaffiliation{zju}{College of Computer Science and Technology, Zhejiang University, Hangzhou, China}
  \icmlaffiliation{zju2}{College of Software Technology, Zhejiang University, Ningbo, China}
  \icmlaffiliation{sgh}{Singapore General Hospital, Singapore}
  \icmlaffiliation{nuh}{National University Hospital, Singapore}

  \icmlcorrespondingauthor{Jiaqi Zhu}{jiaqi77@nus.edu.sg}
  \icmlkeywords{Home Care, AI for Healthcare}
  \vskip 0.3in
]

\printAffiliationsAndNotice{}  % no special notice (required even if empty)

\begin{abstract}
Generative AI is reshaping healthcare, yet most existing advances rely on hospital-grade devices, which limits their accessibility and potential for health management outside clinical settings.
% With the proliferation of wearables, mobile sensors, and telemedicine, healthcare is shifting toward the home, giving rise to the emerging field of \textbf{Diagnosis-It-Yourself (DIY) at home}, \emph{i.e.}, home care.
With the proliferation of portable devices and telemedicine, healthcare is shifting toward home-based Diagnosis-It-Yourself (DIY) care.
Despite this promise, several distinctive challenges remain:
(i) home-collected data are heterogeneous, exacerbated by the absence of standardized large-scale datasets;
(ii) models require adaptation to variable task demands and evolving individual conditions;
(iii) the broad spectrum of home care tasks lacks a unified benchmark for systematic evaluation.
In this paper, we present \textbf{\Suite}, a comprehensive framework designed to address these challenges through a tailored dataset, model, and benchmark.
% \blue{We first introduce \textbf{\Engine}, an LLM-powered data engine 
% with human-in-the-loop verification 
% tailored for real-world home care scenarios.}
We first curate \textbf{\Dataset}, a large-scale multimodal dataset capturing diverse real-world home care scenarios.
Building on this, we propose \textbf{\Name}, an adaptive foundation model for home-based health management, powered by the novel Hybrid Hyper Low-Rank Adaptation 
technique.
% , which integrates expert mixtures with hypernetwork-driven modulation to balance cross-task generalization and instance-level personalization.
Finally, we establish \textbf{\Benchmark}, the first benchmark to evaluate foundation models on home care tasks.
Extensive experiments demonstrate that \Name delivers state-of-the-art performance over both general-purpose and medical-specific baselines on 11 home care tasks in both open-QA and closed-QA settings, laying the groundwork for the next generation of personalized health management at home.
% AI-driven, personalized, and scalable health management at home.
\end{abstract}
\vspace{-20pt}
\vspace{-1mm}
\section{Introduction}
\vspace{-1mm}
\begin{figure*}
\centering
\includegraphics[width=0.82\textwidth]{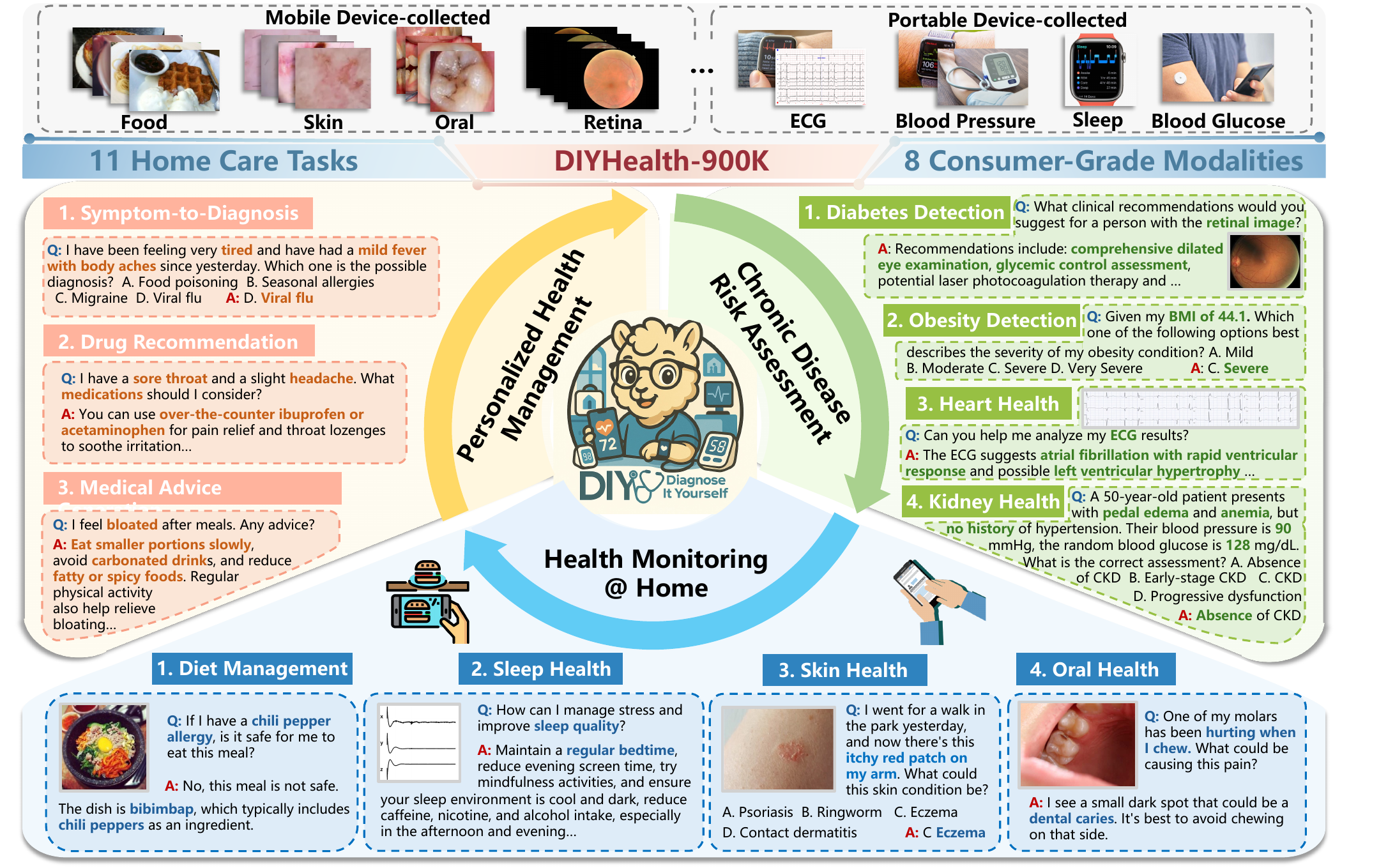}
  \caption{\scalebox{0.9}{Overview of \Suite, integrating \Dataset, \Name, and \Benchmark towards health management at home.}}
  % across 11 home care tasks
  \label{fig:framework}
  \vspace{-15pt}
\end{figure*}

% kp: for the model novelty concern, highlight in key places (introduction, methodology, experimental results - e.g., Fig. 4) to avoid being considered limited in novelty, or incremental as just enhancing MoELoRA to some extent
%% 1) existing works either use HyperLoRA or MoELoRA - each has limitations, etc.
%% 2) different from them, we strike a balance between these 2 mechanisms and hence have the advantages of both -> confirm our model novelty
%% 3) this innovative methodology design also brings benefits in performance, as affirmed in the results shown in Fig. 4

Generative AI has advanced rapidly from large language models (LLMs)~\citep{brown2020language,openai2023gpt4} to multimodal architectures~\citep{radford2021learning}, driving a broad impact across diverse domains.
Within healthcare, these advances have enabled the interpretation of complex clinical data, the integration of heterogeneous modalities, and the provision of decision support, opening new opportunities to enhance diagnosis, treatment, and patient care~\citep{wang2015singa,lin2025healthgpt,qiu2024llm,chen2025generative}.

% Recent studies have begun to investigate medical foundation models that leverage large-scale clinical data to improve tasks such as radiology report generation~\citep{thawkar2023xraygpt}, medical image interpretation~\citep{rajpurkar2023current}, and clinical question answering~\citep{li2025eyecaregpt,xie2025heartcare}.
% They achieve strong performance by capturing domain-specific knowledge and adapting to a broad range of medical tasks, representing a significant step toward AI-assisted healthcare.
% Nevertheless, current efforts remain predominantly clinical-centric: they are trained on hospital-grade data, tailored for professional use, and optimized for environments where high-quality imaging, electronic health records, and expert annotations are readily available.
Recent studies have explored medical foundation models trained on large-scale clinical data for tasks such as radiology report generation~\citep{thawkar2023xraygpt}, medical image interpretation~\citep{rajpurkar2023current}, and clinical question answering~\citep{li2025eyecaregpt,xie2025heartcare}. While these models achieve strong performance by capturing domain-specific knowledge across diverse medical tasks, they remain largely clinical-centric, relying on hospital-grade data and expert annotations and targeting professional healthcare settings.

Although foundation models have achieved considerable success in clinical contexts, their application to the home care setting has remained largely \textbf{unexplored} to date. The home care setting holds significant promise for collecting multimodal health data and enabling convenient inference beyond clinical environments, facilitated by the widespread adoption of smartphones, wearables, and home sensors~\citep{zaidan2018survey,kruzan2023perceived}.
Despite this opportunity, home care introduces distinct challenges that differ fundamentally from those encountered in clinical settings.
To begin with, data collected at home often originates from consumer-grade devices and self-reported inputs, resulting in heterogeneous and lower-quality signals; the absence of standardized large-scale datasets further constrains the systematic development of foundation models in this context.
Further, whereas population-level models may suffice in hospital settings, home care demands adaptation to highly variable personal health baselines and evolving individual conditions.
Finally, home care spans a wide spectrum of tasks, from personalized health management to daily health monitoring, yet currently lacks a unified benchmark for performance evaluation across such diverse applications.

To systematically unlock the potential for health management at home while addressing the aforementioned key challenges of data accessibility, personalization, and task diversity, we introduce \textbf{\Suite}\footnote{\blue{Please refer to Appendix~\ref{sec:appendix-related-work} for a discussion of related work.}}, a comprehensive ecosystem that integrates a large-scale multimodal dataset curated for home care, an adaptive foundation model designed to accommodate individual variability, and a unified benchmark spanning diverse tasks in everyday health management and monitoring.

Within the ecosystem,
we propose three core components.
We first construct \textbf{\Dataset}, a multimodal dataset curated via \textbf{\Engine} to aggregate heterogeneous inputs from home environments under rigorous quality control.
To address individual variability, we propose \textbf{\Name}, an adaptive foundation model that employs a novel Parameter-Efficient Fine-Tuning (PEFT) technique, Hybrid Hyper Low-Rank Adaptation (H$^2$LoRA), which combines shared low-rank expert mixtures for efficient cross-task knowledge sharing with hypernetwork-driven adaptation for instance-aware personalization.
Finally, to enable systematic evaluation, we establish \textbf{\Benchmark}, a unified benchmark guided by a multi-dimensional evaluation protocol spanning both open-QA and closed-QA, thereby capturing the dual requirements of home care: adaptive dialogue for personalized advice and structured reasoning for decision support.
Collectively, these components lay the groundwork for accessible, intelligent health management at home,
% beyond clinical settings,
advancing inclusive AI for everyday well-being.
Our contributions are as follows:
\begin{itemize}[itemsep=0mm,leftmargin=4mm]
\vspace{-2mm}
\item We curate \Dataset, a large-scale multimodal dataset derived from everyday devices to reflect the complexity and diversity of real-world home-care scenarios.
% \item We 

\item We propose \Name, an adaptive foundation model for home-based health management, powered by the innovative H$^2$LoRA mechanism that ensures personalized representations and robust generalization.

\item We introduce \Benchmark, the first unified benchmark for evaluating foundation models in non-clinical settings, spanning tasks from daily health monitoring to chronic disease risk assessment and personalized health management, reflecting the diverse needs of home care.

\item Extensive experiments on \Benchmark demonstrate that \Name consistently outperforms both state-of-the-art generalist and medical-specific baselines across diverse home care tasks, affirming its effectiveness in facilitating personalized health management at home.
\end{itemize}
% \input{secs/2.related_work}

% \vspace{-13pt}
\section{\Dataset}
\label{sec:dataset}
\vspace{-1mm}
\subsection{\Engine}
% kp: should change to "\Engine" for consistency?

\label{sec:data-engineering}
\blue{To construct a reliable data resource for home care, we develop an LLM-powered data synthesis engine with structured human-in-the-loop verification to support multiple modalities and diverse task formulations in a unified manner, as illustrated in Figure~\ref{fig:data_engine}. Instead of generating task-specific datasets in isolation, the engine enforces cross-task semantic consistency and scenario realism through a modular yet tightly coupled design comprising four components.
% elaborated below.

\begin{figure}[t]
    \centering
    \includegraphics[width=\linewidth]{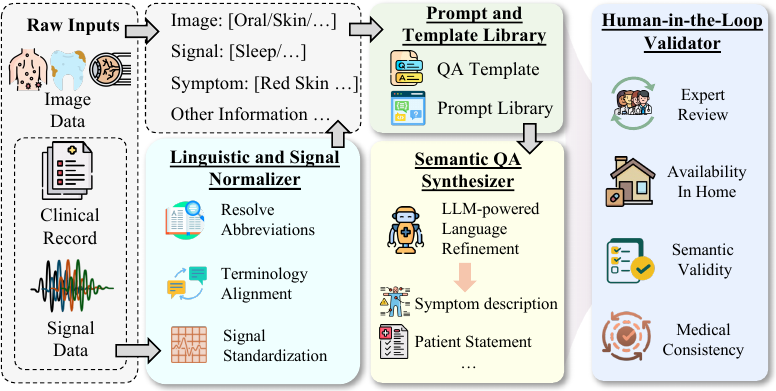}
    \caption{\blue{Illustration of \Engine}}
    \label{fig:data_engine}
    \vspace{-6mm}
\end{figure}

(i) \textit{Linguistic and Signal Normalizer}:
% We standardize heterogeneous raw data by addressing common issues such as medical text abbreviations and preprocessing of physiological signals (e.g., ECG).
We introduce a normalization process that harmonizes heterogeneous raw inputs across modalities, including textual symptom descriptions and physiological signals, by resolving medical abbreviations, aligning terminology, and standardizing signal preprocessing. This process establishes a shared semantic and statistical basis for downstream synthesis.

(ii) \textit{Prompt and Template Library}:
% We design a principled library of prompts and templates to guide QA pair generation. The library spans both text-only QA and VQA tasks, supports open-QA (free-form answers) and closed-QA (multiple-choice),
% and incorporates diverse user perspectives (first-person and third-person), simulating realistic home care interactions.
We design a structured prompt library that encodes task objectives, modality constraints, and user perspectives. The library supports both text-only QA and VQA, accommodates open-QA and closed-QA formats, and incorporates first- and third-person narratives, enabling the synthesis of realistic and diverse home-care interactions while maintaining semantic alignment across tasks.

(iii) \textit{Semantic QA Synthesizer}:
% We employ Claude 3 Haiku to automatically synthesize large-scale QA pairs, guided by the prompt schema to maintain semantic consistency across tasks.
Leveraging Claude 3 Haiku, we automatically generate large-scale QA pairs conditioned on the shared prompt schema. This design ensures that synthesized samples across different tasks and modalities adhere to consistent clinical semantics, rather than producing isolated task-specific distributions.

(iv) \textit{Human-in-the-Loop Validator}:
% To guarantee quality, we randomly sample 10\% of the automatically generated QA pairs for inspection by human reviewers. Medical professionals focus on semantic validity, medical consistency, and format standardization. Each entry undergoes two rounds of independent review, providing fine-grained data quality control and ensuring reliability.
We adopt a human-in-the-loop validation strategy to supervise and refine the automatic QA generation pipeline. Automatically generated QA pairs are incorporated into an expert review, with emphasis on availability in home care settings, semantic validity, and medical consistency. Each reviewed entry undergoes multiple rounds of independent assessment, forming a quality control loop that reduces systematic errors from automated generation and ensures the overall integrity of the dataset.}
% \vspace{-1mm}

\subsection{Task Landscape and Data Curation in Home Care}
\label{sec:task&data}
To reflect the diversity of real-world home care scenarios, we categorize the tasks in the \Dataset dataset into three groups: (i) \textit{Personalized Health Management}, which includes core tasks such as symptom-based diagnosis, drug recommendation, and medical advice generation; (ii) \textit{Chronic Disease Risk Assessment}, which targets conditions such as diabetes, obesity, cardiovascular, and kidney health through self-reported symptoms and home-acquired signals; and (iii) \textit{Daily Health Monitoring}, which encompasses dietary intake, sleep, skin, and oral assessments.
All tasks are curated or adapted to incorporate multimodal input, real-world variability, and nonclinical supervision, ensuring their relevance to home care environments. 

\begin{figure}[t]
    \begin{center}
    \includegraphics[width=0.45\textwidth]{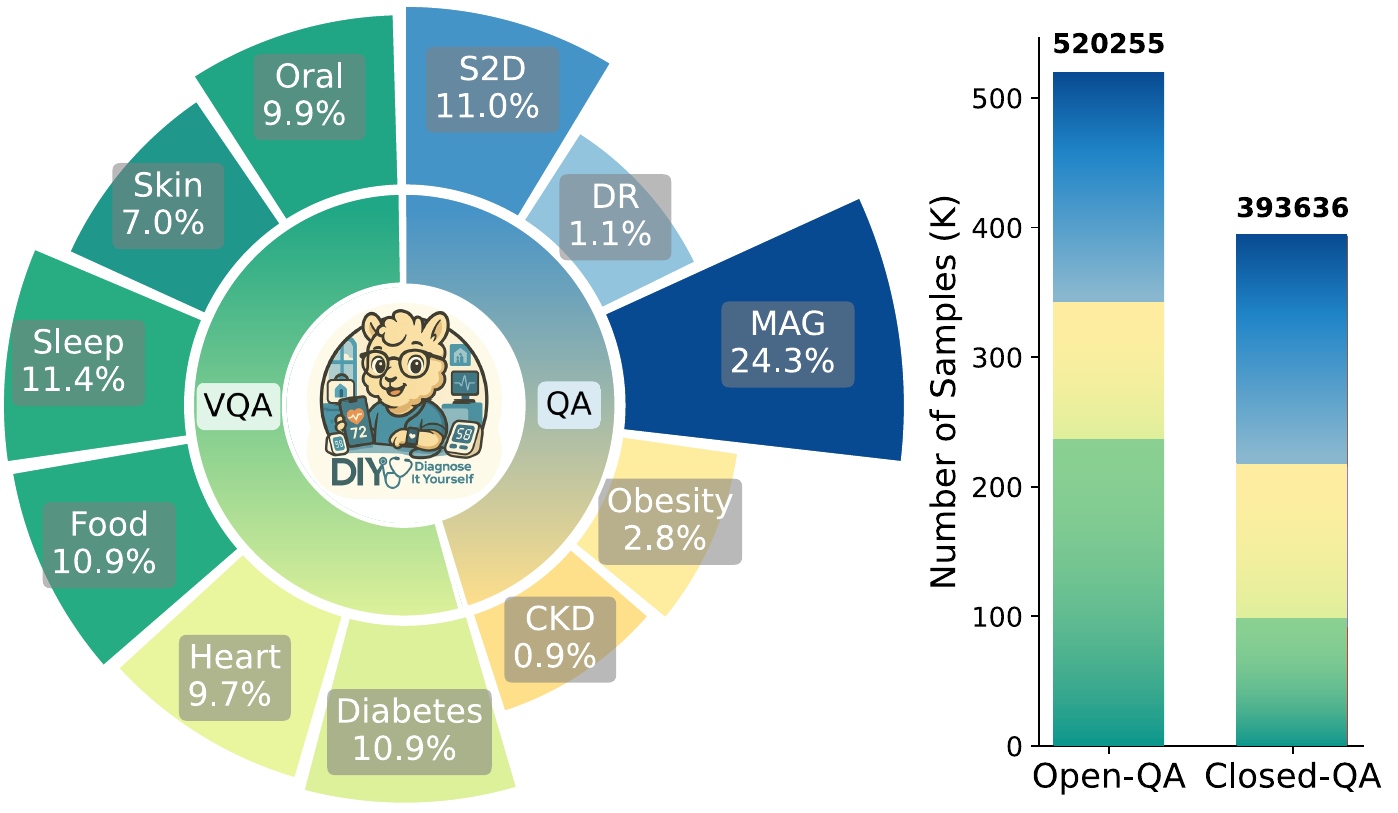}
        \caption{Data statistics of \Dataset}
        \vspace{-6mm}
        \label{fig:dataset-stats}
    \end{center}
\end{figure}

% To support these tasks, we construct \Dataset by collecting and integrating \blue{private datasets from three organizations}
% %% cs: our own dataset: HPB, AUSNUT and NUS
% with 20 publicly available data sources,
% such as Kaggle, PhysioNet, and Figshare.
% Each dataset focuses on a specific medical task and contains data ranging from patient demographics, vital signs, and laboratory test results to medical conversations, questionnaires, and other clinically relevant records. Each task is aligned with one or more source datasets that have been adapted to the home care setting.
% The selection prioritizes modalities commonly observed in practice, including natural language symptom descriptions, wearable-derived signals (for heart and sleep health), and mobile-captured images (for dietary intake, skin conditions, and oral health). Data statistics are summarized in Figure~\ref{fig:dataset-stats}, with task abbreviations defined in Table~\ref{tab:home_tasks}.
% Details on task design and dataset construction are provided in Appendix~\ref{sec:appendix-dataset}.

To support these tasks, we construct \Dataset by integrating \blue{private datasets from three organizations} with 20 publicly available sources, including Kaggle, PhysioNet, and Figshare. Each dataset focuses on a specific medical task and covers diverse data types, such as demographics, vital signs, laboratory results, medical conversations, and questionnaires. Each task is aligned with one or more source datasets that have been adapted to the home care setting via \Engine. We prioritize commonly observed modalities, including natural language symptom descriptions, wearable-derived signals (e.g., heart and sleep), and mobile-captured images (e.g., diet, skin, and oral health). Dataset statistics are summarized in Figure~\ref{fig:dataset-stats}, with task abbreviations in Table~\ref{tab:home_tasks}, and additional details in Appendix~\ref{sec:appendix-dataset}.
\vspace{-13pt}
\section{\Name}
\label{sec:model}
\vspace{-1mm}
\subsection{Multimodal Perception Unification}
\label{subsec:multimodal perception unification}
Home care scenarios involve heterogeneous data sources, such as food images, skin images, and textual symptom descriptions, among others. To enable consistent reasoning across such diverse modalities, \Name designs a multimodal perception unification mechanism that projects visual and textual inputs into a shared semantic space.

\noindent
\textbf{Visual Encoding.} 
Given an input image $\mathcal{I} \in \mathbb{R}^{H \times W \times 3}$, we employ a pretrained vision encoder $\mathcal{E}_v(\cdot)$
to extract a sequence of patch-level representations:
\begin{align}
\mathcal{V} = \mathcal{E}_v(\mathcal{I}) \in \mathbb{R}^{L_v \times d_v}
\end{align}
where $L_v$ denotes the number of visual tokens and $d_v$ represents the visual embedding dimension.

\noindent
\textbf{Textual Encoding.} 
For a textual input $\mathcal{T} = \{t_1, \ldots, t_{|\mathcal{T}|}\}$, where $t_i\in \mathcal{V}_{txt}$, and $\mathcal{V}_{txt}$ represents the vocabulary of the backbone language model, we use a pretrained tokenizer and embedding layer $\mathcal{E}_t(\cdot)$. 
This yields:
\begin{align}
\mathcal{U} = \mathcal{E}_t(\mathcal{T}) \in \mathbb{R}^{L_t \times d}
\end{align}
where $L_t$ denotes the token length and $d$ is the language embedding dimension.

\noindent
\textbf{Modality Projection and Unification.} 
To align heterogeneous modalities within a shared semantic space, we introduce a learnable projection function $\mathcal{P}_v: \mathbb{R}^{d_v} \to \mathbb{R}^{d}$ that maps visual embeddings into the language embedding space. The unified multimodal representation is formulated as:
\begin{align}
\mathcal{Z} = \big[ \mathcal{P}_v(\mathcal{V}); \mathcal{U} \big] \in \mathbb{R}^{(L_v+L_t) \times d}
\end{align}
where $[\,;\,]$ denotes token concatenation.
% and $d$ is the shared embedding dimension in \Name.
This process defines a heterogeneous-to-homogeneous interface:
$\Phi: (\mathcal{I}, \mathcal{T}) \mapsto \mathcal{Z}$, which ensures that both visual and textual home care signals are embedded within a coherent semantic space. The resulting unified representation $\mathcal{Z}$ is subsequently provided as input to the backbone language model $\mathcal{M}_{\text{LLM}}$ for downstream adaptation and generation.

\begin{figure*}
\centering
\vspace{-2mm}
\includegraphics[width=0.88\textwidth]{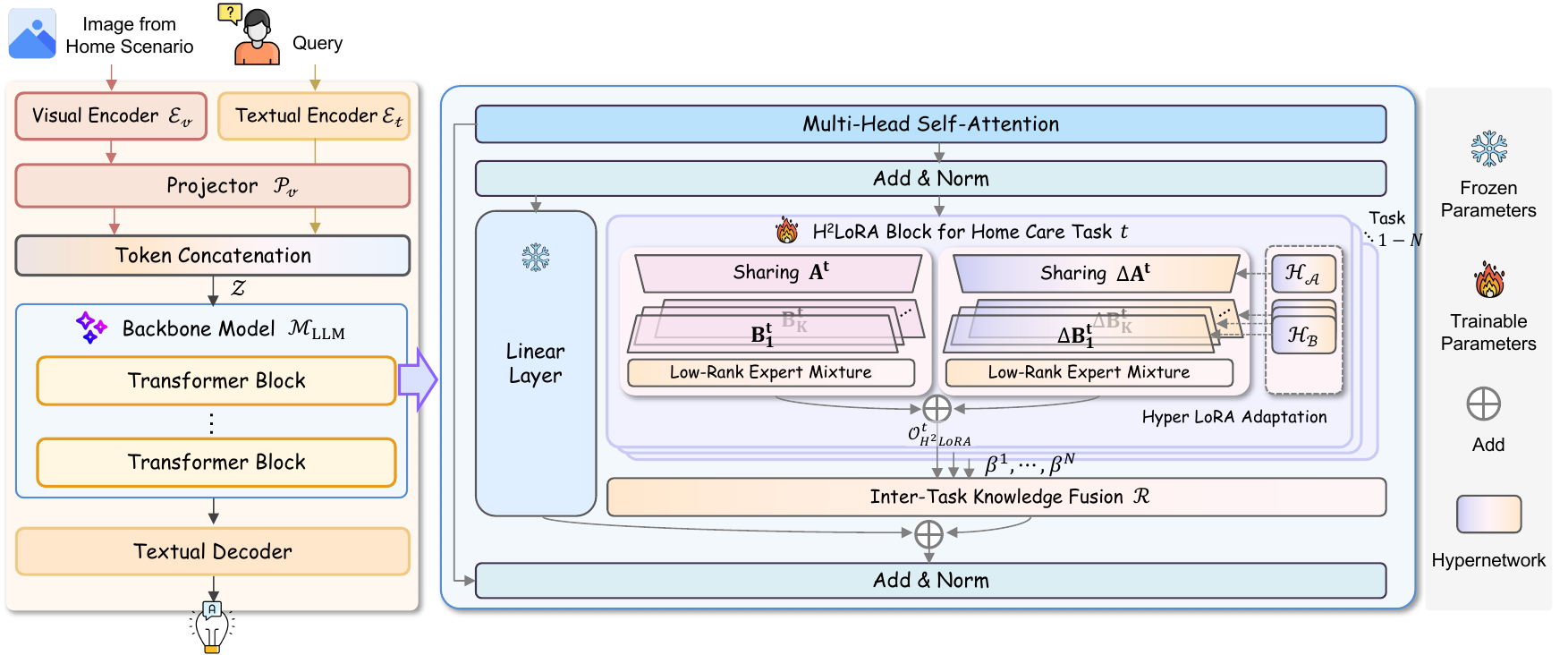}
\vspace{-2mm}
\caption{Model architecture of \Name, where H$^2$LoRA integrates Shared Low-Rank Expert Mixture with Hyper LoRA Adaptation
to balance task generalization and instance-level personalization.}
\vspace{-16pt}
\label{fig:model}
\end{figure*}

\subsection{Hybrid Hyper Low-Rank Adaptation}
\label{subsec:hybrid hyper low-rank adaptation}

While LLMs provide strong general reasoning capabilities, they generally lack the domain- and task-specific specialization required in home care scenarios.
PEFT techniques~\citep{ding2023parameter}, particularly Low-Rank Adaptation (LoRA)~\citep{hu2022lora}, offer a scalable approach by introducing trainable low-rank adapters into otherwise frozen pretrained weights. 
However, conventional LoRA strategies face inherent limitations: (i) allocating a distinct adapter for each task restricts cross-task knowledge sharing, whereas (ii) relying on a fully shared adapter overlooks the fine-grained task-specific distinctions.
To overcome these limitations, we propose Hybrid Hyper Low-Rank Adaptation (H$^2$LoRA), a
novel mechanism designed to enable efficient parameter sharing while retaining adaptive task-specific specialization.

Given the unified multimodal embedding $\mathcal{Z}$, the backbone language model $\mathcal{M}_{\text{LLM}}$ produces task-aware outputs $\mathcal{O}$ by combining frozen parameters $\Theta$ with task-adaptive parameters $\Theta_{H^2L}$ introduced through our H$^2$LoRA mechanism:
\begin{equation}
\small
\mathcal{O}_{H^2LoRA} = \mathcal{M}_{\text{LLM}}(\mathcal{Z}; \Theta, \Theta_{H^2L}), \,\, \Theta_{H^2L} = \{ \mathcal{A}, \mathcal{B}, \mathcal{R} \}
\end{equation}
where $\mathcal{A} = \{\mathbf{A}^t,\Delta\mathbf{A}^t\}_{t=1}^N$ and $\mathcal{B} = \{\mathbf{B}^t,\Delta\mathbf{B}^t\}_{t=1}^N$ denote the task-specific low-rank parameters associated with the home care tasks $T=\{1,\cdots,N\}$.
The routing parameters $\mathcal{R}$ further integrate task-level outputs, as detailed in Section~\ref{subsec:inter-task knowledge fusion}.
To achieve cross-task knowledge sharing and task-level specialization simultaneously, the task-specific pair $(\mathbf{A}^t, \mathbf{B}^t)$ and $(\Delta\mathbf{A}^t, \Delta\mathbf{B}^t)$, $t\in T$ are internally structured through two complementary mechanisms: \textit{Shared Low-Rank Expert Mixture} and \textit{Hyper LoRA Adaptation}.

\textbf{Shared Low-Rank Expert Mixture.}
At the task level, consider a weight matrix $\Theta \in \mathbb{R}^{d_{\text{out}} \times d_{\text{in}}}$ in the backbone model $\mathcal{M}_{\text{LLM}}$. 
H$^2$LoRA augments this parameter with a shared low-rank projection $\mathbf{A}^t \in \mathbb{R}^{d_{\text{out}} \times r}$ and a set of $K$ expert matrices $\{\mathbf{B}^t_1,\ldots,\mathbf{B}^t_K\}$, where $r \ll \min(d_{\text{out}}, d_{\text{in}})$.
Inspired by the Mixture of Experts (MoE) paradigm, a task embedding $\mathcal{Z} \in \mathbb{R}^{d}$ is processed by a routing layer to generate expert weights $\mathcal{W}^t \in \mathbb{R}^K$, typically normalized with a softmax function.
To interface with the low-rank structure, these weights are expanded along the rank dimension as
\begin{equation}
\vspace{-5pt}
\hat{\mathcal{W}}^{t} = K\mathcal{W}^t/r \otimes \mathbf{1}_r
% \vspace{-pt}
\end{equation}
where $\otimes$ denotes the replication operation.
The task-adaptive projection is obtained as a combination of experts:
\begin{equation}
\vspace{-5pt}
\mathbf{B}^t = \hat{\mathcal{W}}^{t} \odot \text{Concat}(\mathbf{B}^t_1, \ldots, \mathbf{B}^t_K)
\label{eq:B_concat}
% \vspace{-5pt}
\end{equation}
where $\odot$ denotes element-wise multiplication.
This design employs $\mathbf{A}^t$ as a shared anchor across $K$ expert matrices $\mathbf{B}^t_k$, encouraging subspace alignment. Meanwhile, the MoE-driven mixture of $\mathbf{B}^t$ matrices provides the flexibility required for task-level specialization. During the subsequent multi-task training phase (Sec.~\ref{subsec:training pipeline} Stage 4), this structure further facilitates cross-task knowledge sharing.
By integrating shared LoRA with expert routing, the shared low-rank expert mixture strikes a principled balance between efficiency and expressivity, providing greater adaptation capacity than either fully shared or fully task-isolated alternatives.

\textbf{Hyper LoRA Adaptation.}
H$^2$LoRA further incorporates hyper LoRA adaptation to capture the task-aware, instance-specific variations that frequently arise in home care scenarios, such as 
personalization across patients.
In this design, both the shared projection $\mathbf{A}^t$ and the expert matrices $\mathbf{B}^t_k$ are equipped with instance-dependent offsets, dynamically generated by dedicated hypernetworks:
\begin{align}
\Delta \mathbf{A}^t = \mathcal{H}_A(\mathcal{Z}), 
\,\, 
\Delta \mathbf{B}^t_k = \mathcal{H}_B(\mathcal{Z}), \,\, k=1,\ldots,K
\end{align}
The offset for MoE-driven mixture $\Delta\mathbf{B}^{t}$ is then computed via Eq.(\ref{eq:B_concat}).
Thus, the task-level output of H$^2$LoRA is
$\mathcal{O}^t_{H^2LoRA}=\mathcal{Z}\mathbf{A}^{t}\mathbf{B}^{t}+\mathcal{Z}\Delta\mathbf{A}^{t}\Delta\mathbf{B}^{t}$.
By conditioning the hypernetworks on $\mathcal{Z}$, this formulation renders the adaptation instance-aware, enabling flexible personalization that extends beyond coarse task-level specialization and better reflects the heterogeneity of home care contexts.

\vspace{-1mm}
\subsection{Inter-Task Knowledge Fusion}
\label{subsec:inter-task knowledge fusion}
\vspace{-1mm}

While H$^2$LoRA equips each task with a specialized adapter, healthcare tasks in home care scenarios are generally strongly correlated.
For instance, dietary patterns directly affect the risks of diabetes and obesity, while early symptom recognition provides essential context for subsequent drug recommendation and personalized advice generation~\citep{hu2011globalization,mozaffarian2016dietary}.
To exploit such inter-dependencies, we introduce an inter-task knowledge fusion mechanism, where a global soft-MoE router $\mathcal{R}$ dynamically integrates the outputs from the $N$ task-specific H$^2$LoRA blocks.
Specifically, the router $\mathcal{R}$ assigns mixture weights $\beta$ conditioned on the shared embedding $\mathcal{Z}$:
\begin{align}
\beta =(\beta^1, \ldots, \beta^N)=\mathcal{R}(\mathcal{Z}), \quad \beta^t \geq 0, \; \sum_{t=1}^N \beta^t = 1
\end{align}
The overall H$^2$LoRA update is then expressed as: $\mathcal{O}_{H^2LoRA}=\mathcal{Z}\Theta+\sum_{t=1}^N\beta^t\mathcal{O}^t_{H^2LoRA}$, achieving a principled balance among efficiency, specialization, and personalization.
Within each task, the shared $\mathbf{A}^t$ promotes common representation learning, the expert matrices $\mathbf{B}^t_k$ capture task-specific signals, and hypernetwork-driven offsets provide instance-aware modulation.
% for fine-grained variation.
Beyond task-level adaptation, the inter-task fusion layer treats each H$^2$LoRA block as an expert and leverages the global soft router $\mathcal{R}$ to integrate them contextually, thereby exploiting correlations across healthcare tasks.
\blue{As shown in Figure~\ref{fig:model}, this integration enables coherent, contextually grounded, and personalized health responses across diverse scenarios while adhering to strict parameter budgets.
% Prior approaches account for only partial aspects: 
In comparison, MoELoRA~\citep{luo2024moelora} emphasizes feature-level expert diversity with limited cross-task coordination, whereas HyperLoRA~\citep{lv2024hyperlora} directly generates task-specific LoRA weights via a hypernetwork, which can complicate optimization and limit task-wise parameter sharing.}
% H$^2$LoRA broadens the design space by integrating expert mixtures, a residual hypernetwork-driven modulation, and cross-task soft fusion.

\vspace{-1mm}
\subsection{Training Pipeline}
\label{subsec:training pipeline}
\vspace{-1mm}

To optimize \Name, we design a four-stage training pipeline that progressively aligns modalities, adapts the backbone to the medical domain, specializes task-level experts, and integrates them into a unified framework.

\vspace{-1mm}
\textbf{Stage 1: Cross-Modal Alignment.} 
Using PubMedVision~\citep{chen2024towards} and LLaVA-558k~\citep{liu2024improved}, we begin by training the projector $\mathcal{P}_v$, which maps visual embeddings into the shared semantic space, ensuring robust alignment between visual and language representations.

\vspace{-1mm}
\textbf{Stage 2: Medical Domain Adaptation.} 
We then perform supervised fine-tuning by jointly training the projector $\mathcal{P}_v$ and the backbone $\mathcal{M}_{\text{LLM}}(\cdot; \Theta)$ on our curated \Dataset dataset covering 11 tasks.
To balance efficiency and coverage, we uniformly sample 10\% of training data per task to derive a fixed subset, allowing the backbone to acquire medical knowledge while reducing overfitting risk. 

\vspace{-1mm}
\textbf{Stage 3: Task-Specific Expert Training.} 
For each task, we train a dedicated H$^2$LoRA block, parameterized by $(\mathbf{A}^t,\Delta\mathbf{A}^t,\mathbf{B}^t,\Delta\mathbf{B}^t)$,
on its respective subset of \Dataset, while keeping all other parameters fixed.
After individual training,
we introduce a hard-MoE layer that activates a single expert at a time, and jointly optimize all experts on the full multi-task training set of \Dataset, ensuring integration under shared supervision.

\vspace{-1mm}
\textbf{Stage 4: Cross-Task Knowledge Transfer.} 
Finally, we fine-tune the task-specific H$^2$LoRA expert blocks while replacing the hard-MoE layer with a global soft-MoE router $\mathcal{R}$ on the multi-task training set of \Dataset.
This stage facilitates cross-task knowledge transfer, enabling the model to exploit inter-task correlations while maintaining parameter efficiency.

\vspace{-2mm}
\section{\Benchmark}
\label{sec:benchmark}
\vspace{-1mm}
\textbf{Standardized Benchmark for Home Care AI.}
To enable a rigorous and fair evaluation of foundation models in non-clinical settings, we introduce \Benchmark, a benchmark dedicated to 11 real-world home care tasks defined in \Dataset. These tasks span three major categories: personalized health management, chronic disease risk assessment, and daily health monitoring.
The tasks are formulated in both open-QA and closed-QA formats, reflecting the dual requirements of home care: naturalistic dialogue for adaptive advice and structured reasoning for actionable decision support.
\blue{\Benchmark is derived from the test set of \Dataset and comprises 12,167 examples in total. For each task, we randomly sample 1\% of the data for evaluation, with a minimum of 1,000 examples for small datasets to ensure statistical reliability and adequate task coverage.
% For small datasets, a minimum of 1,000 samples is used to ensure statistical reliability and sufficient task coverage.
% To ensure representativeness, 
The samples are balanced across task categories, input modalities, and disease categories, resulting in a representative and standardized evaluation setting.
% basis for evaluation in home-based health management.
As the first benchmark designed for home care AI, \Benchmark bridges the gap left by existing hospital-centric evaluations and lays a foundation for assessing personalized health assistance beyond clinical settings.}

\textbf{Multi-Dimensional Evaluation Protocol.}
We evaluate both general-domain LVLMs and medical-specific LVLMs as baselines, establishing reference performance through a multi-dimensional evaluation suite.
For closed-QA, we adopt accuracy (ACC) as the primary measure of diagnostic precision and complement it with Matthews Correlation Coefficient (MCC).
For open-QA, we employ two complementary groups of metrics. \textit{Content-level} metrics, including F1-RadGraph (F1-Rad)~\citep{yu2023evaluating}
and F1-BioBERT (F1-Bio)~\citep{lee2020biobert}, quantify the semantic and biomedical fidelity of generated responses.
\textit{Language-level} metrics, namely BLEU~\citep{papineni2002bleu} and ROUGE-L (RL)~\citep{lin2003automatic}, assess surface-level fluency and textual overlap with ground-truth answers,
providing a balanced assessment for medical applications. Please refer to Appendix~\ref{sec:appendix-evaluation-metrics} for details of evaluation metrics.

\vspace{-1mm}
\section{Experiments}
\label{sec:exps}
\vspace{-1mm}
\subsection{Experimental Setup}
\textbf{Data Details.}
Following the \Benchmark protocol, we evaluate \Name on \Dataset, a multimodal QA dataset of approximately 900K samples spanning 11 home care tasks, split into training and test sets at a 99:1 ratio with strict user-level separation to prevent data leakage.

\vspace{-1mm}
\textbf{Baselines.}
We evaluate \Name against a broad set of baselines, including state-of-the-art generalist models (e.g., LLaVA-1.5~\citep{liu2023visual}, InstructBLIP~\citep{dai2023instructblip}, Llama 3.2~\citep{dubey2024llama},  Yi-VL~\citep{young2024yi}, InternVL3~\citep{zhu2025internvl3}, Qwen2.5-VL~\citep{bai2025qwen2}, Qwen3-VL~\citep{yang2025qwen3}, Gemma 3~\citep{team2025gemma}, Claude 3 Haiku~\citep{anthropic2024claude} and GPT-4o Mini~\citep{achiam2023gpt}) and medical-specific models (e.g., LLaVA-Med v1.5~\citep{li2023llava}, Med-Flamingo~\citep{moor2023med}, HuatuoGPT-Vision~\citep{chen2024huatuogpt}, MedGemma~\citep{sellergren2025medgemma}, HealthGPT~\citep{lin2025healthgpt}, Med-R1~\citep{lai2025med}, Lingshu~\citep{xu2025lingshu}, and MedVLM-R1~\citep{pan2025medvlm}).
Experiments are conducted on 11 home care tasks under open-QA settings, and 10 tasks are included for closed-QA, as the drug recommendation task is inherently multi-label and thus better suited to open-QA format.
Please refer to Appendix~\ref{sec:appendix-implementation-setting} for \Name implementation.

% User Feedback, Subgroup Study and Inter-rater Agreement Analysis are reported in Appendix~\ref{sec:appendix-exps}.

\begin{table*}[ht]
\centering
\setlength{\tabcolsep}{4.2pt}
\renewcommand{\arraystretch}{1}
\caption{Comparison of \Name with baselines under \textit{closed-QA} settings in \Benchmark.}
\vspace{-2mm}
\label{tab:exp-main-close}
\resizebox{1.0\textwidth}{!}{
\begin{threeparttable}
{\setlength{\tabcolsep}{3pt}
\begin{tabular}{l*{1}{cc}{cc|}*{3}{cc}{cc|}*{4}{cc}{|cc}}
\toprule
\multirow{2}{*}{Model} &
\multicolumn{2}{c}{S2D} &
\multicolumn{2}{c|}{MAG} &
\multicolumn{2}{c}{Diabetes} &
\multicolumn{2}{c}{Obesity} &
\multicolumn{2}{c}{Heart} &
\multicolumn{2}{c|}{CKD} &
\multicolumn{2}{c}{Food} &
\multicolumn{2}{c}{Sleep} &
\multicolumn{2}{c}{Skin} &
\multicolumn{2}{c}{Oral} &
\multicolumn{2}{|c}{Avg.}\\
\cmidrule(lr){2-3}\cmidrule(lr){4-5}\cmidrule(lr){6-7}\cmidrule(lr){8-9}\cmidrule(lr){10-11}\cmidrule(lr){12-13}\cmidrule(lr){14-15}\cmidrule(lr){16-17}\cmidrule(lr){18-19}\cmidrule(lr){20-21}\cmidrule(lr){22-23}
 & ACC & MCC & ACC & MCC & ACC & MCC & ACC & MCC & ACC & MCC & ACC & MCC & ACC & MCC& ACC & MCC & ACC & MCC& ACC & MCC& ACC & MCC\\
\midrule
\multicolumn{23}{c}{\textit{General Domain Models}} \\
\midrule
LLaVA-1.5-7B& 52.36 & 39.22 & 36.74 & 19.13 & 58.90 & 39.72 & 61.90 & 47.97 & \underline{48.94} & \underline{32.79} & 81.44 & 75.27 & 80.69 & 74.29 & 21.35 & -0.56 & 40.21 & 16.68 & 52.35 & 32.29&52.87&35.39\\
InstructBLIP-7B& 25.67 & 4.44 & 4.87 & 0.89 & 5.52 & 5.38 & 15.40 & 6.52 & 0.18 & 2.60 & 5.15 & 4.01 & 42.92 & 35.26 & 9.42 & 0.65 & 14.17 & 2.89 & 21.37 & 5.43 &14.35&8.54\\
Llama 3.2-11B& 60.99 & 49.89 & 46.48 & 33.66 & 64.23 & 47.03 & 54.12 & 39.58 & 25.35 & 6.96 & 56.91 & 48.42 & 74.03 & 66.96 & 20.96 & 2.53 & 38.54 & 17.55 & 41.03 & 25.97&47.42&33.12\\
Yi-VL-6B& 75.56 & 67.39 & 48.49 & 33.49 & 75.27 & 59.01 & 65.47 & 52.78 & 48.59 & 31.71 & 73.20 & 64.38 & 76.39 & 68.63 & 25.38 & 7.65 & 45.83 & 19.12 & 69.23 & 51.75&59.46&46.08\\
InternVL3-8B& \underline{85.42} & \underline{80.54} & \textbf{59.73} & \underline{48.16} & 88.61 & 79.94 & 70.92 & 59.61 & 37.15 & 16.29 & 72.81 & \underline{91.88} & 85.41 & 80.53 & 21.54 & 2.10 & 74.38 & 51.45 & \underline{91.88} & \underline{85.44}&68.79&59.59\\
Qwen2.5-VL-7B& 46.82 & 29.56 & 57.89 & 46.82 & 70.28 & 52.03 & 58.01 & 43.28 & 33.80 & 12.69 & 64.74 & 55.10 & 73.18 & 64.58 & 20.38 & 3.78 & 54.79 & 27.87 & 47.44 & 26.94&51.77&35.50\\
Qwen3-VL-8B & 77.82 & 70.57 & 47.32 & 32.75 & 86.12 & 75.32 & 69.67 & 58.10 & 30.99 & 8.10 & 81.44 & 75.39 & 82.19 & 76.50 & 23.65 & 5.01 & 62.71 & 36.14 & 80.56 & 68.11 & 64.25 & 50.60 \\
Gemma 3-4B& 66.94 & 56.32 & 46.14 & 30.47 & 67.08 & 47.6 & 65.94 & 53.18 & 34.68 & 12.56 & 72.81 & 66.45 & 86.05 & 81.41 & 15.96 & -4.25 & 48.13 & 27.06 & 66.45 & 54.28&57.02&42.51\\
Claude 3 Haiku& 29.57 & 6.12 & 40.77 & 24.67 & 73.67 & 56.17 & 60.65 & 46.16 & 43.84 & 25.98 & 79.38 & 72.72 & 54.72 & 41.65 & 33.65 & 17.77 & 54.79 & 28.02 & 78.21 & 65.01&54.26&38.98\\
GPT-4o Mini& 72.07 & 62.85 & 59.23 & 47.93 & 75.98 & 60.63 & 62.36 & 50.36 & 13.03 & -5.17 & 70.97 & 68.59 & \underline{86.48} & \underline{82.34} & 26.73 & 9.03 & 60.63 & 32.67 & 68.59 & 50.26&59.61&45.95\\
\midrule
\multicolumn{23}{c}{\textit{Medical Domain Models}} \\
\midrule
LLaVA-Med v1.5-7B & 68.17 & 58.36 & 32.55 & 15.28 & 35.59 & 27.00 & 58.48 & 45.85 & 7.22 & -6.96 & 48.04 & 39.72 & 37.12 & 22.08 & 20.19 & 0.08 & 41.88 & 21.82 & 46.37 & 30.96 &38.73&26.08\\
Med-Flamingo-7B& 28.54 & 6.04 & 16.78 & 0.97 & 11.74 & 5.47 & 17.42 & 3.60 & 20.42 & -3.50 & 10.84 & 13.46 & 28.11 & 6.10 & 18.85 & -0.48 & 3.33 & -2.03 & 13.46 & 4.00&16.95&3.36\\
HuatuoGPT-Vision-7B& 81.11 & 74.80 & 53.69 & 40.53 & 87.19 & 77.80 & 67.96 & 56.04 & 42.78 & 25.39 & 80.00 & 73.48 & 85.19 & 80.28 & 22.12 & 1.71 & 76.46 & 56.42 & 90.81 & 83.79 &68.08&58.76\\
MedGemma-4B& 61.40 & 49.26 & 53.69 & 40.35 & 76.16 & 61.71 & 70.30 & 58.68 & \underline{48.94} & 32.66 & 70.00 & 86.32 & 70.82 & 61.03 & 15.58 & -4.46 & 58.13 & 31.85 & 86.32 & 76.09&61.13&49.35\\
HealthGPT-3.8B & 77.41 & 70.13 & 54.70 & 41.55 & 89.50 & 81.14 & 71.85 & 61.08 & 40.14 & 20.20 & 82.68 & 77.06 & 75.11 & 66.79 & 25.96 & 9.45 & \underline{85.42} & \underline{66.88} & 87.82 & 78.79 &68.50&58.38\\
Med-R1-2B& 77.41 & 69.87 & 48.83 & 33.88 & 84.16 & 72.20 & 65.94 & 52.23 & 37.32 & 17.19 & \underline{84.33} & 79.21 & 83.26 & 77.63 & \underline{46.73} & \underline{34.00} & 64.58 & 38.06 & 90.60 & 83.78&67.80&56.94\\
Lingshu-7B &80.08 & 73.79 & 57.72 & 46.21 & \underline{89.86} & \underline{82.02} & \underline{72.63} & \underline{61.63} & 40.49 & 20.39 & 78.44 & 91.24 & 82.62 & 76.85 & 30.96 & 13.64 & 83.33 & 63.93 & 91.24 & 84.57&\underline{70.74}&\underline{61.43}\\
MedVLM-R1-2B &75.36 & 67.45 & 37.75 & 21.22 & 64.95 & 44.69 & 59.72 & 43.65 & 27.64 & 9.50 & 69.82 & 80.34 & 84.33 & 79.37 & 25.96 & 9.00 & 48.75 & 22.22 & 80.34 & 66.47&57.46&44.39\\
\rowcolor{blue!10} \Name-3.8B & \textbf{97.74} & \textbf{96.98} & \textbf{59.73} & \textbf{48.26} & \textbf{95.02} & \textbf{90.76} & \textbf{85.23} & \textbf{78.89} & \textbf{83.10} & \textbf{77.57} & \textbf{99.73} & \textbf{99.57} & \textbf{97.85} & \textbf{97.14} & \textbf{51.90} & \textbf{40.02} & \textbf{98.13} & \textbf{95.21} & \textbf{99.57} & \textbf{99.19}&\textbf{86.80}&\textbf{82.36}\\
\bottomrule
\end{tabular}
\vspace{-2mm}
}
\end{threeparttable}
}
\end{table*}

\begin{table*}[ht]
\centering
\setlength{\tabcolsep}{4.2pt}
\renewcommand{\arraystretch}{1}
\caption{Comparison of \Name with baselines under \textit{open-QA} settings in \Benchmark.}
\vspace{-2mm}
\label{tab:exp-main-open}
\resizebox{1.0\textwidth}{!}{
\begin{threeparttable}
{\setlength{\tabcolsep}{3pt}
\begin{tabular}{l*{2}{cc}{cc|}*{3}{cc}{cc|}*{4}{cc}{|cc}}
\toprule
\multirow{2}{*}{Model} &
\multicolumn{2}{c}{S2D} &
\multicolumn{2}{c}{DR} &
\multicolumn{2}{c|}{MAG} &
\multicolumn{2}{c}{Diabetes} &
\multicolumn{2}{c}{Obesity} &
\multicolumn{2}{c}{Heart} &
\multicolumn{2}{c|}{CKD} &
\multicolumn{2}{c}{Food} &
\multicolumn{2}{c}{Sleep} &
\multicolumn{2}{c}{Skin} &
\multicolumn{2}{c}{Oral}&
\multicolumn{2}{|c}{Avg.}\\
\cmidrule(lr){2-3}\cmidrule(lr){4-5}\cmidrule(lr){6-7}\cmidrule(lr){8-9}\cmidrule(lr){10-11}\cmidrule(lr){12-13}\cmidrule(lr){14-15}\cmidrule(lr){16-17}\cmidrule(lr){18-19}\cmidrule(lr){20-21}\cmidrule(lr){22-23}\cmidrule(lr){24-25}
 & F1-Bio & RL & F1-Bio & RL & F1-Bio & RL & F1-Bio & RL & F1-Bio & RL & F1-Bio & RL& F1-Bio & RL& F1-Bio & RL& F1-Bio & RL& F1-Bio & RL& F1-Bio & RL& F1-Bio & RL\\
\midrule
\multicolumn{25}{c}{\textit{General Domain Models}} \\
\midrule
LLaVA-1.5-7B& 68.43 & 5.25 & 63.29 & 3.97 & 78.04 & 34.92 & 76.83 & 14.84 & 72.23 & 11.58 & 72.65 & 11.22 & 74.98 & 11.40 & 66.29 & 3.32 & 73.88 & 11.58 & 77.41 & 23.81 & 79.29 & 22.75&73.03&14.06\\
InstructBLIP-7B& 58.24 & 2.28 & 58.96 & 4.02 & 69.21 & 5.91 & 66.46 & 8.75 & 60.68 & 4.48 & 69.02 & 2.65 & 70.33 & 17.02 & 40.81 & 11.78 & 65.45 & 8.35 & 45.91 & 8.73 & 63.37 & 11.42&60.77&7.76\\
Llama 3.2-11B& 66.22 & 4.67 & 62.17 & 4.07 & 79.94 & 40.98 & 73.20 & 12.09 & 70.51 & 11.56 & 67.53 & 7.71 & 73.56 & 12.19 & 68.87 & 6.55 & 69.37 & 7.27 & 73.25 & 16.46 & 73.85 & 15.89&70.77&12.68\\
Yi-VL-6B& 66.06 & 4.33 & 62.02 & 4.34 & \underline{83.99} & 34.86 & 77.49 & 16.62 & 71.49 & 11.19 & 72.69 & 12.84 & 76.58 & 14.70 & 71.59 & 9.14 & 74.24 & 13.53 & 77.11 & 22.78 & 78.88 & 22.40&73.83&15.16\\
InternVL3-8B& 63.63 & 2.85 & 60.97 & 3.03 & 76.97 & 33.01 & 75.89 & 17.78 & 67.9 & 8.24 & 68.91 & 9.26 & 73.01 & 8.63 & 67.64 & 4.92 & 66.46 & 7.14 & 74.06 & 21.78 & 74.61 & 20.09&70.00&12.43\\
Qwen2.5-VL-7B& 64.90 & 3.14 & 65.01 & 6.76 & 76.36 & 23.86 & 76.06 & 15.31 & 68.90 & 8.40 & 69.97 & 8.75 & 70.57 & 7.00 & \underline{78.00} & \underline{26.61} & 71.03 & 7.00 & 76.57 & 24.09 & 77.66 & 22.32&72.28&13.93\\
Qwen3-VL-8B & 62.31 & 2.67 & 60.36 & 2.74 & 81.28 & 27.08 & 73.67 & 12.31 & 68.95 & 9.59 & 65.82 & 6.78 & 67.19 & 7.18 & 66.59 & 6.30 & 68.57 & 6.54 & 69.59 & 14.24 & 71.35 & 15.24 & 68.70 & 10.06 \\
Gemma 3-4B& 62.48 & 1.84 & 60.57 & 2.14 & 71.17 & 26.14 & 69.55 & 8.14 & 65.87 & 6.18 & 64.31 & 5.43 & 64.38 & 3.17 & 67.66 & 5.44 & 65.14 & 5.06 & 68.28 & 10.09 & 68.74 & 10.30&66.20&7.63\\
Claude 3 Haiku& 67.24 & 4.85 & 63.37 & 4.90 & 79.39 & 35.84 & 78.02 & 17.87 & 72.23 & 11.48 & 72.06 & 12.36 & 72.31 & 11.05 & 67.04 & 3.77 & 73.81 & 10.90 & 77.30 & 22.14 & 78.66 & 22.92&72.86&14.37\\
GPT-4o Mini& 66.19 & 4.24 & 62.15 & 4.02 & 80.92 & \underline{45.17} & 75.58 & 15.09 & 71.01 & 11.90 & 68.57 & 8.73 & 75.51 & 12.64 & 70.66 & 10.14 & 70.84 & 8.44 & 74.92 & 22.87 & 75.08 & 21.09&71.95&14.94\\

\midrule
\multicolumn{25}{c}{\textit{Medical Domain Models}} \\
\midrule
LLaVA-Med v1.5-7B & \underline{72.11} & \underline{9.09} & \underline{65.98} & \underline{7.93} & 76.33 & 24.31 & 77.93 & 17.77 & \underline{74.94} & \underline{15.95} &\underline{74.66} & \underline{18.76} & 77.27 & \underline{17.14} & 70.75 & 4.97 & \underline{75.72} & \underline{19.08} & 78.58 & 26.32 & 79.12 & 24.82 &\underline{74.82}&\underline{16.88}\\
Med-Flamingo-7B& 59.70 & 2.51 & 56.57 & 3.36 & 59.98 & 7.80 & 65.47 & 8.70 & 57.95 & 5.80 & 56.98 & 2.44 & 62.64 & 5.24 & 51.91 & 0.45 & 54.83 & 3.71 & 61.87 & 8.14 & 62.43 & 7.49&59.12&5.06\\
HuatuoGPT-Vision-7B& 67.77 & 4.13 & 64.39 & 4.41 & 71.86 & 13.03 & 77.76 & 15.18 & 72.56 & 10.80 & 70.14 & 8.01 & 73.27 & 8.61 & 65.08 & 2.14 & 72.55 & 8.23 & 77.58 & 21.53 & 80.39 & 26.83&72.12&11.17\\
MedGemma-4B & 65.29 & 5.00 & 61.48 & 4.67 & 79.55 & 34.99 & 72.27 & 11.20 & 67.47 & 8.83 & 68.22 & 9.83 & 68.90 & 8.42 & 64.91 & 3.64 & 68.68 & 7.99 & 69.70 & 13.33 & 71.17 & 15.84 & 68.88 & 11.25 \\
HealthGPT-3.8B& 67.87 & 4.77 & 63.59 & 4.33 & 77.53 & 25.52 & 79.47 & 18.56 & 71.81 & 11.71 & 70.18 & 7.78 & 76.05 & 13.27 & 67.36 & 4.15 & 73.29 & 10.66 & \underline{80.34} & \underline{29.39} & \underline{82.92} & \underline{34.47}&73.67&14.96\\
Med-R1-2B& 64.42 & 3.53 & 61.54 & 3.20 & 76.02 & 27.94 & 74.94 & 14.22 & 68.68 & 9.43 & 63.15 & 5.45 & 74.72 & 10.99 & 66.30 & 3.26 & 65.26 & 5.15 & 75.11 & 20.71 & 77.25 & 20.32&69.76&11.29\\
Lingshu-7B & 68.65 & 5.40 & 64.71 & 5.11 & 74.23 & 18.95 & \underline{79.64} & \underline{19.14} & 72.95 & 13.42 & 71.19 & 10.34 & \underline{77.50} & 16.38 & 67.21 & 3.84 & 73.82 & 10.09 & 79.74 & 27.36 & 81.86 & 31.19&73.77&14.66\\
MedVLM-R1-2B & 64.58 & 3.75 & 61.01 & 3.61 & 77.12 & 26.26 & 74.8 & 13.86 & 70.47 & 11.28 & 68.86 & 9.62 & 75.34 & 13.98 & 69.82 & 8.73 & 72.08 & 10.49 & 75.62 & 20.80 & 76.04 & 18.47&71.43&12.80\\
\rowcolor{blue!10} \Name-3.8B &\textbf{86.45} & \textbf{45.41} & \textbf{75.47} & \textbf{22.10} & \textbf{89.62} & \textbf{60.20} & \textbf{87.45} & \textbf{44.64} & \textbf{84.44} & \textbf{42.97} & \textbf{82.26} & \textbf{33.01} & \textbf{90.57} & \textbf{64.80} & \textbf{97.92} & \textbf{90.28} &\textbf{85.55} & \textbf{43.33} & \textbf{88.31} & \textbf{55.79} & \textbf{92.68} & \textbf{70.65}&\textbf{87.34}&\textbf{52.11}\\
\bottomrule
\end{tabular}
\vspace{-5mm}
}
\end{threeparttable}
}
\end{table*}

\vspace{-5pt}
\subsection{Main Results}
\label{main_results}
The experimental results in closed- and open-QA settings are in Tables~\ref{tab:exp-main-close} and~\ref{tab:exp-main-open}, with supplementary results of F1-Rad and BLEU-1 provided in Appendix~\ref{sec:appendix-exps-main-metric}.
From these results, we derive several key observations.
\textbf{(i) State-of-the-art performance.}
Despite its compact model size, \Name delivers the best performance across all tasks, with an average improvement of 22.7\% in ACC under closed-QA and 16.7\% in F1-Bio under open-QA, substantially surpassing both general and medical domain models.
These results highlight \Name's strong capability in producing accurate choices and generating faithful, coherent responses.
\textbf{(ii) Narrow margin on MAG under closed-QA.} 
\Name's gain over the runner-up (i.e., InstructBLIP) is marginal on MAG task, as MAG is inherently general and depends more on wide medical knowledge than on specialized cues. Thus, models pretrained on large corpora transfer well, raising baselines and narrowing gaps.
Yet, \Name's advantage becomes pronounced on the open-QA setting of MAG, where broad knowledge alone is insufficient and robust medical reasoning is required.
\textbf{(iii) Limitations of existing Med-LVLMs.}
Current models, though pretrained on extensive medical corpora, largely overlook home care scenarios, where learning and prediction depend exclusively on data available outside clinical settings.
This omission results in suboptimal performance on home care tasks.
In contrast, we propose \Suite, 
% which integrates \Dataset, \Name, and \Benchmark, 
offering a unified resource and a strong baseline for advancing health management at home. \textbf{(iv) Heterogeneous task complexity.} \Benchmark reflects the diversity of real-world home-care scenarios, spanning both simple routine queries and tasks requiring deeper reasoning or domain expertise. Tasks such as S2D and CKD in closed-QA and Food and Oral in open-QA are relatively easy and perform well even with small models, whereas more complex tasks, including Sleep, Heart, Obesity, and DR in open-QA, remain challenging. Accordingly, easier tasks serve as preliminary indicators of model readiness for home care, while harder tasks reveal meaningful gaps and substantial opportunities for future improvement.

\subsection{Ablation Study and In-Depth Analysis}
\label{sec:ablation study and in-depth analysis}

\textbf{Ablation Study of H$^2$LoRA.} We propose H$^2$LoRA, which integrates a LoRA expert mixture with Hyper LoRA adaptation to enable \Name to acquire multi-task knowledge while 
retaining instance-aware personalization.
To assess the contribution of each component, we conduct a detailed ablation study, with results shown in Table~\ref{tab:abl_part_lora_components}. 
Relative to w/o H$^2$LoRA, incorporating either the Expert Mixture or Hyper LoRA leads to notable improvements across all metrics, confirming the effectiveness of both mechanisms. Further, combining the two within H$^2$LoRA yields consistent gains, showing that the mechanisms are both effective individually and complementary when integrated. 
Analysis of the number of experts is presented in Appendix~\ref{sec:appendix-exps-sensitivity}.

\begin{figure*}[!th]
    \centering
    \begin{minipage}{0.27\textwidth}
        \centering
        \includegraphics[width=\textwidth]{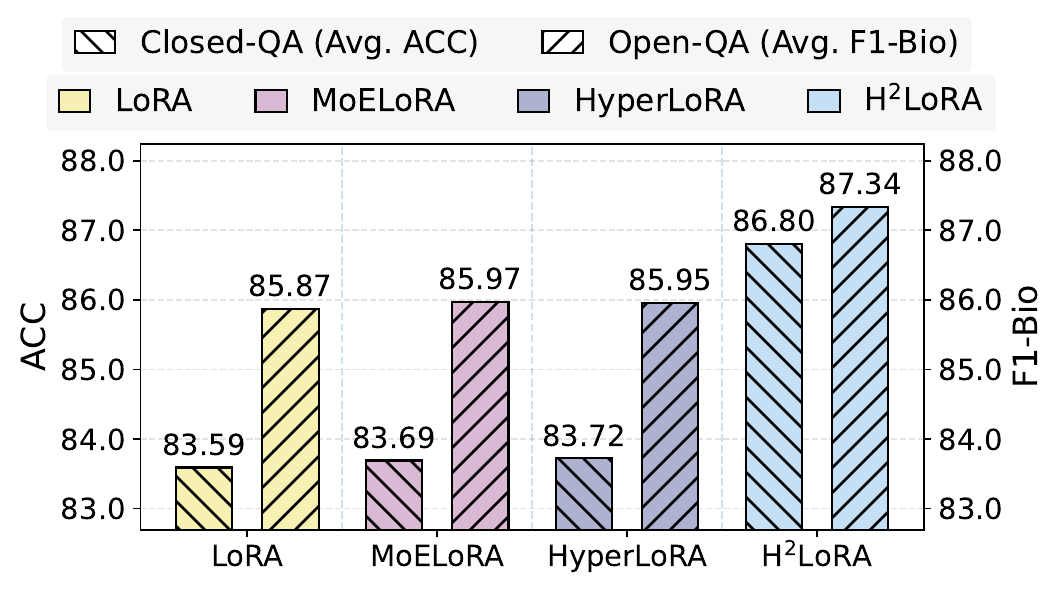}
        \vspace{-15pt}
        \caption{Comparison of H$^2$LoRA with existing LoRA methods.}
        \vspace{-10pt}
        \label{fig:abl_lora_comparison}
    \end{minipage}
    \begin{minipage}{0.45\textwidth}
        \centering
        \includegraphics[width=\textwidth]{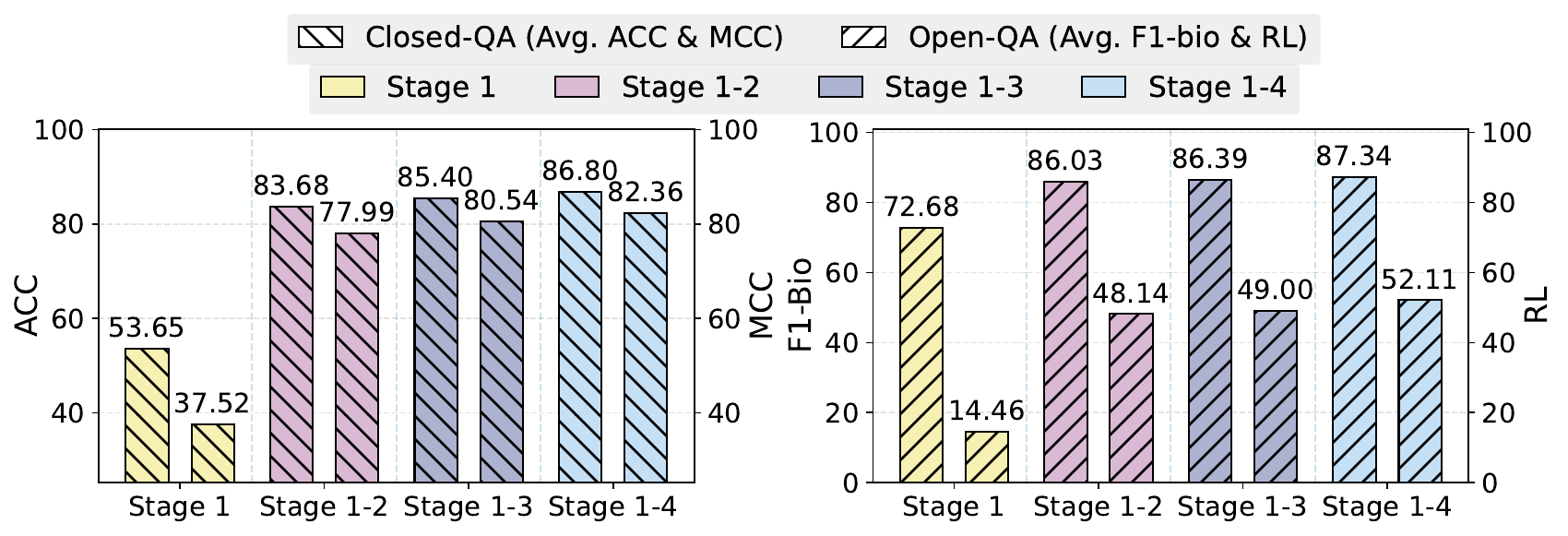}
        \vspace{-15pt}
        \caption{Performance across training stages.}
        \label{fig:abl_train_stage}
    \end{minipage}
    \begin{minipage}{0.25\textwidth}
        \includegraphics[width=0.9\linewidth]{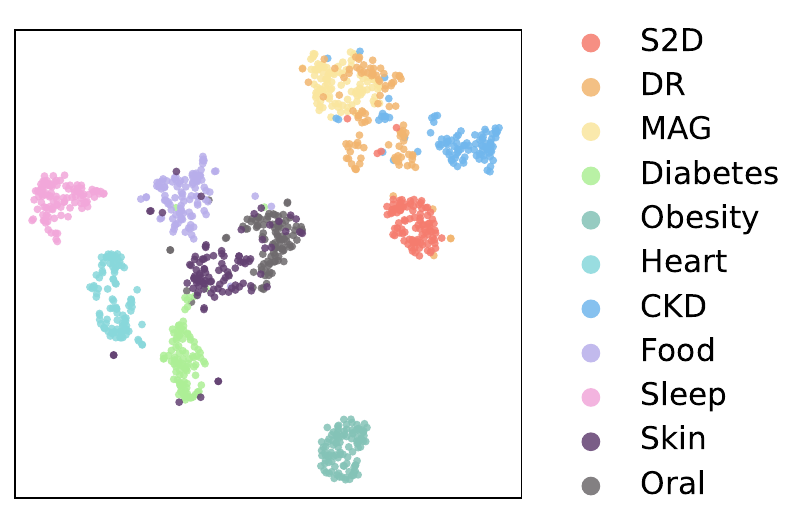}
        \caption{Visualization of representations from \Name.}
        \label{fig:vis_hyper}
        \vspace{-12pt}
    \end{minipage}
\end{figure*}

\textbf{Effectiveness of H$^2$LoRA against Counterparts.} H$^2$LoRA endows an LLM with the ability to achieve 
cross-task generalization while retaining instance-level personalization.
We comprehensively evaluate H$^2$LoRA in comparison with existing PEFT methods, including LoRA~\citep{hu2022lora}, MoELoRA~\citep{luo2024moelora}, and HyperLoRA~\citep{lv2024hyperlora}, with results presented in Figure~\ref{fig:abl_lora_comparison}.
The compared LoRA variants yield only modest performance, with MoELoRA slightly leading on F1-bio and HyperLoRA marginally outperforming on ACC.
In contrast, H$^2$LoRA consistently surpasses 
these LoRA methods,
highlighting its superior effectiveness.

\begin{table*}[t]
\centering
\scriptsize
\setlength{\tabcolsep}{4.2pt}
\caption{Ablation study of H$^2$LoRA within \Name. ``w/o Expert Mixture'' removes the Shared Low-Rank Expert Mixture, while ``w/o Hyper LoRA'' removes the Hyper LoRA Adaptation from the proposed H$^2$LoRA in Sec.~\ref{subsec:hybrid hyper low-rank adaptation}.} % kp: consider referring to Sec 3.2 here for readability
\vspace{-2mm}
\label{tab:abl_part_lora_components}
\renewcommand{\arraystretch}{0.5}
\resizebox{\textwidth}{!}{
\begin{threeparttable}
{\setlength{\tabcolsep}{3pt}
\begin{tabular}{l*{4}{cc}|*{4}{cc}}
\toprule
& \multicolumn{2}{c}{S2D}
 & \multicolumn{2}{c}{Heart}
 & \multicolumn{2}{c}{Food}
 & \multicolumn{2}{c}{Avg.} 
 & \multicolumn{2}{c}{S2D}
  & \multicolumn{2}{c}{Heart}
 & \multicolumn{2}{c}{Food}
 & \multicolumn{2}{c}{Avg.}
 \\
\cmidrule(lr){2-3}\cmidrule(lr){4-5}\cmidrule(lr){6-7}\cmidrule(lr){8-9}
\cmidrule(lr){10-11}\cmidrule(lr){12-13}\cmidrule(lr){14-15}\cmidrule(lr){16-17}
 & ACC & MCC
 & ACC & MCC
 & ACC & MCC
 & ACC & MCC
 & F1-Bio & RL
 & F1-Bio & RL
 & F1-Bio & RL
 & F1-Bio & RL
 \\
\midrule
w/o H$^2$LoRA & 95.28 & 93.71& 64.79 & 53.02  & 94.42 &92.65&84.83&79.79& 85.14 & 40.30 & 81.01 & 30.75 & 96.41 & 83.17  &87.52&51.41\\
w/o Expert Mixture & 97.33 & 96.43 & 80.28 & 73.80 & 97.00&96.01 &91.54&88.75& 85.74&43.22& 81.68 & 31.26 &97.84&89.79&88.42&54.76\\
w/o Hyper LoRA  & 97.33&96.43& 79.93 & 73.30 &97.42&96.57&91.56&88.77 & 85.73&43.27& 81.63  & 31.40 &97.71&89.13&88.36&54.60\\
w/ H$^2$LoRA & \textbf{97.74}&\textbf{96.98}&\textbf{ 83.10 }&\textbf{77.57}&\textbf{97.85}&\textbf{97.14} &\textbf{92.90}&\textbf{90.56}& \textbf{86.45}&\textbf{45.41}&\textbf{82.26}&\textbf{33.01}&\textbf{97.92}&\textbf{90.28}&\textbf{88.88}&\textbf{56.23}\\
\bottomrule
\vspace{-10pt}
\end{tabular}
}
\end{threeparttable}
}
\end{table*}

\begin{figure*}[!th]
    \centering
    % -------- Left: figure with subfigures --------
    \begin{minipage}[t]{0.62\textwidth}
        \centering
        \begin{subfigure}[t]{0.44\textwidth}
            \centering
            \includegraphics[width=\linewidth]{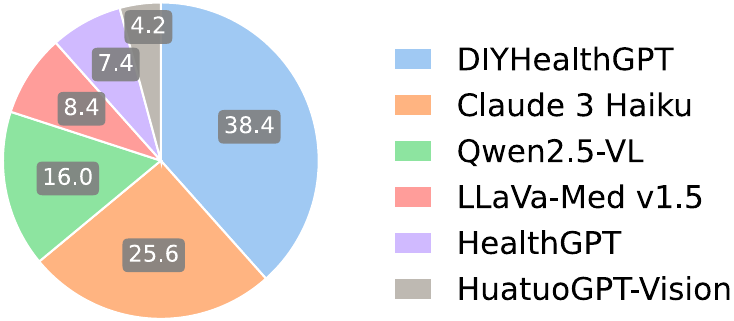}
            \caption{\scalebox{0.94}{Distribution of the first preference.\footnotemark}}
            \label{fig:pie}
        \end{subfigure}
        \hfill
        \begin{subfigure}[t]{0.55\textwidth}
            \centering
            \includegraphics[width=\linewidth]{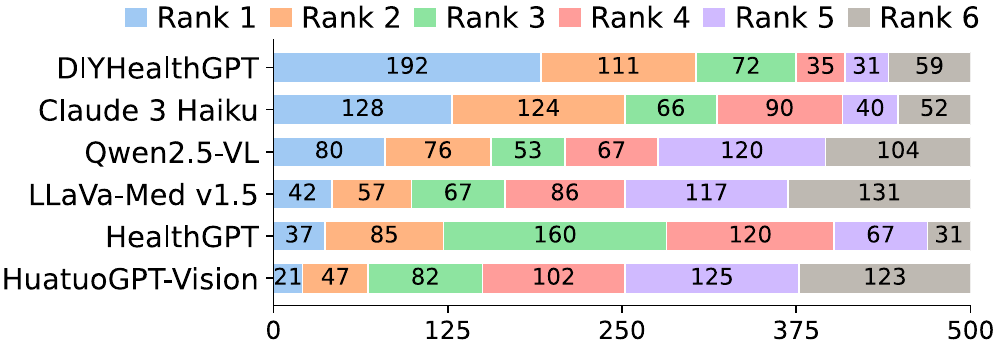}
            \caption{Full ranking landscape across models.}
            \label{fig:rank_stacked}
        \end{subfigure}
        \caption{Results of clinical expert review.}
        \label{fig:expert_review}
    \end{minipage}
    \hfill
    % -------- Right: table --------
    \begin{minipage}[b]{0.35\textwidth}
        \centering
        \scriptsize
        \captionof{table}{Comparison with fine-tuned baselines.}
        \label{tab:fine_tuned_comparison}
        % \vspace{-5pt}
        \resizebox{\textwidth}{!}{
        \begin{tabular}{lcc|cccc}
            \toprule
            \multirow{2}{*}{Model}
            & \multicolumn{2}{c|}{Closed-QA}
            & \multicolumn{2}{c}{Open-QA} \\
            \cmidrule(lr){2-3}\cmidrule(lr){4-5}
            & ACC & MCC & F1-Bio & RL \\
            \midrule
            Gemma 3-4B & 80.96 & 74.42 & 84.72 & 41.59 \\
            LLaVA-Med v1.5-7B & 77.58 & 69.51 & 86.28 & 49.63 \\
            \Name-3.8B & \textbf{86.80} & \textbf{82.36} & \textbf{87.34} & \textbf{52.11} \\
            \bottomrule
        \end{tabular}
        }
    \end{minipage}
\vspace{-15pt}
\end{figure*}
\textbf{Investigation of Training Stages.}
We design a unified four-stage training pipeline that equips \Name with multi-task learning in home-care settings and facilitates effective knowledge transfer. To assess the contribution of each stage, we add them sequentially and report the results in Figure~\ref{fig:abl_train_stage}. As outlined in~\cref{subsec:training pipeline}, the stages correspond to Cross-Modal Alignment $\rightarrow$ Medical Domain Adaptation $\rightarrow$ Task-Specific Expert Training $\rightarrow$ Cross-Task Knowledge Transfer, with each stage introducing a distinct capability.
The progressive performance gains highlight both the necessity and complementarity of all four stages.
In particular, Stage 2 achieves marked improvement through effective medical adaptation, Stage 3 specializes in task-specific patterns, and Stage 4 further enables knowledge sharing across experts rather than focusing on isolated tasks.

% \begin{figure}[t]
%     \centering
% \includegraphics[width=0.42\textwidth]{figs/lora_setting_comparison3.pdf}
% \vspace{-3mm}
%     \caption{Comparison of H$^2$LoRA with existing LoRA methods.}
% \label{fig:abl_lora_comparison}
% \vspace{-2mm}
% \end{figure}

% \begin{figure}[t]
%     \centering
% \includegraphics[width=0.45\textwidth]{figs/exp_train_stage.pdf}
% \vspace{-2mm}
%     \caption{Performance across training stages.}
%     \label{fig:abl_train_stage}
%     \vspace{-4mm}
% \end{figure}

% \begin{figure}[t]
%     \begin{center}
% \includegraphics[width=0.4\textwidth]{figs/embedding_tsne1.pdf}
% \vspace{-2mm}
%     \caption{Visualization of representations from \Name.}
%     \label{fig:vis_hyper}
%     \vspace{-7mm}
%     \end{center}
% \end{figure}

\textbf{\Name's Representation Visualization.}
We visualize the representations derived by \Name in open-QA in Figure~\ref{fig:vis_hyper} using t-SNE~\citep{maaten2008visualizing}.
Distinct clusters emerge for most tasks, reflecting low intra-task variance and large inter-task margins, which indicates that \Name captures task-aware features while preserving individual-level nuances.
An overlap is observed between DR and MAG, which is reasonable since MAG often involves advice related to drug management. This aligns with the performance gains in both closed- and open-QA,
underscoring that \Name yields both task-structured and personalized representations.
% \begin{figure}[t]
%     \begin{center}
% \includegraphics[width=0.4\textwidth]{figs/human_eval.pdf}
% \vspace{-2mm}
%         \caption{Results of the clinical expert review.
%         }
%     \label{fig:expert_review}
%     \vspace{-7mm}
%     \end{center}
% \end{figure}

\textbf{Clinical Expert Review.}
We conduct a clinical expert review by randomly sampling 500 open-QA pairs from \Benchmark.
% and assigning them to clinical experts for assessment of clinical significance. 
Each answer is evaluated by clinical experts according to three criteria: (i) Conciseness: providing direct and succinct answers for non-expert readers 
while avoiding unnecessary details;
(ii) Correctness: ensuring factual accuracy of the response;
(iii) Relevance: evaluating the degree to which the response avoids introducing irrelevant content.
The answers are rated on a 1–6 scale, with 1 indicating the best and 6 the worst.
The results are shown in Figure~\ref{fig:expert_review}. As illustrated in~\cref{fig:expert_review}(a), \Name is the clear first preference of clinical experts, demonstrating the highest clinical utility. Further, \cref{fig:expert_review}(b) shows that \Name concentrates mass in the top ranks, whereas other models exhibit heavier tail ranks. Among competitors, Claude-3 Haiku is the strongest, achieving the largest share of Rank 2 answers, yet it trails \Name in decisive first-preference counts. Overall, clinical experts consistently judged open-QA answers of \Name to be the most faithful and clinically meaningful.
Besides clinical expert review, evaluations via GPT-5 are provided in Appendix~\ref{sec:appendix-exps-gpt5}.
\footnotetext{We select Qwen2.5 as it outperforms Qwen3 in open-QA.}
% \begin{table}[t]
% \centering
% \scriptsize
% \setlength{\tabcolsep}{4.2pt}
% \caption{Comparison with fine-tuned baselines.}
% \label{tab: fine-tuned comparison}
% \resizebox{0.45\textwidth}{!}{
% {\setlength{\tabcolsep}{3pt}
% \begin{tabular}{lcc|cccc}
% \toprule
% \multirow{2}{*}{Model}
% & \multicolumn{2}{c|}{Closed-QA}
%  & \multicolumn{2}{c}{Open-QA}
%  \\
% \cmidrule(lr){2-3}\cmidrule(lr){4-5}
%  & ACC & MCC & F1-Bio & RL
%  \\
% \midrule
% Gemma 3-4B & 80.96 &74.42&84.72&41.59\\
% LLaVA-Med v1.5-7B& 77.58 &69.51& 86.28&49.63 \\
% \Name-3.8B & \textbf{86.80} &\textbf{82.36} & \textbf{87.34} &\textbf{52.11}\\
% \bottomrule
% \end{tabular}
% }
% }
% \vspace{-5mm}
% \end{table}

\vspace{-11pt}
\textbf{Comparison with Fine-tuned Baselines.}
To examine whether the performance gain of \Name is solely due to training on DIYHealth-900K, we conduct fine-tuning experiments on two representative models, Gemma 3-4B and LLaVA-Med v1.5-7B, using exactly the same training data with our limited computational resources. As shown in Table~\ref{tab:fine_tuned_comparison}, \Name-3.8B consistently outperforms Gemma 3-4B and LLaVA-Med v1.5-7B by a substantial margin on the four evaluation metrics, despite LLaVA-Med having a larger model size. These results indicate that the observed improvements cannot be attributed only to the dataset. Instead, the joint effect of \Dataset, the training strategy, and the H$^2$LoRA architecture contributes to the model's effectiveness.

More results on User Feedback, Subgroup Study, and Inter-rater Agreement Analysis are reported in Appendix~\ref{sec:appendix-exps}.

\vspace{-2mm}
\section{Conclusions}
\label{sec:conclusion}
\vspace{-1mm}

The blossoming of generative AI signals new opportunities for
accessible and personalized healthcare beyond traditional clinical settings.
In response, we propose \Suite, a home-based health management framework comprising \Dataset, \Benchmark, and \Name.
\Dataset 
curates diverse multimodal inputs from consumer-grade devices, enabling AI systems to operate in everyday home contexts. 
At the core, \Name leverages our proposed H$^2$LoRA technique to balance cross-task generalization with instance-level personalization,
delivering conversational and personalized support across health domains.
\Benchmark establishes the first dedicated 
evaluation protocol for dynamic, non-clinical home environments, 
ensuring standardized and fair comparison. 
Comprehensive experiments on 11 home care tasks demonstrate that \Name consistently surpasses state-of-the-art 
in both open-QA and closed-QA. 
% Notably, tasks including Sleep, Heart, Obesity, Diabetes, and DR continue to pose significant challenges for both general and medical VLMs. We highlight these tasks as priority areas for future research on home care reasoning and multimodal health understanding.
These components define a new paradigm for non-clinical healthcare AI, rooted in real-world scenarios
and designed for extensibility, usability, and rigorous evaluation.

% \newpage
%% The following are two recommended sections from ICLR this year
% \newpage
% \section*{Impact Statement}

% This work explores multimodal foundation models for home-based health management, with the potential to improve access to personalized health information outside clinical settings. While such systems may support early awareness and self-management, they also pose risks related to patient bias and misinterpretation of model outputs as medical advice. We emphasize that our approach is intended for decision support rather than a replacement for professional clinical diagnosis or treatment. Any real-world deployment should incorporate appropriate human oversight, clear usage disclaimers, and alignment with regulatory and clinical guidelines.

% The code and data for our project are available at \url{https://anonymous.4open.science/r/DIYHealthGPT-codes-E71A}. Detailed descriptions of hyperparameters and experimental settings are provided in Appendix~\ref{sec:appendix-implementation-setting}.


\bibliography{example_paper}
\bibliographystyle{icml2026}

%%%%%%%%%%%%%%%%%%%%%%%%%%%%%%%%%%%%%%%%%%%%%%%%%%%%%%%%%%%%%%%%%%%%%%%%%%%%%%%
%%%%%%%%%%%%%%%%%%%%%%%%%%%%%%%%%%%%%%%%%%%%%%%%%%%%%%%%%%%%%%%%%%%%%%%%%%%%%%%
% APPENDIX
%%%%%%%%%%%%%%%%%%%%%%%%%%%%%%%%%%%%%%%%%%%%%%%%%%%%%%%%%%%%%%%%%%%%%%%%%%%%%%%
%%%%%%%%%%%%%%%%%%%%%%%%%%%%%%%%%%%%%%%%%%%%%%%%%%%%%%%%%%%%%%%%%%%%%%%%%%%%%%%
\newpage
\appendix
\onecolumn
\section*{Appendix}
\label{sec:appendix}

\begin{itemize}
  \item \textbf{\ref{sec:appendix-notation}. Notation Table} \dotfill \pageref{sec:appendix-notation}

  \item \textbf{\ref{sec:appendix-related-work}. Related Work} \dotfill \pageref{sec:appendix-related-work}

  \item \textbf{\ref{sec:appendix-dataset}. Construction Details of \Dataset} \dotfill \pageref{sec:appendix-dataset}
  \begin{itemize}
      \item[$\cdot$] \textit{\ref{sec:appendix-task}. Task Design and Functional Descriptions} \dotfill \pageref{sec:appendix-task}
      \item[$\cdot$] \textit{\ref{sec:appendix-datastats}. Data Sources and Detailed Statistics} \dotfill \pageref{sec:appendix-datastats}
      \item[$\cdot$] \textit{\ref{sec:appendix-prompt}. Prompt Design} \dotfill \pageref{sec:appendix-prompt}
  \end{itemize}

  \item \textbf{\ref{sec:appendix-implementation-setting}. Implementation Details} \dotfill \pageref{sec:appendix-implementation-setting}
  \begin{itemize}
      \item[$\cdot$] \textit{\ref{sec:appendix-implementation-model}. Model Details} \dotfill \pageref{sec:appendix-implementation-model}
      \item[$\cdot$] \textit{\ref{sec:appendix-baseline-details}. Baseline Details} \dotfill \pageref{sec:appendix-baseline-details}
      \item[$\cdot$] \textit{\ref{sec:appendix-implementation-train}. Training Details} \dotfill \pageref{sec:appendix-implementation-train}
      \item[$\cdot$] \textit{\ref{sec:appendix-implementation-environment}. Experimental Environment} \dotfill \pageref{sec:appendix-implementation-environment}
      \item[$\cdot$] \textit{\ref{sec:appendix-evaluation-metrics}. Evaluation Metrics} \dotfill \pageref{sec:appendix-evaluation-metrics}
  \end{itemize}

  \item \textbf{\ref{sec:appendix-exps}. Supplementary Experimental Results} \dotfill \pageref{sec:appendix-exps}
  \begin{itemize}
      \item[$\cdot$] \textit{\ref{sec:appendix-exps-main-metric}. Evaluation of \Name in Terms of Additional Metrics} \dotfill \pageref{sec:appendix-exps-main-metric}
      \item[$\cdot$] \textit{\ref{sec:appendix-exps-gpt5}. Expert Evaluation with GPT-5} \dotfill \pageref{sec:appendix-exps-gpt5}
      \item[$\cdot$] \textit{\ref{sec:appendix-exps-sensitivity}. Hyperparameter Sensitivity Study on Number of Experts} \dotfill \pageref{sec:appendix-exps-sensitivity}
      \item[$\cdot$] \textit{\ref{sec:appendix-user-feedback}. User Feedback Analysis} \dotfill \pageref{sec:appendix-user-feedback}
      \item[$\cdot$] \textit{\ref{sec:appendix-subgroup-study}. Subgroup Study} \dotfill \pageref{sec:appendix-subgroup-study}
      \item[$\cdot$] \textit{\ref{sec:appendix-complete-fine-tuned}. Complete Comparison With Fine-tuned Baselines} \dotfill \pageref{sec:appendix-complete-fine-tuned}
      \item[$\cdot$] \textit{\ref{sec:appendix-inter-rater}. Inter-rater Agreement Analysis} \dotfill \pageref{sec:appendix-inter-rater}
  \end{itemize}

  \item \textbf{\ref{sec:appendix-discussion}. Discussion and Outlook} \dotfill \pageref{sec:appendix-discussion}
  \begin{itemize}
      \item[$\cdot$] \textit{\ref{sec:appendix-impact}. Broader Impact} \dotfill \pageref{sec:appendix-impact}
      \item[$\cdot$] \textit{\ref{sec:appendix-future}. Future Directions} \dotfill \pageref{sec:appendix-future}
  \end{itemize}

  % \item \textbf{\ref{sec:appendix-LLM}. \blue{The Use of LLMs} \red{(ICML does not require)}} \dotfill \pageref{sec:appendix-LLM}
  %% jq todo: consider removing this section

  \item \textbf{\ref{sec:appendix-exps-case}. Case Study} \dotfill \pageref{sec:appendix-exps-case}
\end{itemize}

%% recommended by ICLR this year: authors should describe the precise role of the LLM in a separate section on LLM usage

%%%%%%%%%%%%%%%%%%%%%%%%%%%%%%%%%%%%
\newpage
\section{Notation Table}
\label{sec:appendix-notation}

To provide a comprehensive overview of the notations used throughout the paper, we present a summary of notations in Table~\ref{tab:notation} as a quick reference to facilitate the understanding and recall of each symbol.

%% kp: remember to update the notation table
%% jq: got it, will update once we finish the paper drafting

\begin{table}[ht]
   \small
    \centering
    \renewcommand{\arraystretch}{1.2}
    \caption{Notations.}
    \label{tab:notation}
    % \resizebox{1\columnwidth}{!}{
        \begin{tabular}{c l}
    \toprule[1.5pt]
 Notation & Description\\
  \toprule[0.8pt]
$\mathcal{I} \in \mathbb{R}^{H \times W \times 3}$ & Input image of height $H$, width $W$, and three RGB channels.\\
$\mathcal{T} = \{t_1, \dots, t_{L_t}\}$ & Textual symptom description consisting of $L_t$ tokens.\\
$\mathcal{V}_{txt}$ & Vocabulary of the backbone language model.\\
$\mathcal{E}_v(\cdot)$ & Pretrained vision encoder. \\
$\mathcal{E}_t(\cdot)$ & Pretrained textual encoder. \\
$\mathcal{V} \in \mathbb{R}^{L_v \times d_v}$ & Visual embeddings with $L_v$ tokens, each of dimension $d_v$. \\
$\mathcal{U} \in \mathbb{R}^{L_t \times d}$ & Textual embeddings with $L_t$ tokens, each of dimension $d$. \\
%% kp: change the desc of textual embeddings to "tokens" as well - pls check if ok
%% jq: noted
$\mathcal{P}_v : \mathbb{R}^{d_v} \rightarrow \mathbb{R}^d$ & Learnable projection function aligning visual and textual embeddings. \\
$\mathcal{Z}  \in \mathbb{R}^{(L_v+L_t)\times d}$ & Unified multimodal representation combining visual and textual embeddings. \\
$\mathcal{M}_{LLM}$ & Backbone large language model. \\
$\Theta$ & Frozen pretrained parameters of $\mathcal{M}_{LLM}$. \\
$\Theta_{H^2L} = \{\mathcal{A}, \mathcal{B}, \mathcal{R}\}$ & Task-adaptive parameters introduced in H$^2$LoRA. \\

$N$ & Total number of home care tasks.\\
$T=\{1,\cdots,N\}$ & Set of home care tasks.\\

$\mathcal{A} = \{\mathbf{A}^t,\Delta\mathbf{A}^t\}_{t=1}^N$ & Set of shared projection matrices in H$^2$LoRA.\\
$\mathcal{B} = \{\mathbf{B}^t,\Delta\mathbf{B}^t\}_{t=1}^N$ & Set of expert matrices in H$^2$LoRA.\\
% $\mathbf{A}^t, \mathbf{B}^t$ & Task-specific low-rank parameters for task $t$. \\

$\mathcal{R}$ & Routing parameters for integrating task-level outputs. \\
$\mathbf{B}_k^t$ & The $k$-th expert mixture matrix of task $t$. \\
$K$ & Number of expert mixture matrices per task. \\
$\mathcal{W}^t \in \mathbb{R}^K$ & 
Mixture weights over the $K$ expert matrices $\{\mathbf{B}^t_1,\ldots,\mathbf{B}^t_K\}$ for task $t$.\\
% Weight of $K$ expert matrices $\{\mathbf{B}^t_1,\ldots,\mathbf{B}^t_K\}$ for task $t$. \\
%% kp: consider changing the above to: "Mixture weights over the $K$ expert matrices $\{\mathbf{B}^t_1,\ldots,\mathbf{B}^t_K\}$ for task $t$."?
%% jq: revised
$\hat{\mathcal{W}}^t$ & Expanded expert weight vector for task $t$.\\
% $\mathbf{B}^t$ & MoE-driven expert mixture of task $t$. \\
$\mathcal{H}_A(\cdot)$ & Hypernetwork of the shared projection. \\
$\mathcal{H}_B(\cdot)$ & Hypernetwork of the expert mixture matrix. \\

$\Delta \mathbf{A}^t$ & Instance-aware offset for the shared projection of task $t$. \\
$\Delta \mathbf{B}^t_k$ & Instance-aware offset for the $k$-th expert mixture matrix $\mathbf{B}^t_k$ of task $t$. \\

% $\mathbf{\tilde{A}}^{t}$ & Modulated shared projection for task $t$.\\
% $\mathbf{\tilde{B}}^{t}_k$ & Modulated $k$-th expert matrix for task $t$.\\
% $\mathbf{\tilde{B}}^{t}$ & Aggregated expert mixture for task $t$.\\
$\Delta \mathbf{B}^{t}$ & Instance-aware offset for MoE-driven expert mixture matrix $\mathbf{B}^t$ of task $t$.\\

$\beta=(\beta^1, \ldots, \beta^N)$ & Inter-task fusion mixture weights across $N$ tasks.\\

$\mathcal{O}^t_{H^2LoRA}$ & Output of H$^2$LoRA for task $t$. \\
$\mathcal{O}_{H^2LoRA}$ & Overall output of H$^2$LoRA aggregated across all tasks. \\
    \bottomrule[1.5pt]
        \end{tabular}
        % }
\end{table}

%%%%%%%%%%%%%%%%%%%%%%%%%%%%%%%%%%%%
\section{Related Work}
\label{sec:appendix-related-work}

The development of \Suite is grounded on a broad and evolving body of research spanning pre-LLM healthcare AI, general-purpose LLMs, and emerging medical foundation models.
In this section, we review key advances across these interconnected areas to contextualize our contributions.

\noindent
\textbf{Pre-LLM Healthcare Techniques.}
% Prior to the emergence of LLMs, the application of deep learning in healthcare evolved rapidly, yielding significant advances across both imaging and EHR analysis. 
Before the advent of LLMs, deep learning rapidly advanced healthcare applications, particularly in medical imaging and EHR analysis.
% Fueled by breakthroughs in computer vision, early efforts primarily focused on medical image interpretation, such as MRI, ultrasound, and fundus photography for disease diagnosis and risk prediction. 
Early work---largely driven by progress in computer vision---centered on interpreting medical images such as MRI, ultrasound, and fundus photography for disease diagnosis and risk assessment.
Prominent examples include automated detection of Alzheimer's disease from brain MRI~\citep{brosch2013manifold,helaly2022deep}, segmentation of knee cartilage in osteoarthritis~\citep{prasoon2013deep}, and lesion analysis for conditions such as multiple sclerosis and breast nodules~\citep{yoo2014deep,cheng2016computer}. Convolutional neural networks (CNNs) achieved diagnostic performance comparable to that of clinical experts, including dermatologists classifying skin 
% cancer across large-scale clinical datasets
lesions~\citep{esteva2017dermatologist} and ophthalmologists screening for diabetic retinopathy~\citep{gulshan2016development}. 
In parallel, deep learning began reshaping EHR analysis by enabling predictive modeling over both structured inputs (e.g., diagnoses, laboratory tests) and unstructured clinical narratives. 
Supervised models, such as CNNs and Recurrent Neural Networks (RNNs) equipped with Long Short-Term Memory (LSTM) or Gated Recurrent Unit (GRU) architectures, demonstrated superior performance in tasks including disease onset prediction for congestive heart failure~\citep{cheng2016risk}, disease progression modeling~\citep{pham2016deepcare}, and automated diagnosis and medication recommendation~\citep{choi2016doctor}. 
Unsupervised approaches, including stacked denoising autoencoders~\citep{miotto2016deepa}, restricted Boltzmann machines~\citep{liang2014deep}, and neural embedding methods~\citep{choi2016learning}, facilitated the extraction of latent patient representations for downstream applications such as disease risk stratification and phenotype discovery.
% While these innovations laid a strong foundation for data-driven healthcare, they typically relied on narrowly defined tasks, institution-specific datasets, and lacked generalizable evaluation benchmarks.
%% kp: have rewritten the sentence above - pls check if ok
Although these innovations established a solid foundation for data-driven healthcare, they are typically limited by narrowly scoped tasks, institution-specific datasets, and the absence of standardized benchmarks for evaluation.
%% kp: better move "Pre-LLM Healthcare Techniques" above to the appendix

% Large Language Models 
\noindent
\textbf{LLMs for General Reasoning and Dialogue.}
LLMs have become central to recent advances in natural language understanding and generation, fundamentally advancing the capabilities of machines to perform general-purpose reasoning and engage in coherent, human-like dialogue~\citep{chowdhery2023palm, touvron2023llama}.
Representative models such as GPT-3~\citep{brown2020language} and GPT-4~\citep{openai2023gpt4}, trained on large-scale corpora, exhibit remarkable zero-shot and few-shot generalization across diverse natural language tasks.
Beyond text-only models, the emergence of multimodal LLMs (MLLMs) such as LLaVA-1.5~\citep{liu2023visual}, Llama 3.2~\citep{dubey2024llama}, Qwen2.5-VL~\citep{bai2025qwen2}, and GPT-4o~\citep{achiam2023gpt} further extends these capabilities by enabling joint reasoning across language and visual modalities.
These capabilities stem from their architectural scale---often comprising hundreds of millions to hundreds of billions of parameters---as well as sophisticated pre-training and alignment techniques~\citep{wei2022finetuned, fedus2022switch}.
% , has established LLMs as foundational engines for general AI systems~\citep{wei2022finetuned, fedus2022switch}.
Despite such strengths, 
% while these models demonstrate impressive general reasoning abilities, 
the application of general-purpose LLMs in specialized domains such as healthcare remains constrained by the absence of domain-specific knowledge and the lack of grounding in real-world physiological data, particularly in low-resource and home-based healthcare contexts.

%%jq: what is VLLM? vision large language model? Maybe LVLM?
\noindent
\textbf{Medical Foundation Models.}
% and Med-VLLMs
To bridge the domain gap between general-purpose LLMs and medical applications, a new generation of medical foundation models has emerged.
These include both text-only architectures and multimodal frameworks, referred to as Med-LLMs and Med-LVLMs.
Text-based models such as Med-PaLM~\citep{singhal2023large}, BiomedGPT~\citep{luo2024biomedgpt}, and HuatuoGPT~\citep{chen2024towards} have shown strong performance on clinical question answering (QA) benchmarks by leveraging curated biomedical corpora and large-scale synthetic datasets. BiomedGPT, in particular, achieves a compact yet competitive architecture through cross-modal pre-training and decoder alignment techniques.
In parallel, Med-LVLMs have advanced multimodal medical understanding.
Models such as LLaVA-Med~\citep{li2023llava}, Med-Flamingo~\citep{moor2023med}, MedGemma~\citep{sellergren2025medgemma}, Med-R1~\citep{lai2025med}, Lingshu~\citep{xu2025lingshu}, and MedVLM-R1~\citep{pan2025medvlm} align visual encoders with textual LLMs to enable diagnostic reasoning over imaging data.
HealthGPT~\citep{lin2025healthgpt} extends this direction by supporting multimodal comprehension and generation on multiple medical tasks, while EyecareGPT~\citep{li2025eyecaregpt} devises a resolution mechanism and a layer-wise dense connector to improve ophthalmic visual understanding. These systems demonstrate potential for tasks such as image captioning, visual question answering (VQA), and differential diagnosis.

Despite these advances, the existing Med-LVLMs are trained and evaluated primarily on data from professional medical devices, such as radiographic images and pathology slides, which are not accessible in daily life scenarios. Consequently, their applicability to home-based health management remains limited. The absence of portability, limited adaptability to informal data, and insufficient support for personalized inference underscore the need to redesign medical AI architectures that can operate effectively with consumer-grade devices such as smartphones, wearables, and smart home sensors.
%% kp: this sentence appears not finished: + which motivates our proposal of xxx? this will also agree with the starting of sec2 - provide context for our contributions
To this end, we propose \Suite as an innovative solution for home care health management, comprising three core components, \Dataset, \Benchmark, and \Name, which are elaborated in Sections~\ref{sec:dataset},~\ref{sec:model}, and~\ref{sec:benchmark}, respectively.

%% baselines:
% We evaluate \Name against a broad set of baselines, including state-of-the-art generalist models (e.g., LLaVA-1.5~\citep{liu2023visual},InstructBLIP~\citep{dai2023instructblip}, Llama 3.2~\citep{dubey2024llama},  Yi-VL~\citep{young2024yi}, InternVL3~\citep{zhu2025internvl3},  Qwen2.5-VL~\citep{bai2025qwen2}, Gemma 3~\citep{team2025gemma},Claude 3 Haiku~\citep{anthropic2024claude} and GPT-4o Mini~\citep{achiam2023gpt}) and medical-specific models (e.g., LLaVA-Med v1.5~\citep{li2023llava}, Med-Flamingo~\citep{moor2023med}, HuatuoGPT-Vision~\citep{chen2024huatuogpt}, MedGemma~\citep{sellergren2025medgemma}, HealthGPT~\citep{lin2025healthgpt}, Med-R1~\citep{lai2025med}, Lingshu~\citep{xu2025lingshu}, and MedVLM-R1~\citep{pan2025medvlm}).

%% kp: can copy the whole related work section here; reduce the contents in the main paper then

%%%%%%%%%%%%%%%%%%%%%%%%%%%%%%%%%%%%
% \section{Preliminaries}
% \label{sec:appendix-preliminaries}

% \subsection{LVLM}

% \subsection{LoRA}

%%%%%%%%%%%%%%%%%%%%%%%%%%%%%%%%%%%

\section{Construction Details of \Dataset}
\label{sec:appendix-dataset}

\subsection{Task Design and Functional Descriptions}
\label{sec:appendix-task}

A detailed overview of the tasks included in the \Dataset dataset is provided in Table~\ref{tab:home_tasks}.
To reflect the complexity and diversity of real-world home care scenarios, we categorize the tasks into three major groups:
personalized health management, chronic disease risk assessment, and daily health monitoring. 
The first category, personalized health management, encompasses core tasks such as symptom-based diagnosis, drug recommendation, and tailored medical advice generation, which are essential for supporting early clinical decision-making.
In the context of chronic conditions, \Dataset includes risk assessments for diabetes, obesity, cardiovascular disease, and kidney health, drawing on both self-reported symptoms and home-acquired signals. 
Daily health monitoring tasks address routine wellness dimensions, such as dietary intake analysis, sleep quality evaluation, skin condition assessment, and oral health screening. 
For each task, we curate or adapt relevant datasets to align with the characteristics of home care, with particular emphasis on multimodal inputs, real-world variability, and nonclinical supervision, to ensure their applicability to AI models deployed in home environments.

\begin{table}[ht]
% \small
\centering
    \renewcommand{\arraystretch}{1.3}
\caption{Home care task design in \Dataset with corresponding functional descriptions.} 
%%jq: where is the dataset?
\label{tab:home_tasks}
\resizebox{1\columnwidth}{!}{
  \begin{tabular}{l|c|l}
    \toprule
    \multicolumn{1}{c|}{Task Category} & Home Care Task & \multicolumn{1}{c}{Task Description} \\
    \midrule
    \multirow{3}{*}{Personalized Health Management} & Symptom-to-Diagnosis (S2D) & Predict potential diagnoses based on symptom descriptions \\ \cline{2-3}
    & Drug Recommendation (DR) & Suggest appropriate medications based on symptoms and suspected conditions\\\cline{2-3}
    & Medical Advice Generation (MAG) & Generate personalized medical advice tailored to individual health concerns\\
    \midrule
    \multirow{4}{*}{Chronic Disease Risk Assessment}
     & Diabetes Detection (Diabetes) & Assess for potential diabetes risks based on retinal fundus images\\\cline{2-3}
     %% kp: mention from retina images?
     & Obesity Detection (Obesity) & Identify obesity risk through lifestyle, dietary, and physical measurements \\\cline{2-3}
     & Heart Health (Heart) & Monitor cardiovascular health and detect early signs of heart conditions\\\cline{2-3}
     & 
     % Chronic Kidney Disease Detection
     Kidney Health (CKD)
      & Detect risks of chronic kidney disease through symptoms and home test inputs\\
     \midrule
     %CKD that could be used for experiments if necessary(https://www.kaggle.com/datasets/rabieelkharoua/chronic-kidney-disease-dataset-analysis, https://archive.ics.uci.edu/dataset/336/chronic+kidney+disease)
     \multirow{4}{*}{Daily Health Monitoring} & Diet Management (Food) & Manage dietary habits by analyzing intake and providing dietary recommendations\\\cline{2-3}
     & Sleep Health (Sleep) & Evaluate sleep quality and patterns to provide improvement suggestions\\\cline{2-3}
     & Skin Health (Skin) & Assess skin conditions and detect abnormalities from patient inputs or images\\
     \cline{2-3}
     & Oral Health (Oral) & Screen for common oral issues and provide oral health recommendations\\
     % \cline{2-3}
     % & Eye Health & \\
  \bottomrule
\end{tabular}
}
\end{table}
%% kp: perhaps we should change this table to a figure, maybe show in the dataset subsection with data details, etc.

\subsection{Public Data Sources}
\label{sec:appendix-datastats}

The task-specific public data sources 
% and corresponding statistics
of \Dataset are summarized in Table~\ref{tab:data_stats}. 
To fully leverage real-world data, we curate \Dataset by integrating private datasets from three partner organizations with 20 publicly available data sources selected for their alignment with the target QA tasks.
The private datasets complement the public sources by covering additional real-world scenarios and task variations, while adhering to the same data modality and home-care constraints.
Due to confidentiality agreements, we do not disclose further details of these private sources.
% Importantly, 
All included data are readily obtainable in home settings, ensuring that the dataset reflects scenarios accessible to end users without reliance on specialized clinical equipment.
For instance, in VQA tasks such as diet management, oral health, skin health, and diabetes detection~\citep{naz2024clinical,rogers2021evaluation}, the images consist of everyday photographs that can be captured using mobile devices, rather than specialized medical imaging.
In heart health and sleep health tasks, the physiological signals are collected from portable, home-use devices, ensuring that the data reflect measurements users can realistically obtain outside clinical environments.
Specifically, we develop a translation script to convert the raw physiological signals into images suitable for VQA tasks. For the heart health task, the data are derived from ECG recordings. Due to the hardware limitations of portable devices (e.g., Apple Watch)~\cite{li2022using}, only Lead-I signals are available; therefore, we extract the Lead-I channel from the original 12-lead ECG data. The extracted signals are subsequently processed using NeuroKit2's~\footnote{\url{https://github.com/neuropsychology/NeuroKit}} clean function to perform noise filtering, artifact removal, and baseline correction. For recordings exceeding 10 seconds, a high-quality 10-second segment is selected for downstream processing~\citep{wagner2020ptb}.
For the sleep health task, which involves triaxial accelerometry data, we first extract a 30-second segment for each sample~\citep{wang2024addressing}. Each segment is then subjected to a cleaning procedure to reduce noise and eliminate artifacts~\citep{moscato2022wrist}.

For text-only QA tasks, including drug recommendation, symptom-to-diagnosis reasoning, medical advice generation, obesity detection, and kidney health, we construct home-care–oriented datasets through a carefully designed multi-stage pipeline. In the first stage, we extract candidate samples from a variety of publicly available medical and health-related datasets. We then apply a filtering procedure to exclude records, descriptions, or cases that are clearly irrelevant or impractical in a home setting (e.g., those requiring hospital-grade imaging or laboratory tests). After curating this subset, we generate corresponding question–answer pairs. Because many of the original data sources, such as questionnaires, are not naturally phrased as 
% human 
questions, we employ an LLM to rephrase them into fluent, conversational question formats. 
For VQA tasks covering diabetes detection, diet management, skin health, and oral health, we utilize publicly available image datasets and derive questions from their associated labels. These questions are likewise rephrased into fluent natural language by an LLM.
Finally, human experts validate both generated questions and their associated answers to ensure correctness.

% \todo{can mention detailed task-specific data processing here}
% \todo{describe the processing of the signal data of sleep and heart tasks}

\begin{table}[htbp]
\centering
\small
\renewcommand{\arraystretch}{1.4}
\caption{Public Data sources for each task in \Dataset.}
% \resizebox{1\columnwidth}{!}{
\begin{tabular}{lcc}
\toprule
Task & Type & Data Source \\
\midrule
Symptom-to-Diagnosis (S2D)    & QA& DDXPlus~\tablefootnote{\url{https://figshare.com/articles/dataset/DDXPlus_Dataset_English_/22687585}}\\
Drug Recommendation (DR) & QA & MIMIC-III~\tablefootnote{\url{https://physionet.org/content/mimiciii/1.4/}} and MIMIC-IV~\tablefootnote{\url{https://physionet.org/content/mimiciv/3.0/}}\\
Medical Advice Generation (MAG)           & QA & \makecell{MedQuAD~\tablefootnote{\url{https://www.kaggle.com/datasets/pythonafroz/medquad-medical-question-answer-for-ai-research}},
MedQA-USMLE~\tablefootnote{\url{https://www.kaggle.com/datasets/moaaztameer/medqa-usmle}},\\
PubMedQA~\tablefootnote{\url{https://pubmedqa.github.io/}}, MedAlpaca~\tablefootnote{\url{https://github.com/kbressem/medAlpaca}}, and MedMCQA~\tablefootnote{\url{https://medmcqa.github.io/}}
}
\\
Diabetes Detection (Diabetes) & VQA & Messidor Diabetic Retinopathy~\tablefootnote{\url{https://www.kaggle.com/datasets/ascanipek/eyepacs-aptos-messidor-diabetic-retinopathy/data}}\\
Obesity Detection (Obesity)  & QA & ObesityDataSet~\tablefootnote{\url{https://www.kaggle.com/datasets/aravindpcoder/obesity-or-cvd-risk-classifyregressorcluster/data}}\\
Heart Health (Heart)             & VQA & PTB~\tablefootnote{\url{https://www.physionet.org/content/ptbdb/1.0.0/}}, PTB-XL~\tablefootnote{\url{https://physionet.org/content/ptb-xl/1.0.3/}}, and ECG-arrhythmia~\tablefootnote{\url{https://physionet.org/content/ecg-arrhythmia/1.0.0/}}\\
Kidney Health (CKD)                 & QA & CKD Source Dataset1~\tablefootnote{\url{https://www.kaggle.com/datasets/rabieelkharoua/chronic-kidney-disease-dataset-analysis?resource=download}} and CKD Source Dataset2~\tablefootnote{\url{https://archive.ics.uci.edu/dataset/336/chronic+kidney+disease}} \\
Diet Management (Food)  & VQA & Food-101~\tablefootnote{\url{ http://data.vision.ee.ethz.ch/cvl/food-101.tar.gz}}\\
Sleep Health (Sleep)  & VQA & Dreamt~\tablefootnote{\url{https://physionet.org/content/dreamt/2.1.0/}} and Applewatch~\tablefootnote{\url{https://physionet.org/content/sleep-accel/1.0.0/}}\\
Skin Health (Skin)              & VQA & Fitzpatrick17k~\tablefootnote{\url{https://github.com/mattgroh/fitzpatrick17k}}
% ~\citep{groh2021evaluating}
\\
Oral Health (Oral)  & VQA & Oral Diseases\tablefootnote{\url{https://www.kaggle.com/datasets/salmansajid05/oral-diseases}}\\
\midrule
Total &  QA\&VQA & 20 Publicly Available Data Sources\\
\bottomrule
\end{tabular}
\label{tab:data_stats}
% }
\end{table}

\begin{figure}[!ht]
\centering
\includegraphics[width=0.9\textwidth]{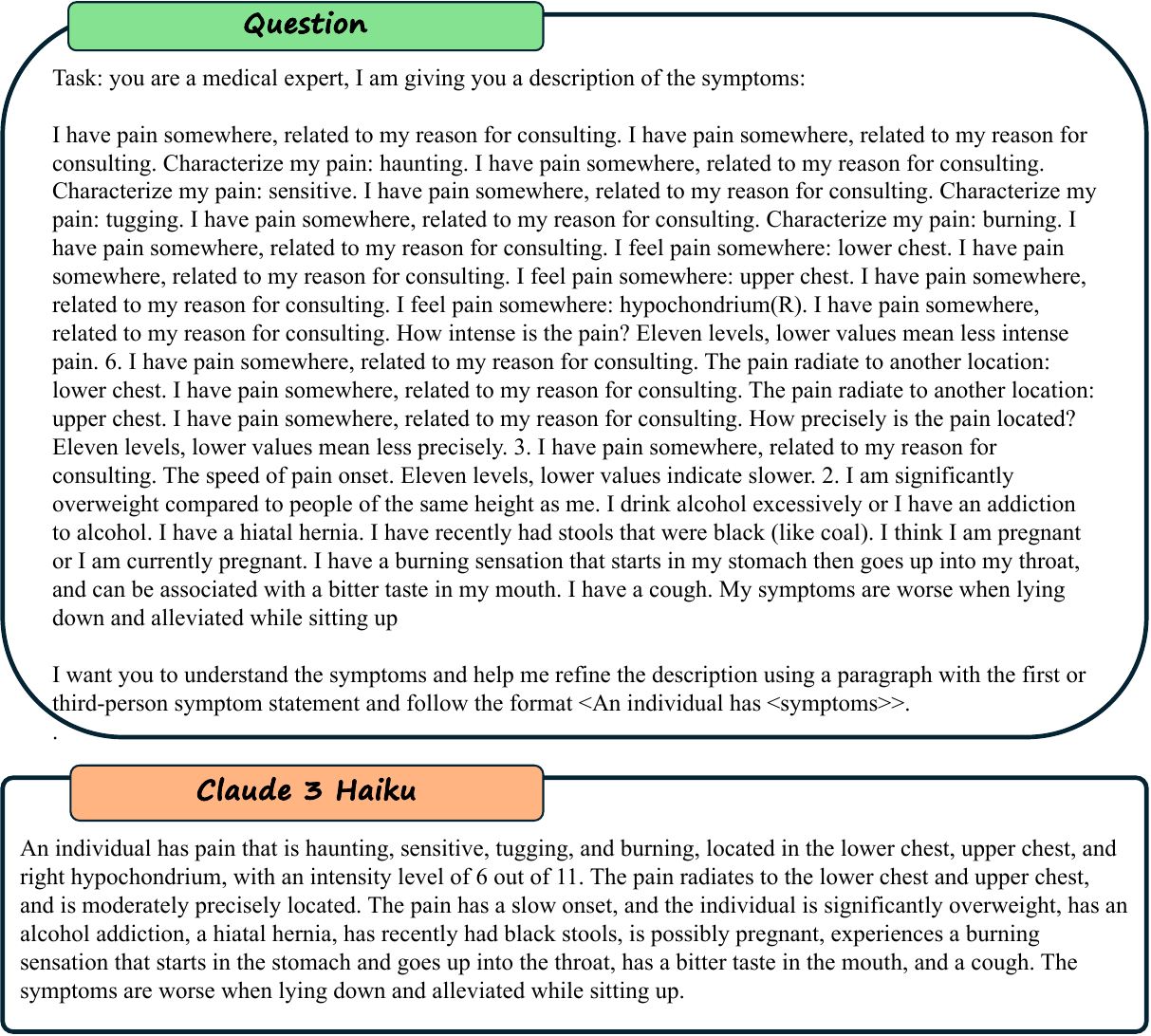}
\caption{An example prompt to rephrase raw data into natural human conversational expressions for the S2D task in the open-QA setting.}
\label{fig:prompt_design}
\end{figure}

\subsection{Prompt Design}
\label{sec:appendix-prompt}
%% we have both first-person and third-person QA pairs? and IMO, we use claude for both scenarios, not just one of them?
Claude 3 Haiku is employed to rephrase the source data of each dataset into first-person or third-person patient statements and descriptions. As an example, in the S2D open-QA setting, we illustrate the prompt design in Figure~\ref{fig:prompt_design}. 
% Notably,
To ensure reliability, we incorporate a human-in-the-loop validator to monitor and guarantee the quality of the rephrased data.

% \begin{figure}
% \centering
% \includegraphics[width=0.9\textwidth]{figs/Prompt_rephrase.pdf}
%   \vspace{-2mm}
%   \caption{Example of the prompts to rephrase the source data and generate third-person patient statements.
%   }
%   \vspace{-5mm}
%   \label{fig:prompt_design}
% \end{figure}

\section{Implementation Details}
\label{sec:appendix-implementation-setting}

\subsection{Model Details}
\label{sec:appendix-implementation-model}

\textbf{Details on \Name}. \Name is built based on Phi-3-mini~\citep{abdin2024phi}, a publicly available, pre-trained, lightweight LLM with 3.8B parameters. We employ CLIP-L/14~\citep{radford2021learning} as the visual feature extractor to encode image representations. The extracted visual features are projected into a shared semantic space through two-layer MLPs, aligning them with text tokens. This alignment bridges the modality gap between vision and language, and the fused representations are subsequently fed into the LLM to support multimodal understanding and coherent language generation. The number of expert matrices per task $K$ is 4 in the experiments.
The ranks of Low-Rank Expert and Hyper LoRA are 16 and 8, respectively. The model is optimized using a next-token prediction objective with the cross-entropy loss~\citep{liu2023visual}.
% The number of experts for H$^2$LoRA is 4 in the experiments. The ranks of Low-Rank Expert and Hyper LoRA are 16 and 8, respectively. $\alpha$ of Low-Rank Expert and Hyper LoRA is 64 and 32, respectively.

\textbf{Implementations on Training Stage 3}. During task-specific expert training in Stage 3, each task $t$ is assigned an H$^2$LoRA block with parameters $(\mathbf{A}^t,\Delta\mathbf{A}^t,\mathbf{B}^t,\Delta\mathbf{B}^t)$, 
which are trained independently using task-specific data, while other parameters remain frozen. Each block is optimized using the same hyperparameter settings across tasks.
The Shared Low-Rank Expert Mixture and Hyper LoRA Adaptation modules within the H$^2$LoRA block are inserted into the linear layers of the backbone LLM as lightweight training layers. Shared Low-Rank Expert Mixture is implemented by a shared matrix $\mathbf{A}^t$ and multiple matrices $\mathbf{B}^t_k$, with a router that aggregates the output of matrices $\mathbf{B}^t_k$. Hyper LoRA follows a similar structure, but its parameters are dynamically generated by hypernetworks $\mathcal{H}_A$ and $\mathcal{H}_B$ rather than learned directly through backward propagation. The hypernetworks learn how to generate parameters conditioned on the instance input.
Overall, this implementation allows the model to acquire task-level specialization and parameter-efficient adaptation before cross-task fusion in Stage 4.

\subsection{Baseline Details}
\label{sec:appendix-baseline-details}

We provide detailed descriptions of all baseline models evaluated in this study below.

\begin{itemize}[itemsep=0mm,leftmargin=4mm]
    \item LLaVA-1.5~\citep{liu2023visual} leverages GPT-4 to generate multimodal instruction-following data and 
    % instruction
    %% kp: "instruction" above is duplicate?
    tunes an end-to-end vision-language model, enabling general-purpose visual language understanding.
    \item InstructBLIP~\citep{dai2023instructblip} enhances BLIP-2~\citep{li2023blip} by tuning on 26 multimodal datasets. InstructBLIP further introduces an instruction-aware Query Transformer to capture informative features tailored to the provided instruction.
    \item Llama 3.2~\citep{dubey2024llama} is part of Meta's multimodal LLM series that integrates vision and language in efficient models for both edge deployment and general-purpose AI. Llama 3.2 Vision Instruct is used for comparison.
    \item Yi-VL~\citep{young2024yi} is an open-source multimodal extension of the Yi LLM that integrates vision and language, supporting image understanding, text recognition, and multi-round visual question answering.
    \item InternVL3~\citep{zhu2025internvl3} is an open-source multimodal LLM natively pre-trained on both text and multimodal data in a unified framework, enhanced with variable visual position encoding, post-training, and test-time scaling.
    \item Qwen2.5-VL~\citep{bai2025qwen2} is Alibaba's vision-language model that combines native-resolution visual understanding, object localization, document parsing, and long-video comprehension with general language capabilities.
    \item Qwen3-VL~\citep{yang2025qwen3} is a large-scale multimodal foundation model developed by Alibaba, capable of jointly understanding visual and textual inputs. It extends the Qwen language model family with strong vision–language alignment, enabling tasks such as visual question answering, image understanding, and multimodal reasoning across diverse domains.
    \item Gemma 3~\citep{team2025gemma} is Google's open-weight multimodal LLM that supports multilingual, visual understanding, and advanced reasoning capabilities. 
    \item Claude 3 Haiku~\citep{anthropic2024claude} is Anthropic's fastest and most lightweight Claude 3 model, optimized for efficiency and low-latency reasoning while preserving strong language and comprehension abilities.
    \item GPT-4o Mini~\citep{achiam2023gpt} is a lightweight, cost-efficient variant of GPT-4o. GPT-4o Mini is optimized for speed while retaining multimodal reasoning across text, vision, and audio.
    \item LLaVA-Med v1.5~\citep{li2023llava} is a biomedical LVLM trained on PubMed~\citep{zhang2023large} figure-caption pairs and GPT-4–generated instructions, enabling multimodal conversations.
    \item Med-Flamingo~\citep{moor2023med} is a multimodal few-shot learner built on OpenFlamingo-9B~\citep{awadalla2023openflamingo} and further pre-trained on medical image-text data, enabling generative medical VQA and few-shot adaptation.
    \item HuatuoGPT-Vision~\citep{chen2024huatuogpt} is a medical LVLM trained on refined PubMedVision, a denoised and reformatted medical VQA dataset curated with GPT-4V.
    \item MedGemma~\citep{sellergren2025medgemma} is Google's domain-adapted variant of Gemma that incorporates biomedical knowledge to support medical text understanding and vision-language tasks.
    \item HealthGPT~\citep{lin2025healthgpt} is a unified medical vision-language model trained with the H-LoRA technique, hierarchical visual perception, and a three-stage learning strategy on the VL-Health dataset.
    \item Med-R1~\citep{lai2025med} is an RL-enhanced medical vision-language model that employs Group Relative Policy Optimization to improve reasoning quality, generalization, and reliability across diverse medical imaging tasks.
    \item Lingshu~\citep{xu2025lingshu} is a medical LVLM trained on a 
    % richly 
    curated multimodal dataset with multi-stage learning and reinforcement learning for enhanced reasoning.
    \item MedVLM-R1~\citep{pan2025medvlm} is a medical LVLM trained with reinforcement learning to generate explicit, human-interpretable reasoning paths.
\end{itemize}

\begin{table}[ht]
\centering
\small
\renewcommand{\arraystretch}{1.2}
\caption{Summary of hyperparameter settings used for training \Name.}
\resizebox{0.6\columnwidth}{!}{
\begin{tabular}{l|r|r|r|r}
\toprule
Hyperparameter & Stage 1 & Stage 2 & Stage 3 & Stage 4 \\
\midrule
Optimizer & AdamW& AdamW& AdamW& AdamW\\
Adapter LR & 1e-4 & 2e-5& /& / \\
Learning Rate & / &1e-5&1e-5&1e-5\\
Global Batch Size & 64& 32& 32& 32\\
Weight Decay & 0.01& 0.1& 0.1& 0.01\\
Dropout Rate & 0 &0.05&0.05&0.05\\
LR Scheduler & Constant& Warm Up&  Warm Up& Constant\\
Max Sequence Length & 2048& 2048& 2048& 2048\\
\bottomrule
\end{tabular}
\label{tab:train-implementation}
}
\end{table}

\subsection{Training Details}
\label{sec:appendix-implementation-train}

We display the detailed hyperparameter configurations for \Name's four-stage training process. The specific settings used are listed in Table~\ref{tab:train-implementation}. These hyperparameters are set up following prior studies~\citep{lin2025healthgpt, liu2023visual,li2023llava}. The architectural parameters are listed in Table~\ref{tab: architectural-parameters}.

\begin{table}[ht]
\centering
\small
\renewcommand{\arraystretch}{1}
\caption{Architectural parameters of \Name.}
\begin{tabular}{l|l|l}
\toprule
\textbf{Module} & \textbf{Parameter} & \textbf{Value / Shape} \\
\midrule
\textbf{Tokenizer}
 & Vocabulary size & 32,064 \\
\midrule
\multirow{5}{*}{\textbf{Backbone}}
 & Hidden size & 3,072 \\
 & \# Transformer blocks & 32 \\
 & Self-attn projections & qkv: (3072, 9216), o: (3072, 3072 )\\
 & MLP& gate$\_$up: (3072, 16384); down: (8192, 3072) \\
\midrule
\multirow{4}{*}{\textbf{CLIP}}
 & Input resolution & $336 \times 336$ \\
 & Patch size & $14 \times 14$ \\
 & Embedding dim & 1{,}024 \\
 & \# Transformer blocks & 24 \\
\midrule
\multirow{3}{*}{\textbf{Expert Mixture of H$^2$LoRA}}
 & LoRA rank & 16 \\
 & Shared matrix $\mathbf{A}$ (up / down) & (3072,16) \;/\; (8192,16) \\
 & Matrix $\mathbf{B}$ (up / down) & (16,16384)\;/\; (16,3072) \\
\midrule
\multirow{4}{*}{\textbf{Hyper LoRA of H$^2$LoRA}}
 & Rank & 8\\
 & Input & (3072, 8)\\
 & Up-path generator (down / up) & (8, 24576) \;/\; (8, 131072) \\
 & Down-path generator (down / up) & (8, 65536) \;/\; (8, 24576)\\
\bottomrule
\end{tabular}
\label{tab: architectural-parameters}
\end{table}

\subsection{Experimental Environment}
\label{sec:appendix-implementation-environment}
%% jq: pls update the following information
All experiments are conducted on a server equipped with an AMD EPYC 9334 CPU @ 2.7 GHz (32 cores), 128 GB of memory, and an NVIDIA H100 NVL with CUDA 12.9.
The operating system is Ubuntu 24.04 running Linux kernel 6.8.0-79-generic.
\subsection{Evaluation Metrics}
\label{sec:appendix-evaluation-metrics}
\subsubsection{Evaluation Metrics for Closed-QA}
\begin{itemize}[leftmargin=*]
\item \textbf{Accuracy.} Accuracy (ACC) measures the proportion of predictions that exactly match the ground-truth labels. It is a standard metric for classification tasks and provides a straightforward indicator of overall correctness.
\begin{equation}
    ACC= \frac{TP + TN}{TP + TN + FP + FN}
\end{equation}
where TP, TN, FP, and FN denote the numbers of true positives, true negatives, false positives, and false negatives, respectively.
\item \textbf{Matthews Correlation Coefficient.} Matthews Correlation Coefficient (MCC) measures the correlation between the predicted and true labels across all classes by considering the full confusion matrix. Unlike accuracy, which may be inflated under class imbalance, MCC provides a balanced evaluation by jointly accounting for true positives, false positives, false negatives, and inter-class misclassifications. It can be interpreted as a generalized correlation coefficient that reflects how well the prediction distribution aligns with the true label distribution. A value of 1 indicates perfect prediction, 0 corresponds to no better than random guessing, and -1 indicates complete disagreement between predictions and ground truth.
\begin{equation}
\text{MCC} =
\frac{
\sum_{k}\sum_{l}\sum_{m} C_{kk} C_{lm} - C_{kl} C_{mk}
}{\sqrt{
\left( \sum_{k} T_k P_k \right)
\left( \sum_{k} T_k^2 - \sum_{k}\sum_{l} C_{kl} C_{lk} \right)
\left( \sum_{k} P_k^2 - \sum_{k}\sum_{l} C_{lk} C_{kl} \right)
}
}
\end{equation}
where $C$ is the confusion matrix. $C_{kl}$ is the number of samples with ground truth class k and predicted class l. $T_k = \sum_lC_{kl}$ and $P_k = \sum_lC_{lk}$.
\end{itemize}
\subsubsection{Evaluation Metrics for Open-QA}
\begin{itemize}[leftmargin=*]
\item \textbf{F1-RadGraph.} F1-RadGraph computes entity-level F1 scores based on RadGraph~\citep{jain2021radgraph} evaluation, evaluating the correctness of clinical entity extraction and relation grounding. It is widely used in medical language generation and assesses whether the model generates clinically valid content.
\begin{equation}
\text{F1-RadGraph} = \frac{2 \cdot \text{Precision} \cdot \text{Recall}}{\text{Precision} + \text{Recall}}
\end{equation}
where $\text{Precision} = \frac{TP}{TP + FP}, \text{Recall} = \frac{TP}{TP + FN}$. TP, FP, and FN are computed based on RadGraph entity matching.
\item \textbf{F1-BioBERT.} F1-BioBERT measures the semantic alignment between generated answers and ground truth using BioBERT~\citep{lee2020biobert} embeddings. It captures domain-specific semantic similarity, making it suitable to evaluate models of medical open-QA tasks. The computation of F1-BioBERT is similar to F1-RadGraph, but TP, FP, and FN are computed by semantic matching using BioBERT similarity.
\item \textbf{BLEU.} BLEU computes the precision of n-gram overlaps between predictions and ground truth to assess lexical fidelity.
\begin{equation}
\text{BLEU} = \text{BP} \cdot \exp(\sum^N_{n=1}W_n\log(P_n))
\end{equation}
where $\text{BP}$ is brevity penalty. $P_n$ is the modified n-gram precision. $W_n$ is weight of n-gram precision and $N$ is the maximum n-gram order.
\item \textbf{ROUGE-L.} ROUGE-L evaluates the longest common subsequence between predicted and ground truth answers, reflecting recall-oriented similarity. It captures global structural overlap and complements BLEU's precision-based evaluation.
\begin{equation}
\text{ROUGE-L} = 
\frac{(1 + \beta^2) \cdot R_{lcs} \cdot P_{lcs}}
{R_{lcs} + \beta^2 \cdot P_{lcs}}
\end{equation}
where $R_{lcs} = \frac{\text{LCS}(X,Y)}{|X|}, P_{lcs} = \frac{\text{LCS}(X,Y)}{|Y|}.$ $X$ is the ground truth, $Y$ is the predicted text. $\text{LCS}(X,Y)$ is the length of the longest common subsequence. $\beta$ controls the weighting of recall versus precision.
\end{itemize}
Together, these metrics, ACC, MCC, F1-RadGraph, F1-BioBERT, BLEU, and ROUGE-L, jointly assess correctness, robustness, clinical grounding, semantic fidelity, and textual similarity, offering a balanced and reliable evaluation across the diverse tasks in \Benchmark.
\section{Supplementary Experimental Results}
\label{sec:appendix-exps}

\subsection{Evaluation of \Name in Terms of Additional Metrics}
\label{sec:appendix-exps-main-metric}

In this section, we report detailed results for supplementary evaluation metrics, namely F1-RadGraph (F1-Rad)~\citep{yu2023evaluating} and BLEU-1~\citep{papineni2002bleu}, across 11 home care tasks in the open-QA settings.
The comparative results of \Name against baseline models are shown in Table~\ref{tab:exp_extend_open}. Overall, \Name achieves superior performance on both F1-RadGraph and BLEU-1, underscoring its capacity to generate accurate and contextually appropriate responses to open-ended instructions. An exception is observed in the MAG task, where \Name underperforms relative to LLaVA-Med v1.5. This discrepancy can be partly attributed to the nature of BLEU-1 and F1-RadGraph, which emphasize exact lexical overlap and structural correspondence (e.g., token repetition, entity mentions, and relation extraction). Models such as LLaVA-Med, which are trained and tuned extensively on medical corpora, are naturally more adept at reproducing domain-specific terminology and relation templates.
In contrast, F1-BioBERT and ROUGE-L place greater emphasis on semantic consistency and contextual alignment. On these metrics, \Name consistently attains higher scores, reflecting its ability to capture underlying clinical meaning and preserve semantic fidelity even when surface forms differ. These results suggest that \Name is particularly effective at producing semantically coherent and clinically relevant responses, which are essential for real-world medical advice and patient-facing applications.

\begin{table*}[ht]
\centering
\setlength{\tabcolsep}{4.2pt}
\renewcommand{\arraystretch}{1.15}
\caption{Comparison of \Name with baselines under \textit{open-QA} settings in \Benchmark.} 
\label{tab:exp_extend_open}
\resizebox{1.0\textwidth}{!}{
\begin{threeparttable}
{\setlength{\tabcolsep}{3pt}
\begin{tabular}{l*{2}{cc}{cc|}*{3}{cc}{cc|}*{4}{cc}{|cc}}
\toprule
\multirow{2}{*}{\textbf{Model}} &
\multicolumn{2}{c}{S2D} &
\multicolumn{2}{c}{DR} &
\multicolumn{2}{c|}{MAG} &
\multicolumn{2}{c}{Diabetes} &
\multicolumn{2}{c}{Obesity} &
\multicolumn{2}{c}{Heart} &
\multicolumn{2}{c|}{CKD} &
\multicolumn{2}{c}{Food} &
\multicolumn{2}{c}{Sleep} &
\multicolumn{2}{c}{Skin} &
\multicolumn{2}{c}{Oral}&
\multicolumn{2}{|c}{Avg.}\\
\cmidrule(lr){2-3}\cmidrule(lr){4-5}\cmidrule(lr){6-7}\cmidrule(lr){8-9}\cmidrule(lr){10-11}\cmidrule(lr){12-13}\cmidrule(lr){14-15}\cmidrule(lr){16-17}\cmidrule(lr){18-19}\cmidrule(lr){20-21}\cmidrule(lr){22-23}\cmidrule(lr){24-25}
 & F1-Rad & BLEU-1 & F1-Rad & BLEU-1 & F1-Rad & BLEU-1 & F1-Rad & BLEU-1 & F1-Rad & BLEU-1 & F1-Rad & BLEU-1 & F1-Rad & BLEU-1& F1-Rad & BLEU-1& F1-Rad & BLEU-1& F1-Rad & BLEU-1& F1-Rad & BLEU-1& F1-Rad & BLEU-1\\
\midrule
\multicolumn{25}{c}{\textit{General Domain Models}} \\
\midrule
LLaVA-1.5-7B& 0.84 & 4.53 & 0.21 & 5.44 & 11.90 & 18.58 & 13.59 & 13.71 & 8.87 & 9.99 & 1.48 & 7.77 & 13.73 & 7.96 & 1.18 & 1.16 & 0.62 & 7.49 & 18.16 & 29.31 & 18.38 & 27.68&8.09&12.15\\
InstructBLIP-7B& 0.56 & 1.31 & 0.47 & 4.30 & 0.83 & 0.00 & 4.80 & 6.79 & 3.63 & 2.77 & 0.47 & 5.12 & \underline{16.27} & \underline{16.10} & 1.27 & \underline{5.69} & 0.58 & 6.07 & 9.46 & 1.37 & 10.46 & 5.94&4.44&5.04\\
Llama 3.2-11B & 0.62 & 3.41 & 0.90 & 4.69 & 13.59 & 20.50 & 9.95 & 8.59 & 8.91 & 8.68 & 0.92 & 2.76 & 9.40 & 7.28 & 1.35 & 1.37 & 0.49 & 2.97 & 11.74 & 15.09 & 11.38 & 14.45&6.30&8.16\\
Yi-VL-6B& 0.62 & 2.83 & 0.22 & 4.58 & 12.83 & 17.83 & 14.45 & 16.40 & 8.18 & 8.49 & 2.19 & 8.21 & 13.93 & 9.41 & 1.51 & 3.21 & 0.67 & 9.39 & 18.09 & 28.26 & 16.71 & 26.03&8.13&12.24\\
InternVL3-8B& 0.58 & 2.00 & 0.73 & 3.18 & 10.75 & 13.69 & 15.43 & 13.25 & 7.00 & 5.34 & 1.58 & 5.15 & 10.52 & 5.15 & 1.35 & 1.49 & 0.35 & 3.25 & 15.41 & 16.70 & 14.96 & 14.24&7.15&7.59\\
Qwen2.5-VL-7B& 0.55 & 2.18 & 1.14 & \underline{13.24} & 10.72 & 14.21 & 10.71 & 17.37 & 7.04 & 5.78 & 1.24 & 5.29 & 8.41 & 4.04 & \underline{2.06} & 5.29 & 0.37 & 4.23 & 17.71 & 34.91 & 15.37 & 32.17&6.85&12.61\\
Qwen3-VL-8B & 0.71 & 1.63 & 0.43 & 3.23 & 9.64 & 14.60 & 12.27 & 8.09 & 7.61 & 6.40 & 0.76 & 3.17 & 7.37 & 3.76 & 1.18 & 0.91 & 0.36 & 3.61 & 12.07 & 11.62 & 12.24 & 12.19 & 5.88 & 6.29 \\
Gemma 3-4B& 0.39 & 1.23 & 1.11 & 2.20 & 6.29 & 5.95 & 10.51 & 4.73 & 5.08 & 3.97 & 0.79 & 2.65 & 5.05 & 1.61 & 1.02 & 1.28 & 0.43 & 2.64 & 8.06 & 6.88 & 8.20 & 6.58&4.27&3.61\\
Claude 3 Haiku& \underline{0.96} & 4.20 & 0.83 & 6.55 & 14.45 & 22.39 & 16.69 & 14.70 & 10.48 & 9.46 & 1.17 & 7.81 & 10.14 & 7.03 & 0.93 & 1.46 & 0.51 & 6.99 & 18.62 & 28.83 & 18.74 & 27.16&8.50&12.42\\
GPT-4o Mini& 0.77 & 3.24 & 1.00 & 4.75 & 15.23 & \underline{22.89} & 14.97 & 11.54 & 9.86 & 8.64 & 1.14 & 5.16 & 12.69 & 9.46 & 1.99 & 3.39 & 0.43 & 4.81 & 17.59 & 21.21 & 15.83 & 19.45&8.32&10.41\\
\midrule
\multicolumn{25}{c}{\textit{Medical Domain Models}} \\
\midrule
LLaVA-Med v1.5-7B &  0.86 & \underline{9.49} & 0.30 & 10.88 & \textbf{17.20} & \textbf{28.41} & 15.64 & 18.70 & \underline{11.27} & \underline{15.14} & \underline{3.97} & \underline{12.94} & 8.76 & 14.78 & 1.57 & 1.96 & \underline{1.20} & \underline{14.72} & \underline{22.38} & \underline{40.24} & 17.45 & 36.28&9.14&\underline{18.50}\\
Med-Flamingo-7B& 0.82 & 2.65 & \underline{1.57} & 4.51 & 7.85 & 6.71 & 9.33 & 5.91 & 7.17 & 4.91 & 1.41 & 1.89 & 5.22 & 3.51 & 0.30 & 0.14 & 0.35 & 2.23 & 6.08 & 8.06 & 8.79 & 6.60&4.44&4.28\\
HuatuoGPT-Vision-7B& 0.68 & 3.80 & 0.38 & 6.44 & 9.04 & 13.18 & 14.43 & 16.02 & 7.92 & 10.46 & 0.95 & 6.10 & 7.75 & 6.13 & 0.50 & 0.66 & 0.43 & 5.79 & 17.44 & 26.59 & 23.82 & 29.99&7.58&11.38\\
MedGemma-4B& 0.45 & 3.90 & 0.36 & 6.55 & 10.94 & 15.59 & 11.25 & 8.88 & 6.27 & 6.02 & 1.81 & 4.40 & 8.00 & 5.04 & 1.01 & 0.76 & 0.45 & 4.81 & 11.03 & 10.00 & 13.11 & 14.02&5.88&7.27\\
HealthGPT-3.8B& 0.89 & 4.14 & 0.30 & 5.95 & 14.71 & 20.26 & \underline{18.33} & 18.58 & 10.28 & 9.43 & 0.98 & 8.07 & 13.01 & 8.64 & 1.25 & 1.26 & 0.63 & 8.60 & 21.67 & 38.23 & \underline{29.96} & \underline{39.81}&\underline{10.18}&14.82\\
Med-R1-2B& 0.43 & 2.34 & 0.56 & 3.65 & 11.82 & 17.78 & 14.04 & 8.56 & 7.71 & 6.71 & 0.74 & 2.63 & 10.47 & 7.11 & 0.76 & 1.15 & 0.38 & 2.57 & 14.93 & 17.63 & 14.63 & 20.97&6.95&8.28\\
Lingshu-7B& 0.89 & 4.60 & 0.37 & 6.77 & 13.88 & 18.08 & 18.21 & \underline{20.71} & 9.83 & 11.16 & 1.26 & 8.59 & 13.26 & 12.21 & 1.05 & 1.45 & 0.45 & 7.32 & 19.59 & 36.87 & 23.92 & 38.18&9.34&15.09\\
MedVLM-R1-2B & 0.60 & 2.21 & 0.20 & 3.82 & 13.81 & 17.97 & 13.69 & 8.97 & 9.02 & 7.11 & 1.54 & 3.26 & 14.48 & 8.43 & 1.99 & 2.24 & 0.51 & 5.31 & 14.91 & 20.68 & 12.83 & 19.61&7.60&9.06\\
\rowcolor{blue!10} \Name-3.8B & \textbf{42.97} & \textbf{60.84} & \textbf{15.92} & \textbf{31.63} & \underline{16.01} & 19.25 & \textbf{40.72} & \textbf{57.65} & \textbf{31.87} & \textbf{54.09} & \textbf{12.34} & \textbf{36.31} & \textbf{57.32} & \textbf{66.98} & \textbf{15.73} & \textbf{87.93} & \textbf{3.16} & \textbf{46.24} & \textbf{44.60} & \textbf{63.76} & \textbf{66.89} & \textbf{77.89}&\textbf{30.07}&\textbf{51.76}\\
\bottomrule
\end{tabular}
}
\end{threeparttable}
}
\end{table*}

\subsection{Expert Evaluation with GPT-5}
\label{sec:appendix-exps-gpt5}
\begin{figure}[h]
    \centering
    \begin{subfigure}[t]{0.44\textwidth}
        \centering
        \includegraphics[width=\linewidth]{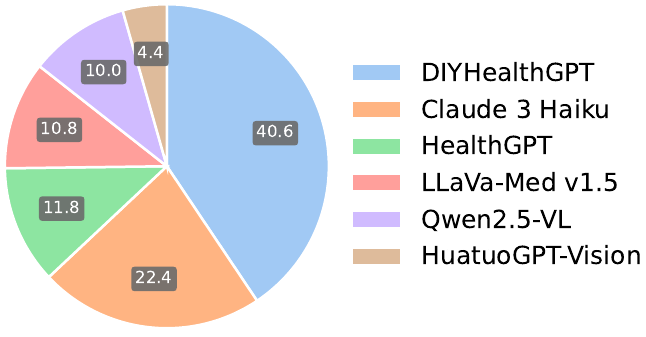}
        % \vspace{-2mm}
        % \caption{Proportion of the first preference from GPT-5.}
        \caption{Distribution of the first preference by GPT-5.}
        \label{fig:pie_gpt}
    \end{subfigure}
    \hfill
    \begin{subfigure}[t]{0.52\textwidth}
        \centering
        \includegraphics[width=\linewidth]{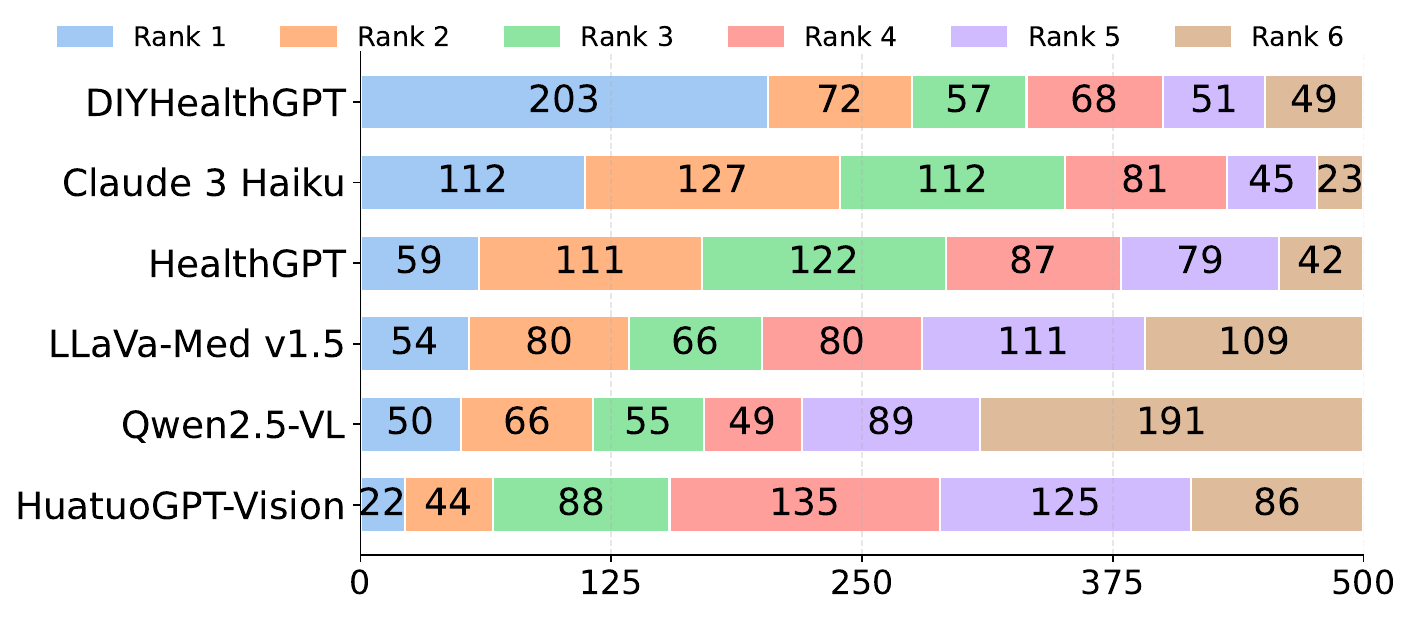}
        % \caption{Distribution of ranks assigned by GPT-5.}
        \caption{Full ranking landscape across models by GPT-5.}
        \label{fig:rank_stacked_gpt}
    \end{subfigure}
    % \caption{Evaluation results from GPT-5.}
    \caption{Results of the GPT-5 review.}
    \label{fig:gpt_eval}
\end{figure}

Beyond clinical expert review, we perform an additional evaluation using GPT-5, following the same sampling procedure and ranking guidelines described in~\cref{sec:ablation study and in-depth analysis}. As shown in Figure~\ref{fig:gpt_eval}, the results are consistent with the clinical expert assessment. GPT-5 favors \Name as the first preference model, with 
% its ranks heavily weighted toward the top ranks. 
rankings concentrated at the top.
Moreover, GPT-5 consistently identifies \Name as providing the most faithful and comprehensive responses.

\subsection{Hyperparameter Sensitivity Study on Number of Experts}
\label{sec:appendix-exps-sensitivity}

\begin{figure}[t]
    \begin{center}
    % \vspace{-7mm}
    \includegraphics[width=0.7\textwidth]{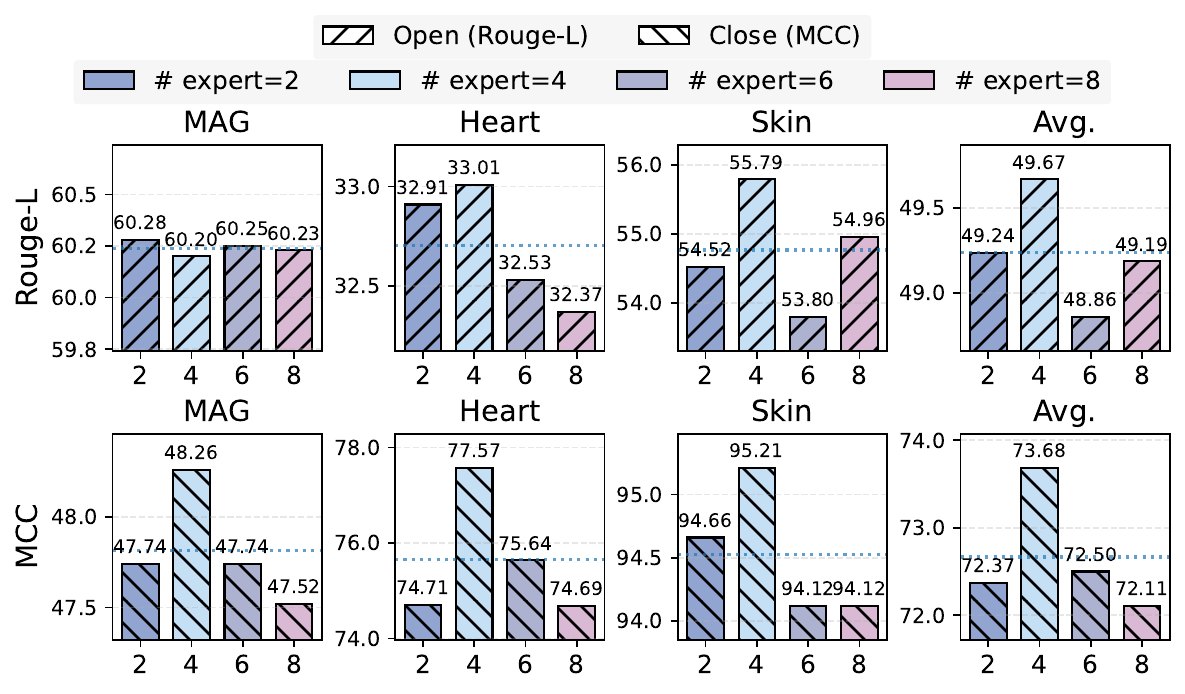}
        % \vspace{-4mm}
        \caption{Sensitivity study on the number of experts in \Name.}
    \label{fig:abl_num_experts}
    \end{center}
\end{figure}

We examine the impact of the number of experts on model performance, using MAG, Heart, and Skin tasks, along with the average performance as representative cases. The results are summarized in Figure~\ref{fig:abl_num_experts}. The Rouge-L remains stable in the closed-QA MAG task, which could be attributed to its reliance on broad knowledge, as discussed in Section~\ref{main_results}. On average, four experts achieve the optimal balance between Rouge-L and MCC.
% both Rouge-L and MCC consistently suggest that employing four experts provides the optimal balance.
% cs: avoid misunderstandi of the number of experts.
\subsection{User Feedback Analysis}
\label{sec:appendix-user-feedback}
To assess real-world usability, we conduct a human-centered user study with 36 participants spanning diverse characteristics, such as age, gender, and self-rated health status. The study comprises two components: answer-level evaluation and system-level evaluation. All items are assessed using a 5-point scale, where 1 indicates strongly disagree and 5 indicates strongly agree.

\begin{figure}[th]
    \centering
    \begin{subfigure}[t]{0.40\textwidth}
        \centering
        \textbf{}\vspace{-2mm}
        \includegraphics[width=\linewidth]{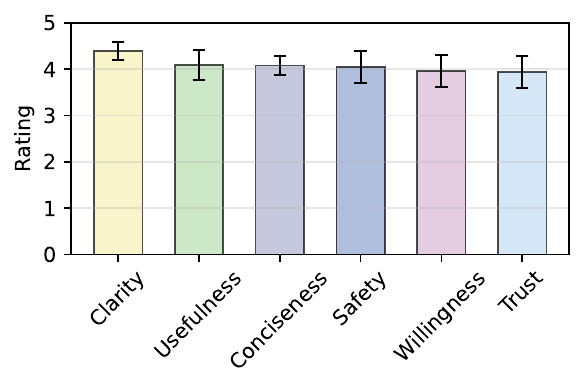}
        % \vspace{-7mm}
        % \caption{Overall ratings of answer-level evaluation.}
        % \label{fig:user_study_answer}
    \end{subfigure}
    \hfill
    \begin{subfigure}[t]{0.59\textwidth}
        \centering
        \textbf{}\vspace{-1mm}
        \includegraphics[width=\linewidth]{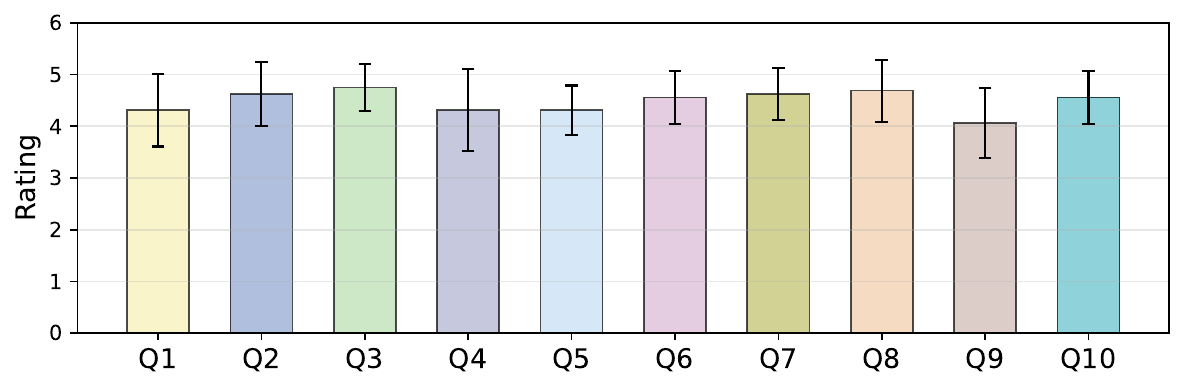}
        % \vspace{2mm}
        % \caption{Overall ratings of system-level evaluation.}
        % \label{fig:user_study_system}
    \end{subfigure}
    \\
    \begin{subfigure}[t]{0.40\textwidth}
        \centering
        \caption{Overall ratings of answer-level evaluation.}
        \label{fig:user_study_answer}
    \end{subfigure}
    \hfill
    \begin{subfigure}[t]{0.59\textwidth}
        \centering
        \caption{Overall ratings of system-level evaluation.}
        \label{fig:user_study_system}
    \end{subfigure}
    \caption{Results of user feedback analysis.}
    \label{fig:user_study}
\end{figure}

For answer-level evaluation, participants are asked to rate the quality of the model-generated answers from \Name across six dimensions related to usability and interaction:
\begin{itemize}[leftmargin=*]
    \item \textbf{Clarity}: This answer is easy to understand.
    \item \textbf{Usefulness}: This answer is useful.
    \item \textbf{Conciseness}: The length is appropriate, not overly long or short.
    \item \textbf{Safety}: It's safe to follow the answer.
    \item \textbf{Willingness}: I would be willing to follow this answer.
    \item \textbf{Trust}: I would trust the answer.
\end{itemize}
As shown in Figure~\ref{fig:user_study_answer}, all six dimensions received scores close to or above 4.0 on average, indicating high perceived clarity, usefulness, conciseness, safety, willingness to follow, and trustworthiness.
%% kp: whether "trustworthiness" or "trust"? - need to be consistent in text and figure 11
This suggests that the generated answers are not only considered reliable and contextually appropriate for practical use but also well aligned with users' expectations in real-world home care scenarios.

To evaluate the overall system usability, we employ the System Usability Scale~\citep{brooke1996sus}, a widely used standardized evaluation for assessing system usability.
%% kp: add a ref to SUS
Following prior studies~\citep{tchemeube2023evaluating, aljamaan2024chatgpt}, participants are asked to complete the following SUS items:
\begin{itemize}[leftmargin=*]
    \item Q1: I think that I would like to use this system frequently.
    \item Q2: I found the system unnecessarily complex.
    \item Q3: I thought the system was easy to use.
    \item Q4: I think that I would need the support of a technical person to be able to use this system.
    \item Q5: I found the various functions in this system were well integrated.
    \item Q6: I thought there was too much inconsistency in this system.
    \item Q7: I would imagine that most people would learn to use this system very quickly.
    \item Q8: I found the system very cumbersome to use. 
    \item Q9: I felt very confident using the system.
    \item Q10: I needed to learn a lot of things before I could get going with this system.
\end{itemize}
The final SUS score is calculated by first converting each rating to an adjusted score and then scaling the total to a 0--100 range. The score is computed as follows:
\begin{equation}
    \text{SUS} = 2.5 \times \sum_{i=1}^{10} score'_i,
\quad
\text{where} \quad
score'_i = 
\begin{cases}
rating_i - 1, & \text{if } i \text{ is odd (positive item)} \\
5 - rating_i, & \text{if } i \text{ is even (negative item)}
\end{cases}
\end{equation}
For visualization clarity, Figure~\ref{fig:user_study_system} presents the ratings of negative items that have been rescaled to the same direction as positive ones (i.e., higher scores indicate better). Based on participants’ responses, \Name achieved a SUS score of 87.03. According to established interpretation guidelines~\citep{bangor2009determining, brooke1996sus}, this score falls in the excellent usability, indicating that \Name is not only effective in generating answers but also user-friendly and has strong potential for practical deployment.

\subsection{Subgroup Study}
\label{sec:appendix-subgroup-study}
To assess the robustness of \Name across different subgroups and to support safe deployment in healthcare scenarios, we conducted a preliminary subgroup analysis. Due to dataset de-identification and the nature of home-care data acquisition, demographic attributes such as race, ethnicity, skin tone, and device type are not consistently annotated, which makes comprehensive subgroup evaluation infeasible at this stage. Therefore, we focus on open-QA tasks where relevant metadata are available and perform subgroup analyses on (1) age and gender subgroup analysis for the CKD and Heart tasks, and (2) BMI and vegetable-consumption level subgroup analysis for the Obesity task.

\begin{table}[t]
    \centering
    \setlength{\tabcolsep}{4pt} 
    \renewcommand{\arraystretch}{1.2} 
    \scriptsize
    \caption{Subgroup results for CKD, heart, and obesity tasks in open-QA settings.}
    \begin{subtable}[t]{0.50\textwidth}
        \centering
        \caption{Subgroup results for CKD and heart tasks.}
        \begin{tabular}{lccc|cc}
            \toprule
            \multirow{2}{*}{Subgroup} & \multirow{2}{*}{Size} & \multicolumn{2}{c|}{CKD}& \multicolumn{2}{c}{Heart}\\
            \cmidrule(lr){3-4}\cmidrule(lr){5-6}
            & & F1-Bio & RL & F1-Bio & RL \\
            \midrule
             Age $<$ 18& 7 &93.08&72.45&81.50&29.67\\
             18$\leq$Age$<$40 & 134 &90.51&65.24&83.05&34.38\\
             41$\leq$Age$<$65 &212&91.64&68.32&82.54&32.95\\
             Age $\geq$ 65 &158&90.06&62.13&82.47&33.42\\
            \midrule
            \midrule
             Gender: Male& 193 &89.64&61.49&82.47&32.90\\
            Gender: Female & 205 &89.93&62.18&82.67&33.62\\
            \bottomrule
        \end{tabular}
        \label{tab: subgroup_CKD_Heart}
    \end{subtable}
    \hfill
    \begin{subtable}[t]{0.47\textwidth}
        \centering
        \scriptsize
        \caption{Subgroup Results for the Obesity Task.}
        \begin{tabular}{lccccc}
            \toprule
            \multirow{2}{*}{Subgroup} & \multirow{2}{*}{Size} & \multicolumn{4}{c}{Obesity}\\
            \cmidrule(lr){3-6}
            & & F1-Bio & RL & F1-Rad & BLEU \\
            \midrule
            BMI $<$ 18.5 & 36 & 85.66&43.11&34.62&52.67\\
            18.5$\leq$BMI$<$ 25 & 57 & 84.77&41.37&35.64&53.97\\
            25$\leq$BMI$<$30& 98& 84.34&45.81&26.95&52.13\\
            BMI $\geq$30& 166& 85.20&47.15&32.89&55.14\\
            \midrule
            \midrule
            Vegetable: 1 & 44& 85.65&49.07&32.23&55.68\\
            Vegetable: 2 & 201& 84.84&45.73&31.92&54.31\\
            Vegetable: 3 & 112&84.85&43.41&31.64&52.97\\
            \bottomrule
        \end{tabular}
        \label{tab: subgroup_obesity}
    \end{subtable}
    \label{tab: subgroup_results}
\end{table}

As shown in Table~\ref{tab: subgroup_results}, across all subgroups, the performance remains stable, with fluctuations within a small margin.
For instance, the variation in F1-Bio across different age subgroups is within 3\%, and the differences between male and female subgroups are below 1\%. Similarly, the obesity task shows only moderate variation across BMI and vegetable-consumption levels.

Regarding the large variation in the RL metric for the Age $<18$ subgroup and the Age $\ge$ 65 subgroup (72.45 vs. 62.13), we note that the Age $<18$ subgroup contains only seven samples, which makes the results statistically unstable and sensitive to individual cases. Therefore, the difference is more likely attributable to insufficient cohort size rather than systematic model bias. Besides, although the F1-Rad for the 25 $\le$ BMI $< 30$ subgroup is much lower than that for the 18.5 $\le$ BMI $< 25$ subgroup (26.95 vs. 35.64), this variation is consistent with clinical expectations rather than indicating model bias: individuals in the overweight subgroup (25~$\le$~BMI~$<$~30) often present subtle or non-specific symptoms, making their cases harder to detect~\citep{brod2018development}. Since F1-Rad is highly sensitive to biomedical terminology, such cases naturally require stronger reasoning and interpretation ability. This suggests an opportunity for future personalized adaptation, rather than revealing a systematic risk.

Overall, the subgroup results indicate that \Name generally maintains consistent performance across different user populations, supporting its potential for real-world deployment.

\subsection{Complete Comparison with Fine-tuned Baselines}
\label{sec:appendix-complete-fine-tuned}
To examine whether the performance gain of \Name is solely due to training on DIYHealth-900K, we conduct fine-tuning experiments on two representative models, Gemma 3-4B and LLaVA-Med v1.5-7B, using exactly the same training data with our limited computational resources. As shown in Table~\ref{appendix: tab: fine-tuned comparison}, \Name-3.8B consistently outperforms Gemma 3-4B by a substantial margin on all six evaluation metrics and even surpasses LLaVA-Med v1.5-7B on ACC, MCC, F1-Bio, and RL, despite LLaVA-Med having a larger model size. These results indicate that the observed improvements cannot be attributed only to the dataset. Instead, the combination of \Dataset, the training strategy, and the H$^2$LoRA architecture contributes to the model's effectiveness.

\begin{table}[h]
\centering
% \scriptsize
\setlength{\tabcolsep}{4.2pt}
\caption{Complete Comparison with fine-tuned baselines.}
\label{appendix: tab: fine-tuned comparison}
% \renewcommand{\arraystretch}{0.4}
% \resizebox{0.7\textwidth}{!}{
% \begin{threeparttable}
{\setlength{\tabcolsep}{3pt}
\begin{tabular}{lcc|cccc}
\toprule
\multirow{2}{*}{Model}
& \multicolumn{2}{c|}{Closed-QA}
 & \multicolumn{4}{c}{Open-QA}
 \\
\cmidrule(lr){2-3}\cmidrule(lr){4-7}
 & ACC & MCC & F1-Bio & RL & F1-Rad & BLEU 
 \\
\midrule
Gemma 3-4B & 80.96 &74.42&84.72&41.59&26.67&45.35\\
LLaVA-Med v1.5-7B& 77.58 &69.51& 86.28&49.63&\textbf{30.85}&\textbf{53.21} \\
\Name-3.8B & \textbf{86.80} &\textbf{82.36} & \textbf{87.34} &\textbf{52.11}&30.07&51.76\\
\bottomrule
\end{tabular}
\vspace{-4mm}
}
% \end{threeparttable}
% }
\end{table}

\subsection{Inter-rater Agreement Analysis}
\label{sec:appendix-inter-rater}
To ensure the reliability of clinical expert review for different models, we conduct a comprehensive inter-rater agreement analysis based on the rankings provided by five independent raters. Each question consists of six model-generated outputs, and raters are asked to assign an ordinal rank from 1 to 6. As the annotations follow an ordinal preference-ranking scheme, we assess the degree of agreement using soft agreement criterion~\citep{stemler2004comparison, fu2012modelling} rather than strict rank matching. Two raters are considered to be in agreement on a given answer if the absolute difference between their assigned ranks is no greater than one, i.e. $r_1$ – $r_2$ $\le$ 1. 

\begin{figure}[ht]
\centering
\vspace{-2mm}
\includegraphics[width=0.8\textwidth]{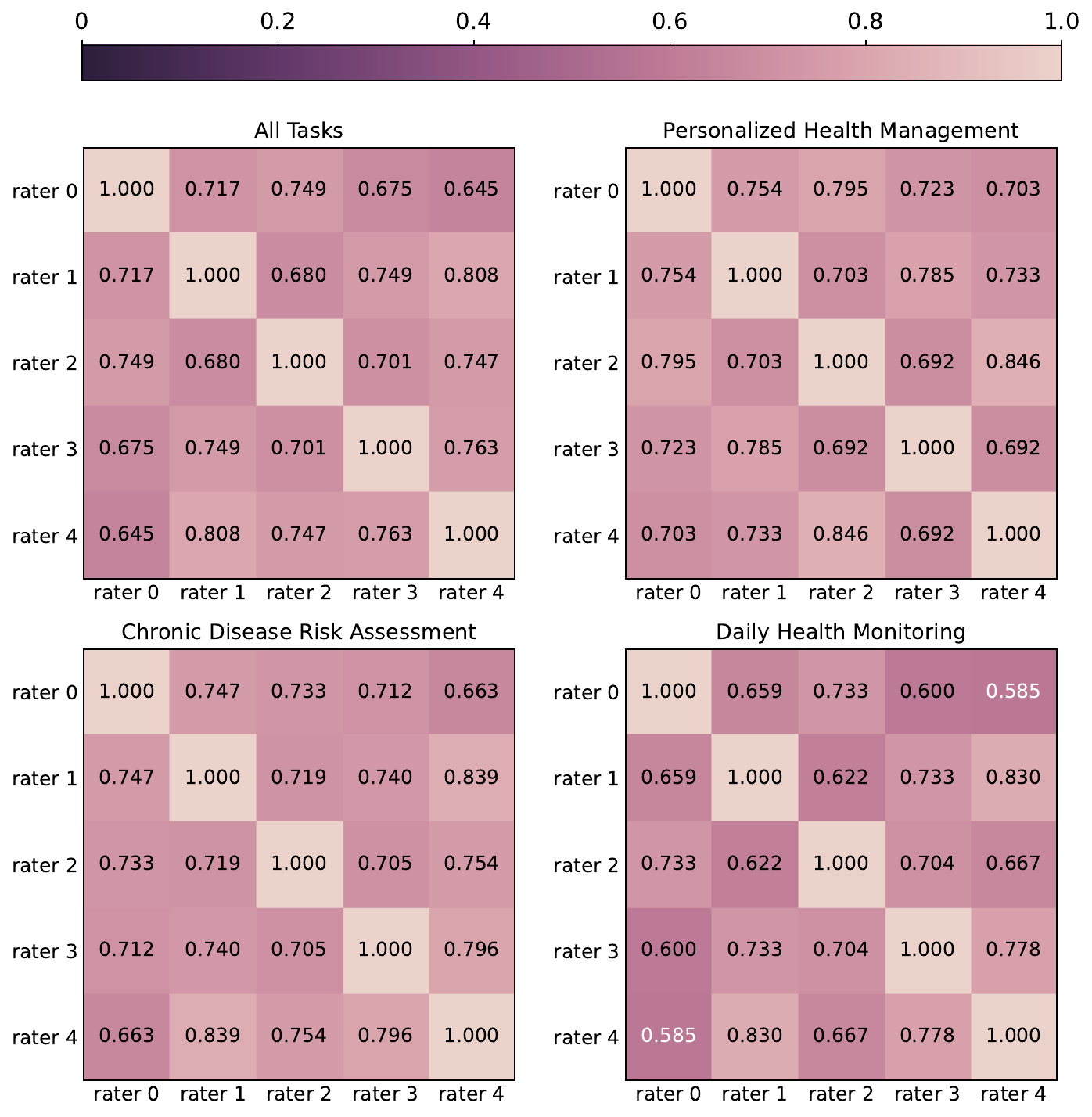}
\vspace{-2mm}
\caption{Results of inter-rater agreement analysis.}
\vspace{-3mm}
\label{fig:inter_rater_analysis}
\end{figure}

Under this soft agreement criterion, we compute Cohen’s kappa score to quantify inter-rater agreement. The results are summarized in Figure~\ref{fig:inter_rater_analysis}. Based on the results, Cohen’s kappa score ranges from [0.645, 0.808], [0.692, 0.846], [0.663, 0.839], [0.585, 0.830] for all tasks, personalized health management tasks, chronic disease risk assessment tasks, and daily health monitoring tasks, respectively. Notably, only one rater pair (rater 0 and rater 4) in the daily health monitoring tasks shows a kappa value of 0.585, which is slightly below but still close to 0.600. Overall, according to the widely adopted interpretation guideline by Landis and Koch~\citep{landis1977measurement}, the majority of kappa values fall within the substantial agreement range from 0.61 to 0.80, indicating a strong level of reliability in expert judgments.

\section{Discussion and Outlook}
\label{sec:appendix-discussion}

\subsection{Broader Impact}
\label{sec:appendix-impact}
Beyond its technical innovations, \Suite has the potential to deliver significant impact across age groups and health domains. For children, it can promote dietary education, support oral hygiene tracking, and facilitate developmental health monitoring. For middle-aged individuals, it can assist with stress management, enable early screening for chronic diseases, and provide guidance for lifestyle optimization. For elderly users, it offers continuous health monitoring, early detection of geriatric syndromes, and chronic disease management.
By reasoning over diverse multimodal signals, including dietary patterns, oral health, visceral functions, dermatological changes, among others, \Suite enables personalized and context-aware health management.
Through these capabilities, \Suite lays the groundwork for a new generation of AI-driven, human-centered healthcare frameworks that are comprehensive, responsive to individual needs, and accessible beyond conventional clinical environments.

\subsection{Future Directions}
\label{sec:appendix-future}
In this work, we envision a future where intelligent health assistants function as daily companions---proactive, trustworthy, and seamlessly integrated into 
% the rhythms of personal health.
personal health routines.
Moving forward, we call for a collaborative agenda spanning AI, medicine, public health, and human-computer interaction to realize this vision.
Promising directions include lifelong personalization through federated and continual learning, integration of clinician oversight to ensure medical alignment, and development of robust privacy-preserving mechanisms for secure large-scale deployment.
We believe these advances have the potential to reshape the global health landscape by augmenting care delivery and expanding broad access to health expertise at home.

%% jq: this part should be consistent with the option selected at the time of submission
%% kp: in submission form "Large Language Models: Yes, to aid or polish writing. Details are described in the paper." - we may need to provide some details
%% jq: maybe can change the option to "Yes, but for none of the above purposes. Details are described in the paper." and can mention we use LLM to (1)construct/rewrite QA pairs in our dataset, (2) evaluate the model output?
% \section{The Use of LLMs}
% \label{sec:appendix-LLM}
% In this work, we employ LLMs in two strictly controlled manners. First, we use LLMs to rephrase source data during dataset construction (see Section~\ref{sec:dataset} and Appendix~\ref{sec:appendix-dataset}), adapting the content to home care while preserving its clinical context.
% Second, we leverage LLMs as 
% % experts to evaluate 
% evaluators to compare
% our model with baselines (see Appendix~\ref{sec:appendix-exps-gpt5}), following clearly defined criteria for consistency and fairness.

%%jq: can mention the corresponding sections

\section{Case Study}
\label{sec:appendix-exps-case}
\begin{figure}[!ht]
\centering
\includegraphics[width=0.9\textwidth]{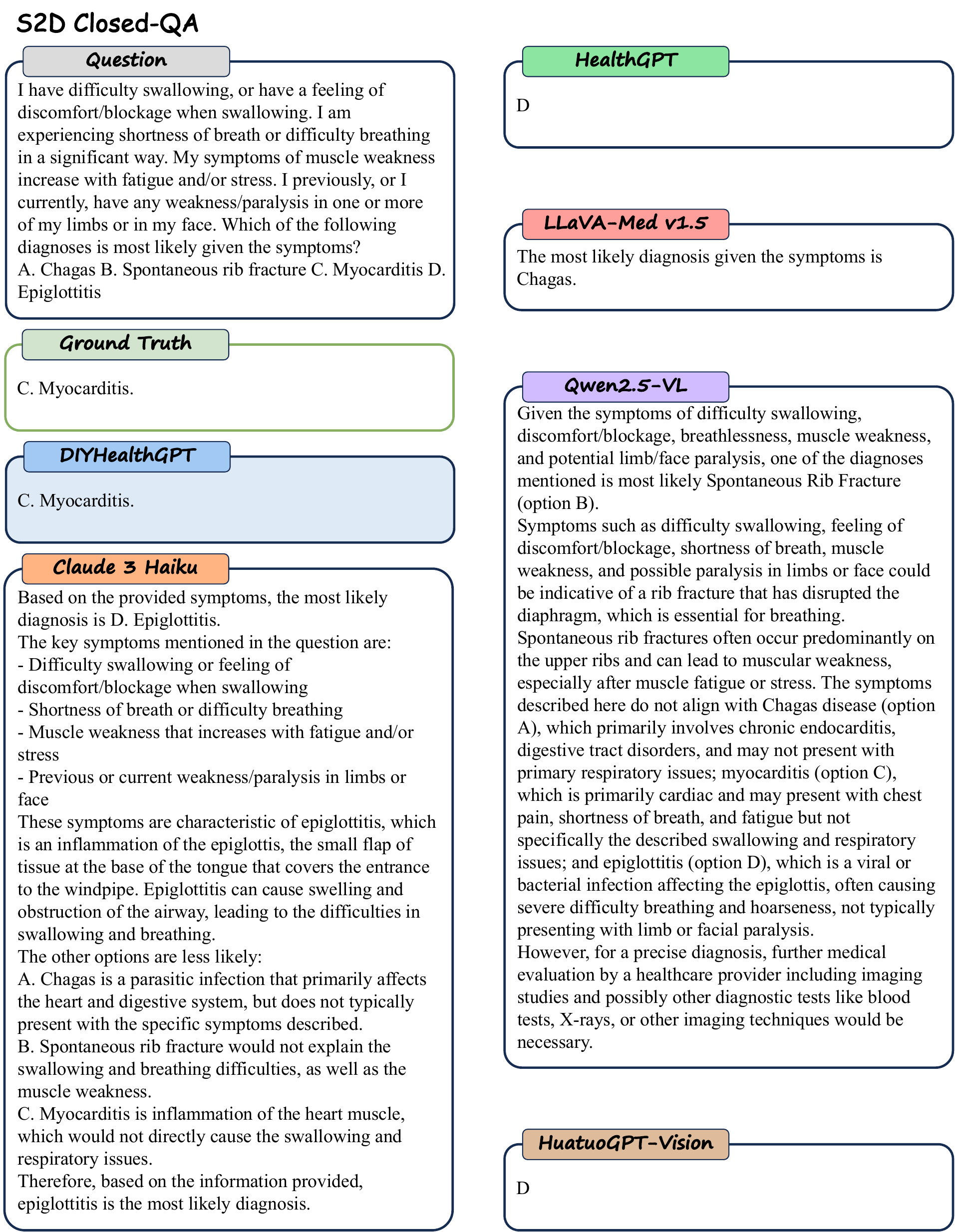}
\caption{A case study of the S2D task under \textit{closed-QA} settings. Some baselines output only the option label, without the corresponding option content.}
\label{fig:case_s2d_close}
\end{figure}

\begin{figure}[!ht]
\centering
\includegraphics[width=\textwidth]{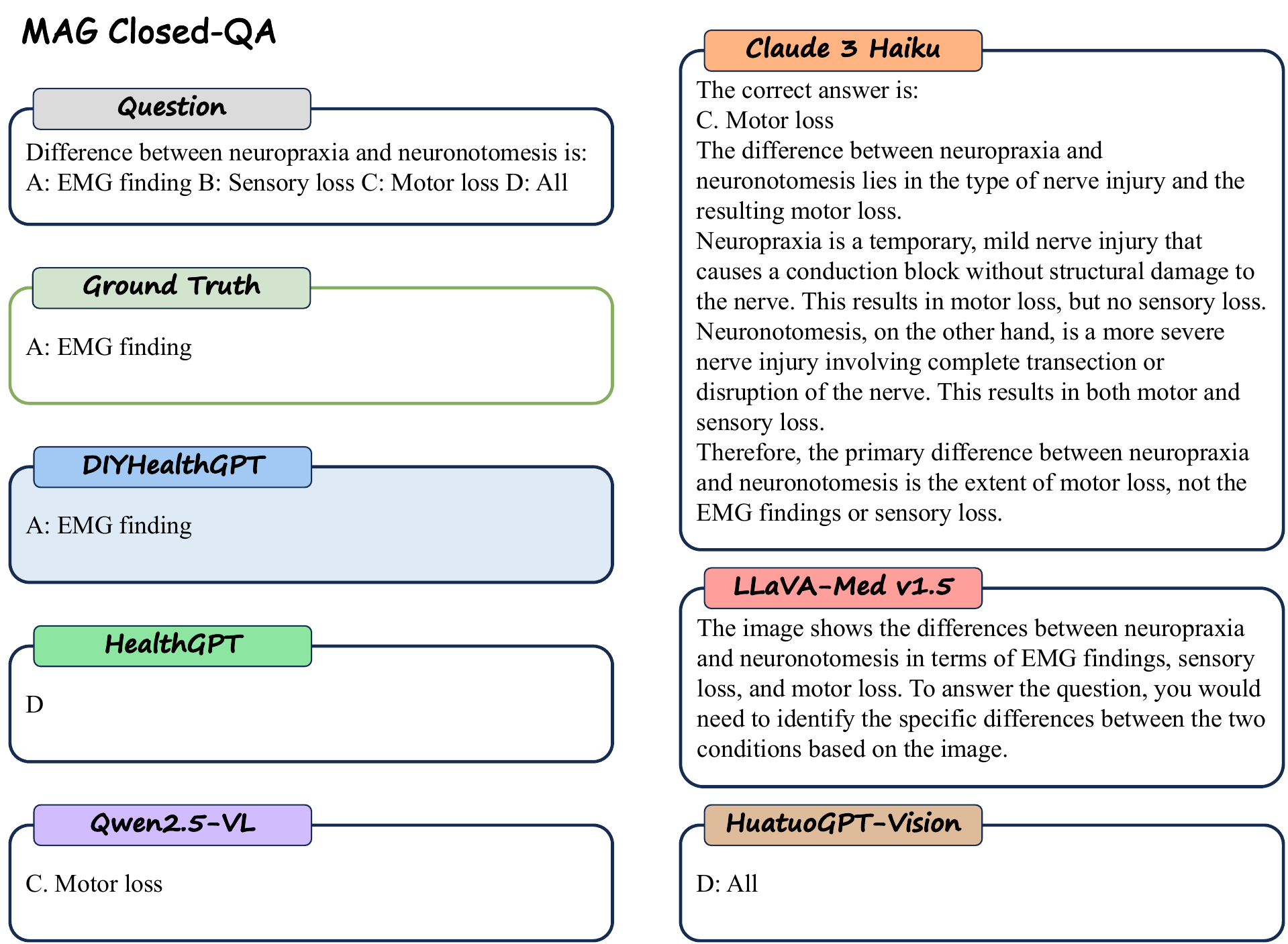}
\caption{A case study of the MAG task under \textit{closed-QA} settings.}
\label{fig:case_mag_close}
\end{figure}

\begin{figure}[!ht]
\centering
\includegraphics[width=\textwidth]{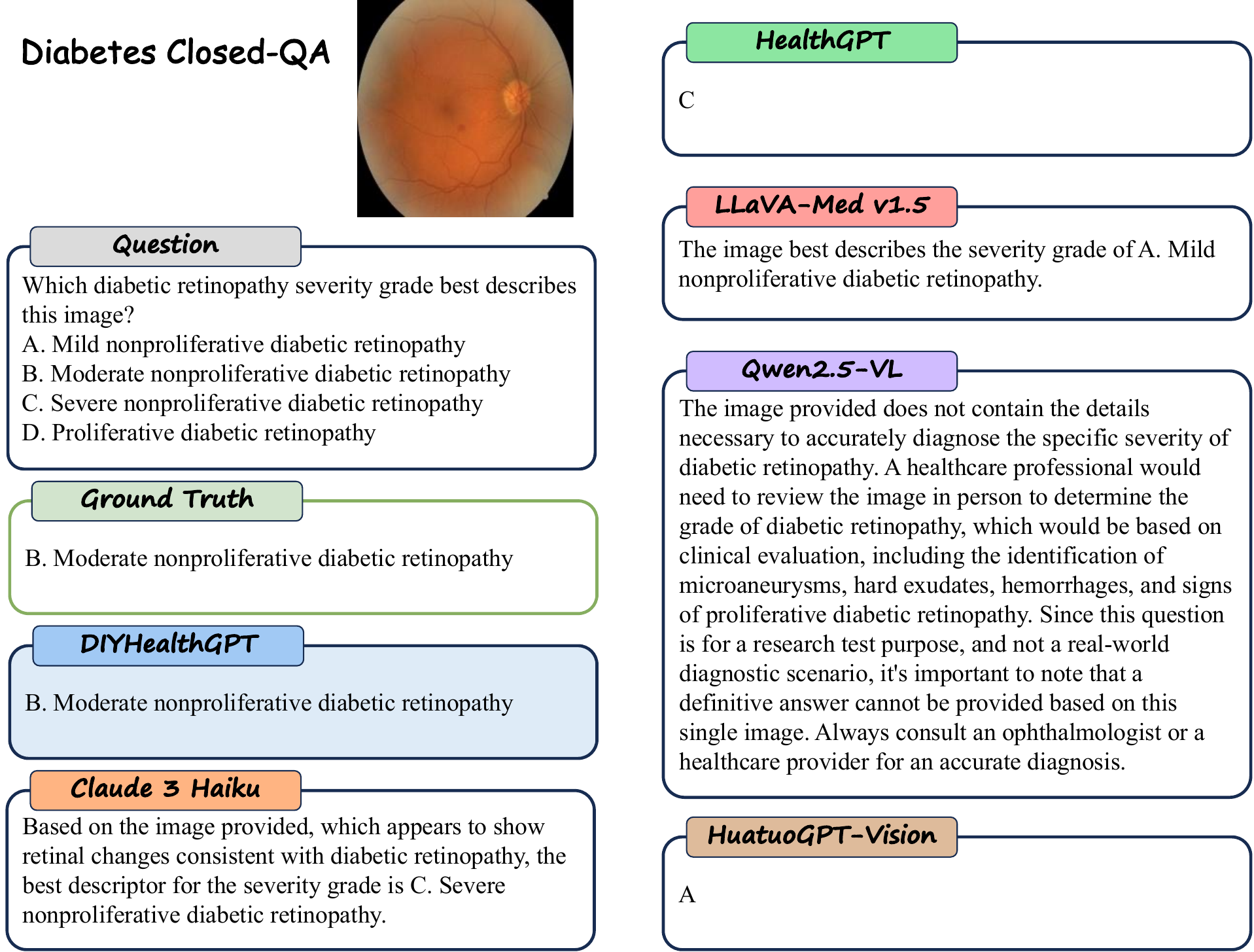}
\caption{A case study of the Diabetes task under \textit{closed-QA} settings.}
\label{fig:case_diabetes_close}
\end{figure}

\begin{figure}[!ht]
\centering
\includegraphics[width=\textwidth]{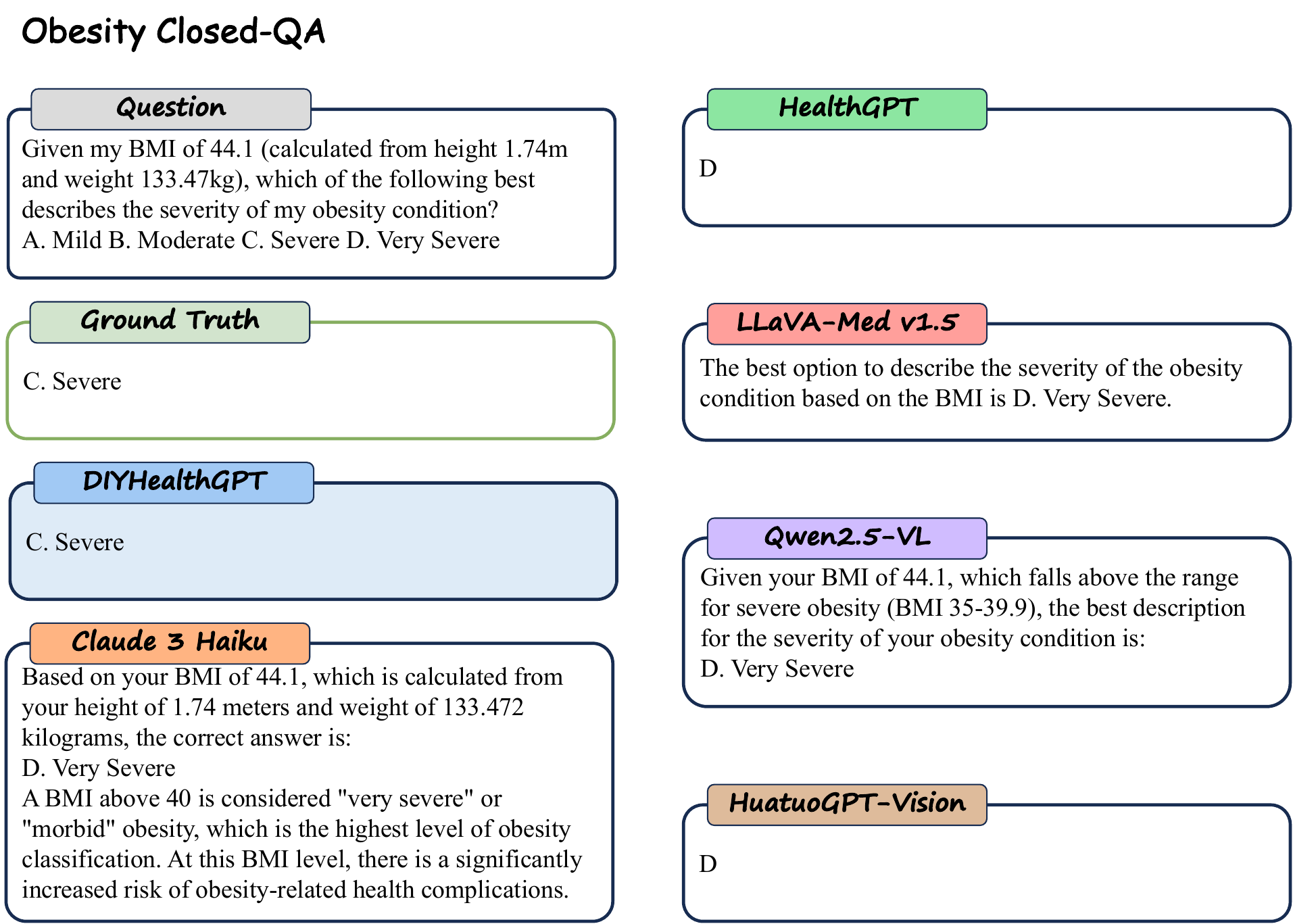}
\caption{A case study of the Obesity task under \textit{closed-QA} settings.}
\label{fig:case_obesity_close}
\end{figure}

\begin{figure}[!ht]
\centering
\includegraphics[width=\textwidth]{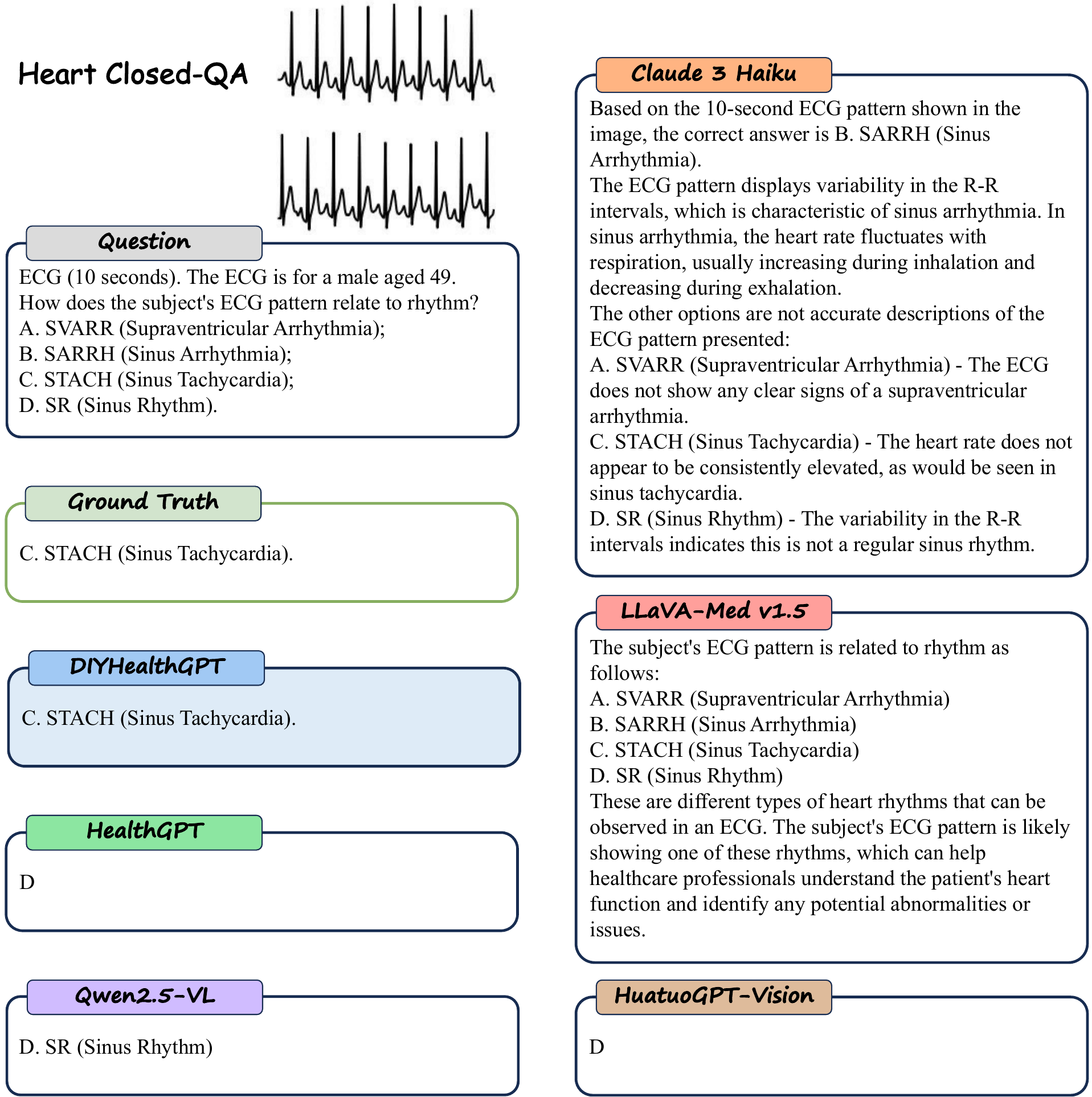}
\caption{A case study of the Heart task under \textit{closed-QA} settings.}
\label{fig:case_heart_close}
\end{figure}

\begin{figure}[!ht]
\centering
\includegraphics[width=\textwidth]{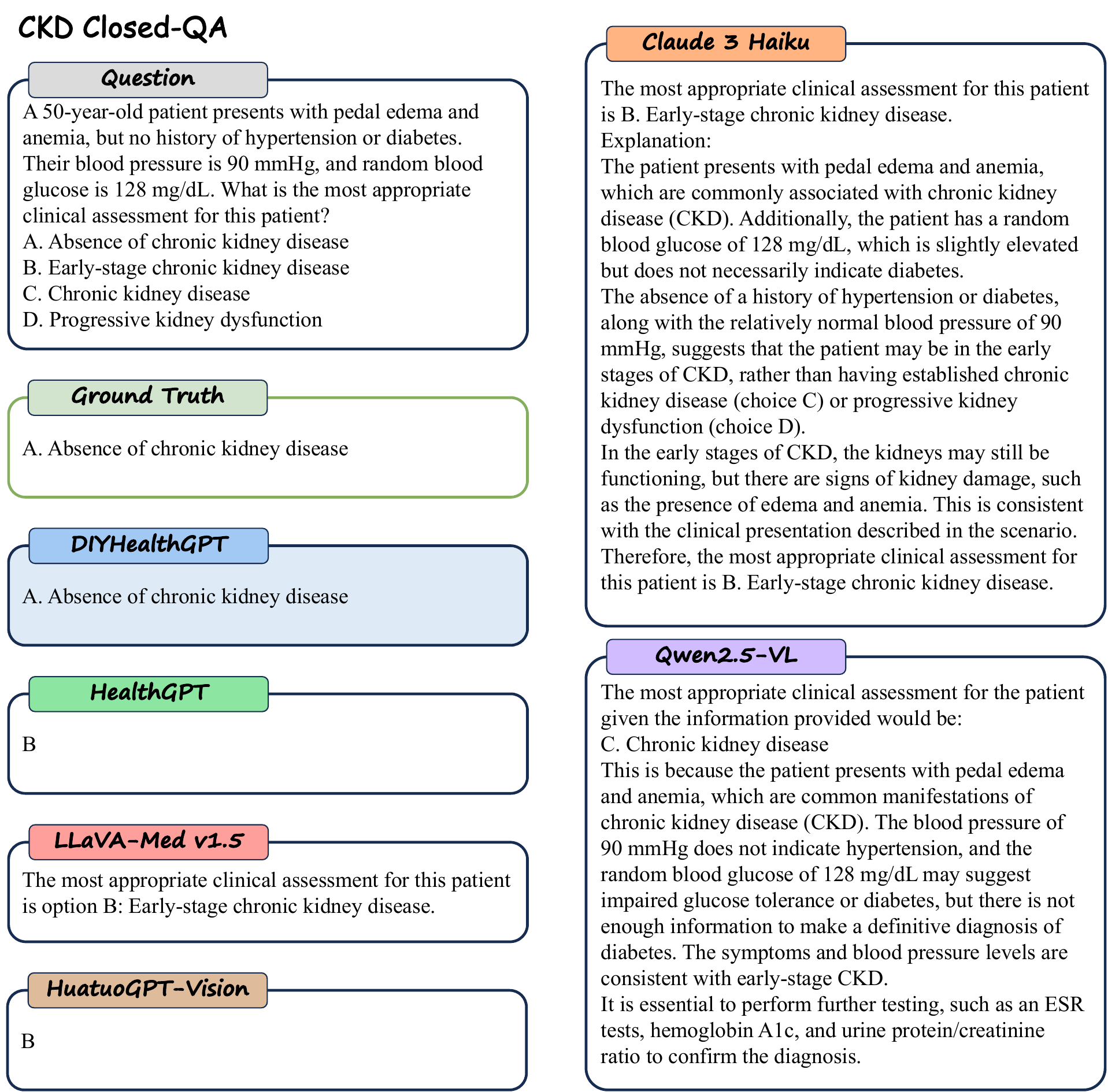}
\caption{A case study of the CKD task under \textit{closed-QA} settings.}
\label{fig:case_ckd_close}
\end{figure}

\begin{figure}[!ht]
\centering
\includegraphics[width=\textwidth]{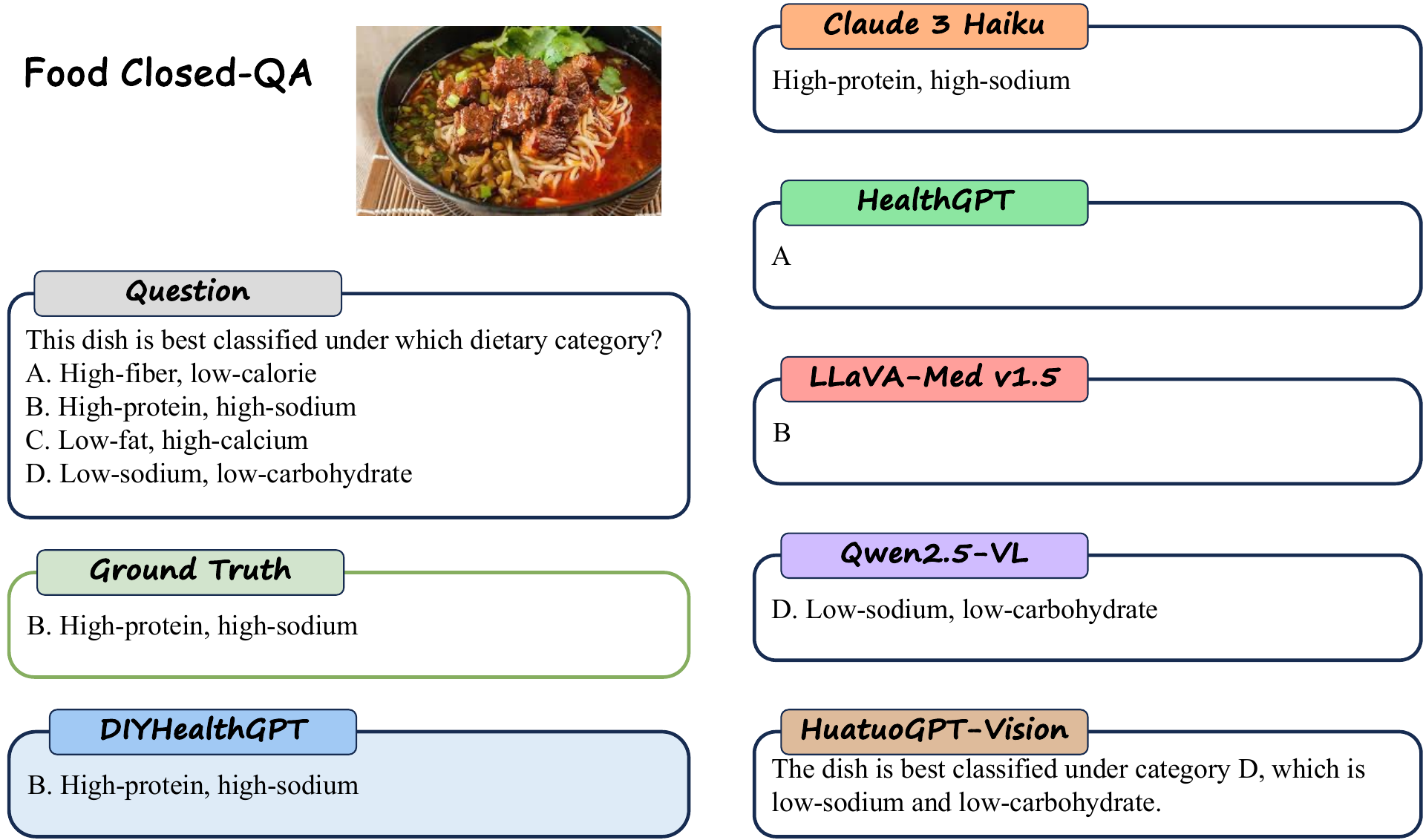}
\caption{A case study of the Food task under \textit{closed-QA} settings.}
\label{fig:case_food_close}
\end{figure}

\begin{figure}[!ht]
\centering
\includegraphics[width=\textwidth]{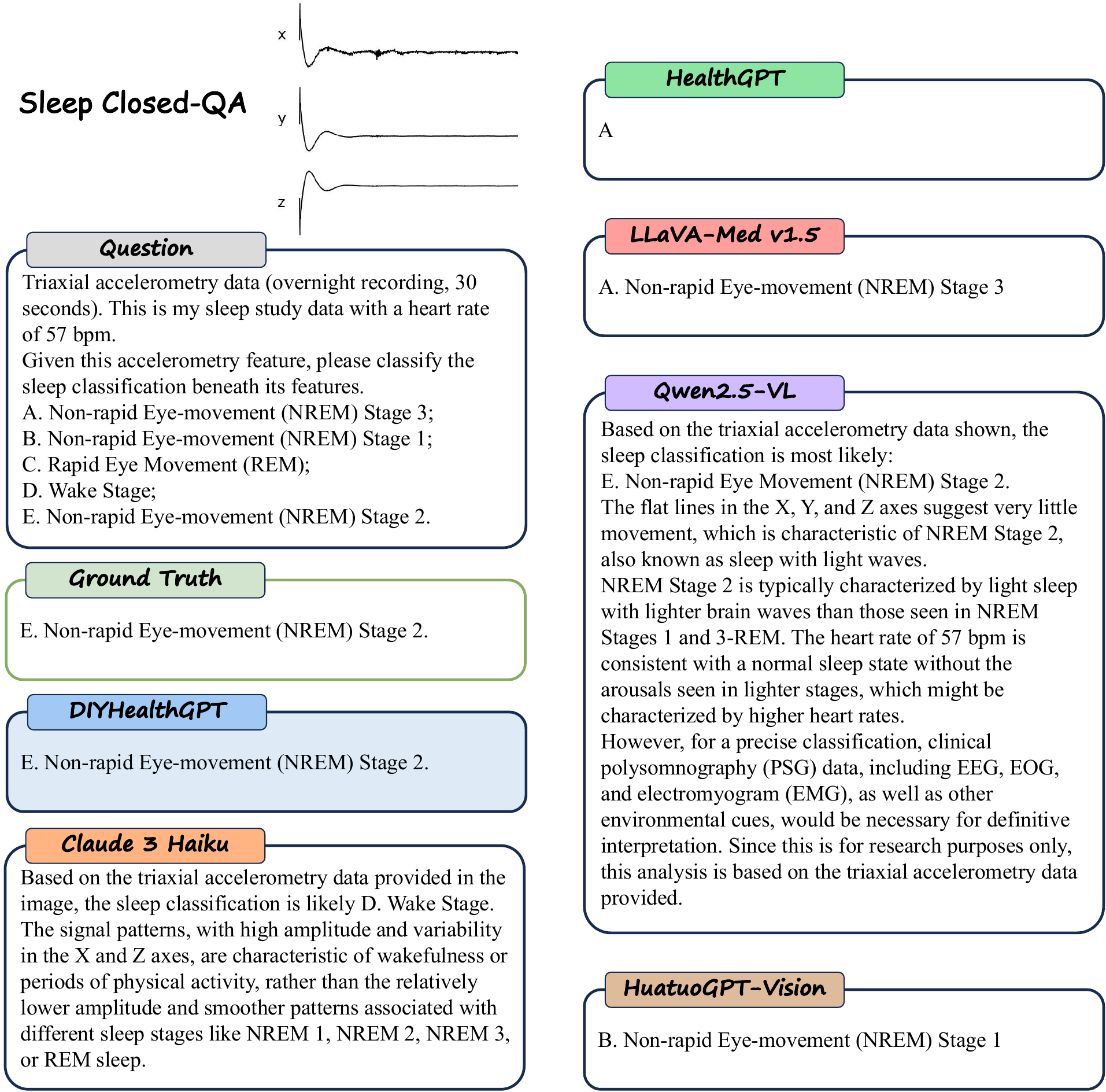}
\caption{A case study of the Sleep task under \textit{closed-QA} settings.}
\label{fig:case_sleep_close}
\end{figure}

\begin{figure}[!ht]
\centering
\includegraphics[width=\textwidth]{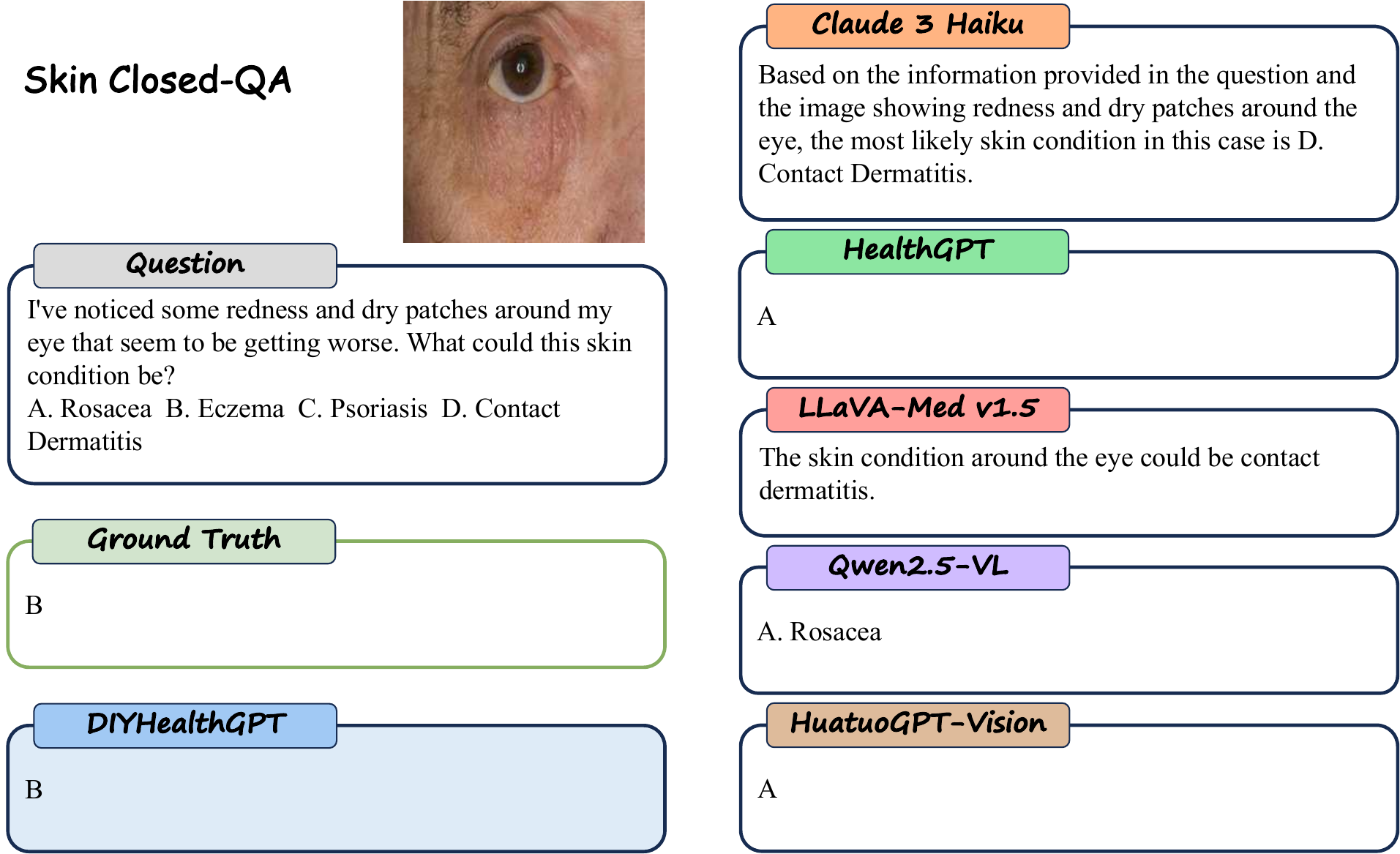}
\caption{A case study of the Skin task under \textit{closed-QA} settings.}
\label{fig:case_skin_close}
\end{figure}

\begin{figure}[!ht]
\centering
\includegraphics[width=\textwidth]{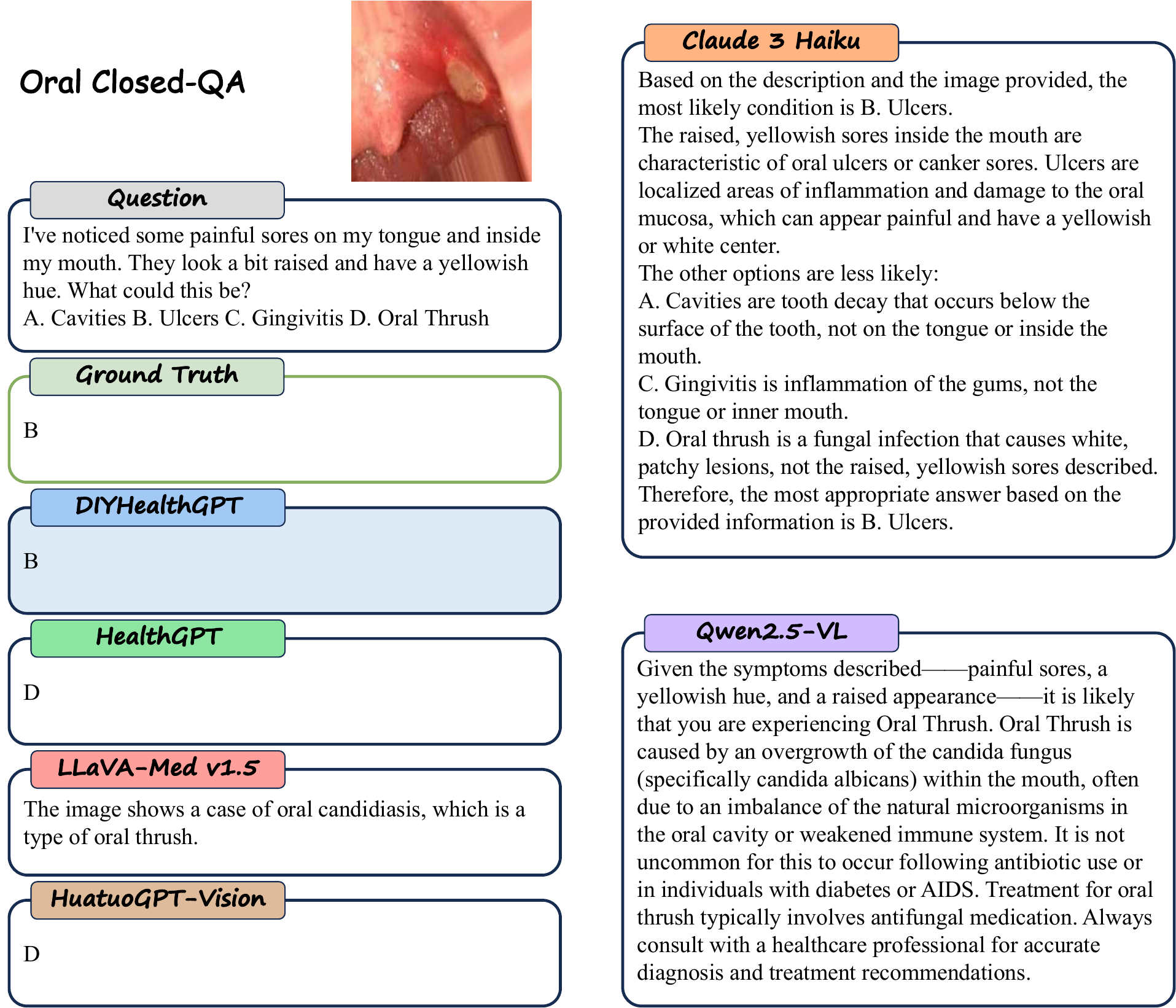}
\caption{A case study of the Oral task under \textit{closed-QA} settings.}
\label{fig:case_oral_close}
\end{figure}

\begin{figure}[!ht]
\centering
\includegraphics[width=\textwidth]{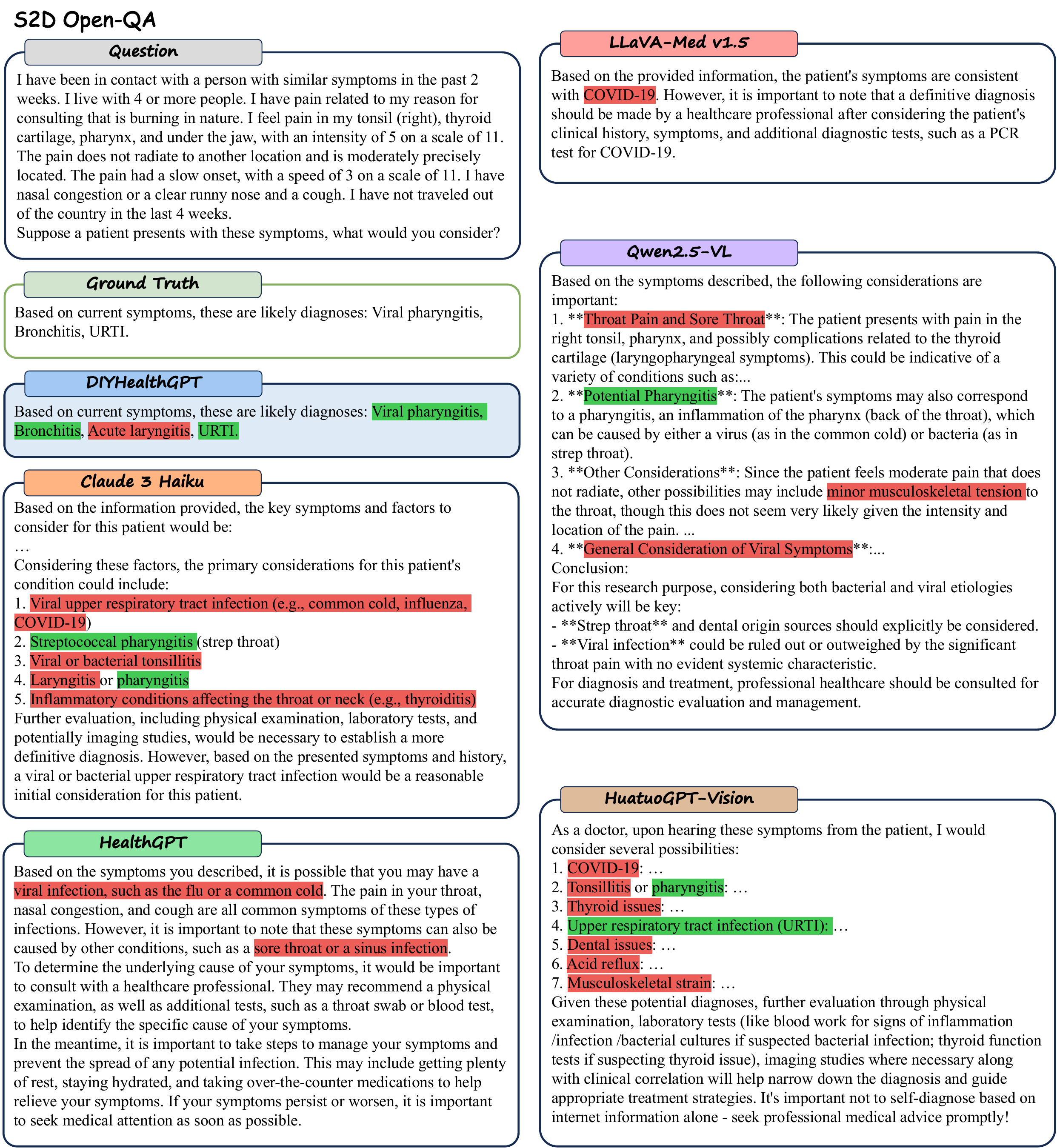}
\caption{A case study of the S2D task under \textit{open-QA} settings.}
\label{fig:case_s2d_open}
\end{figure}

\begin{figure}[!ht]
\centering
\includegraphics[width=\textwidth]{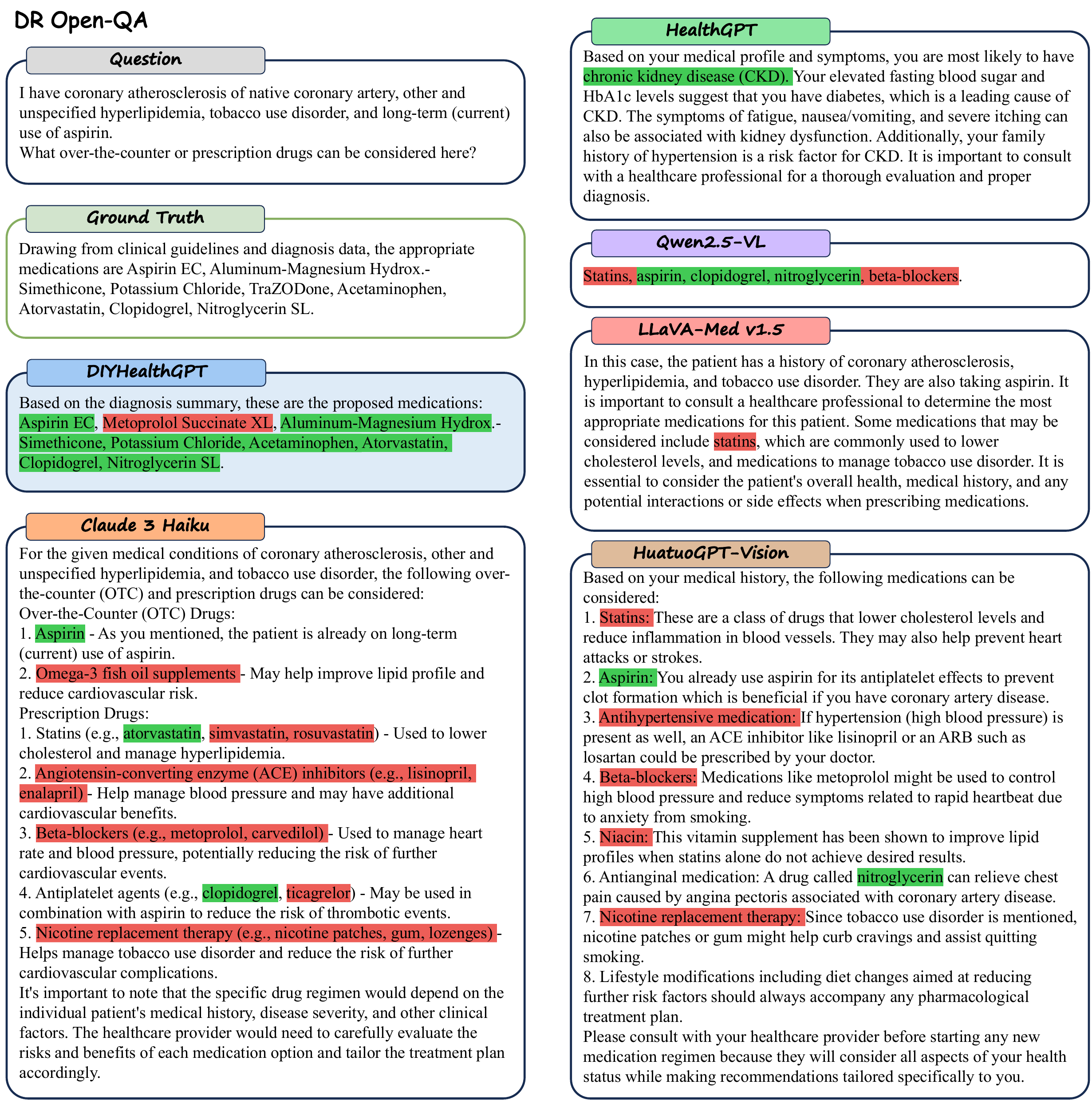}
\caption{A case study of the DR task under \textit{open-QA} settings.}
\label{fig:case_dr_open}
\end{figure}

\begin{figure}[!ht]
\centering
\includegraphics[width=\textwidth]{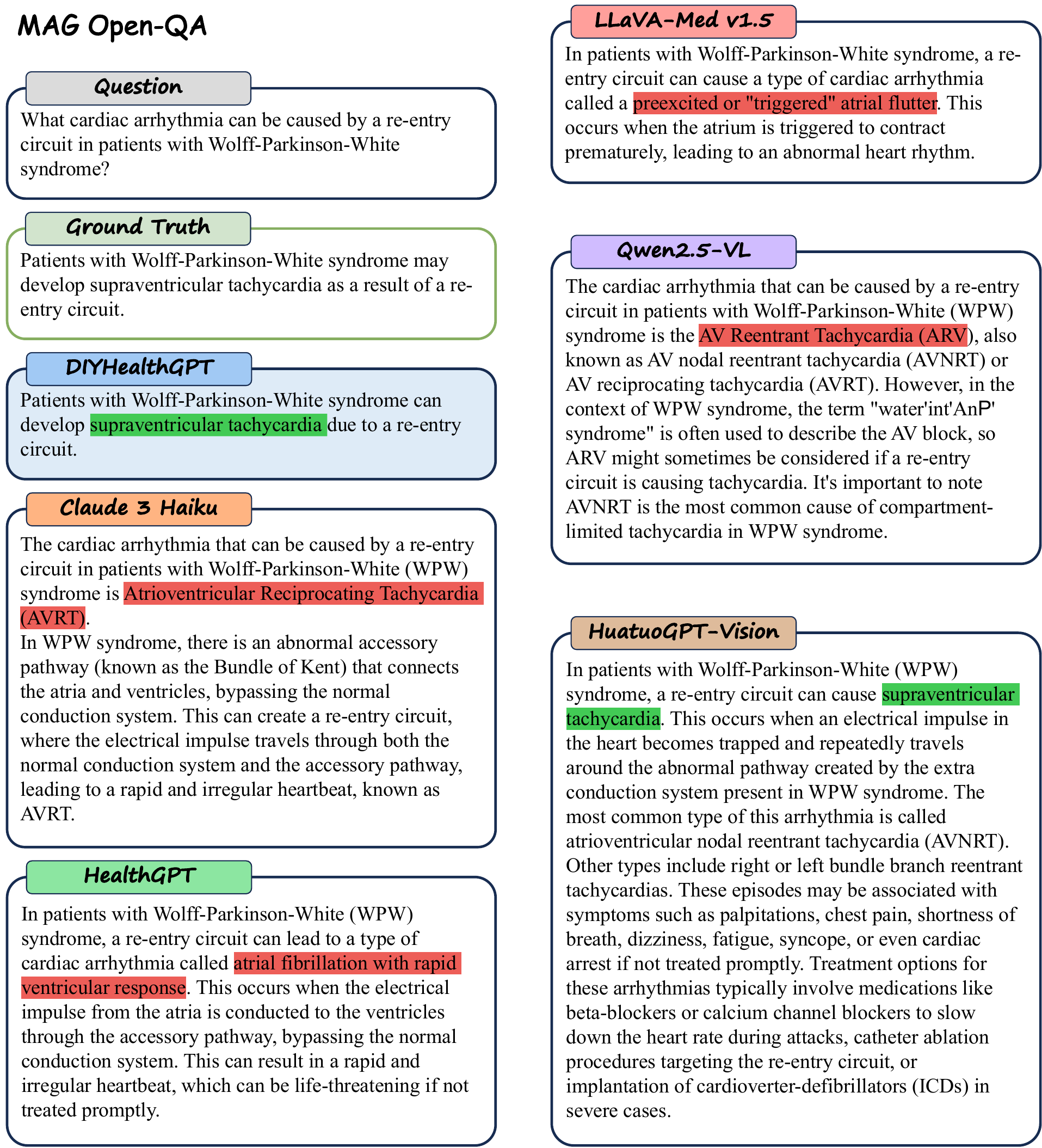}
\caption{A case study of the MAG task under \textit{open-QA} settings.}
\label{fig:case_mag_open}
\end{figure}

\begin{figure}[!ht]
\centering
\includegraphics[width=\textwidth]{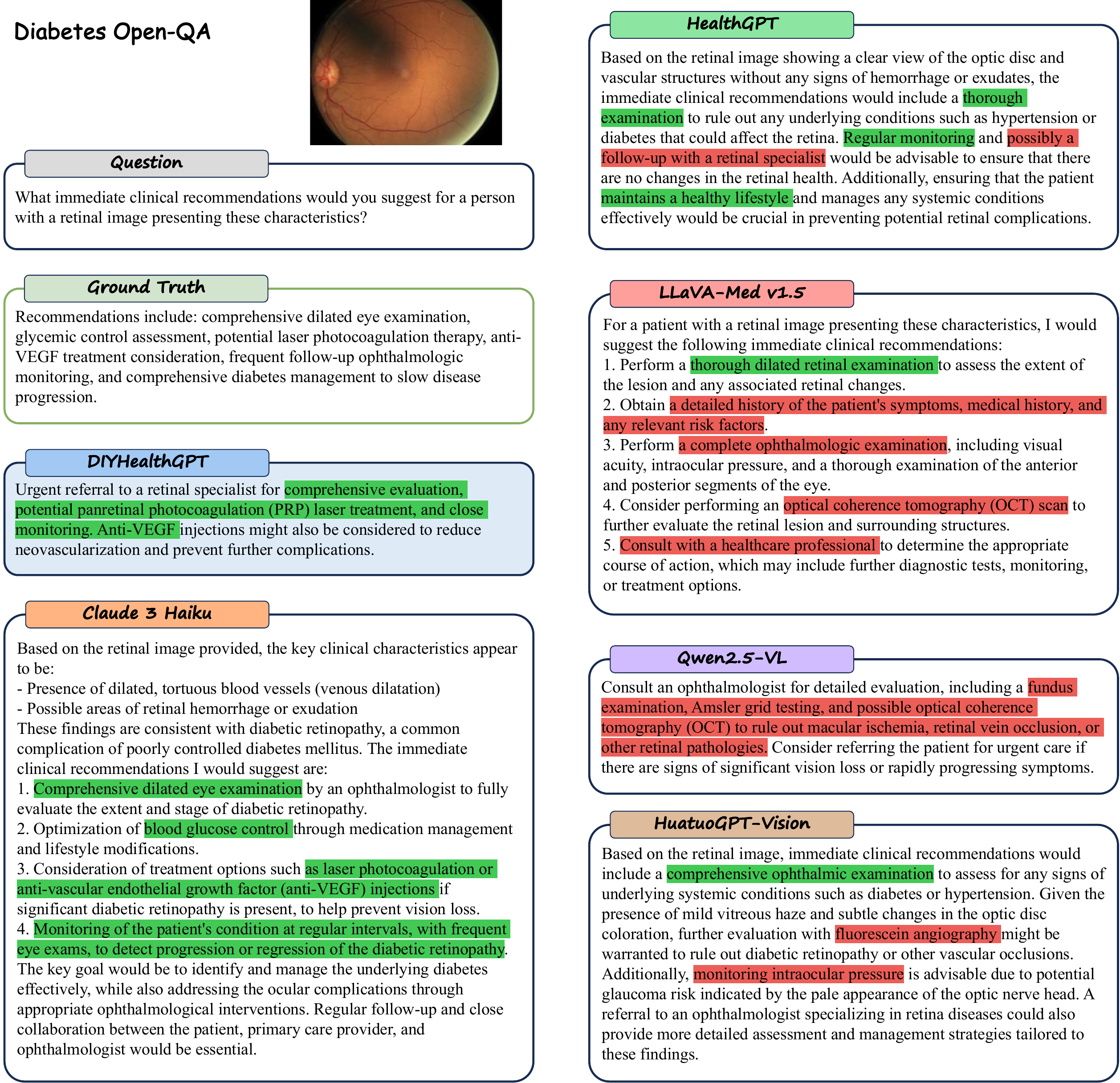}
\caption{A case study of the Diabetes task under \textit{open-QA} settings.}
\label{fig:case_diabetes_open}
\end{figure}

\begin{figure}[!ht]
\centering
\includegraphics[width=\textwidth]{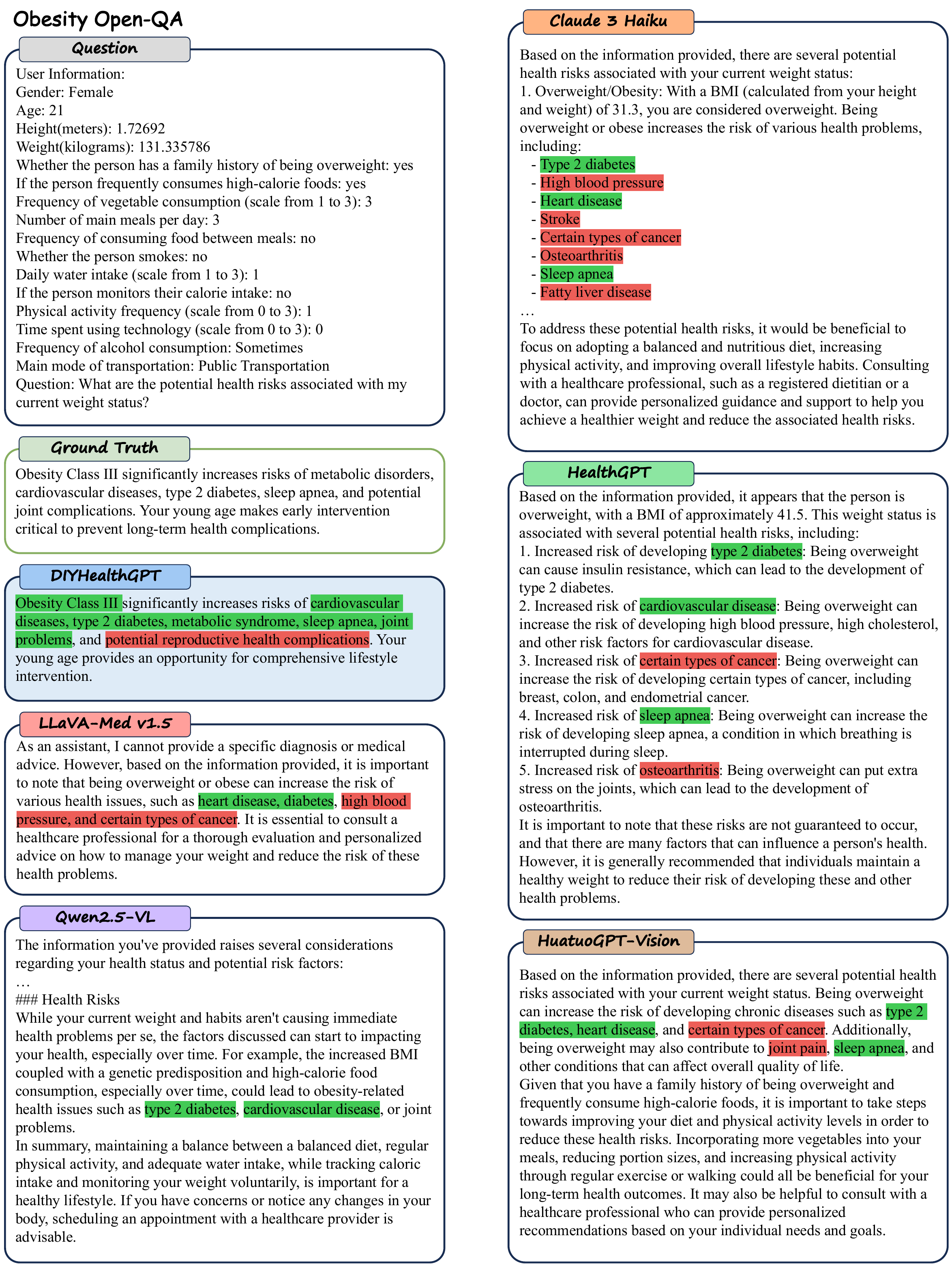}
\caption{A case study of the Obesity task under \textit{open-QA} settings.}
\label{fig:case_obesity_open}
\end{figure}

\begin{figure}[!ht]
\centering
\includegraphics[width=\textwidth]{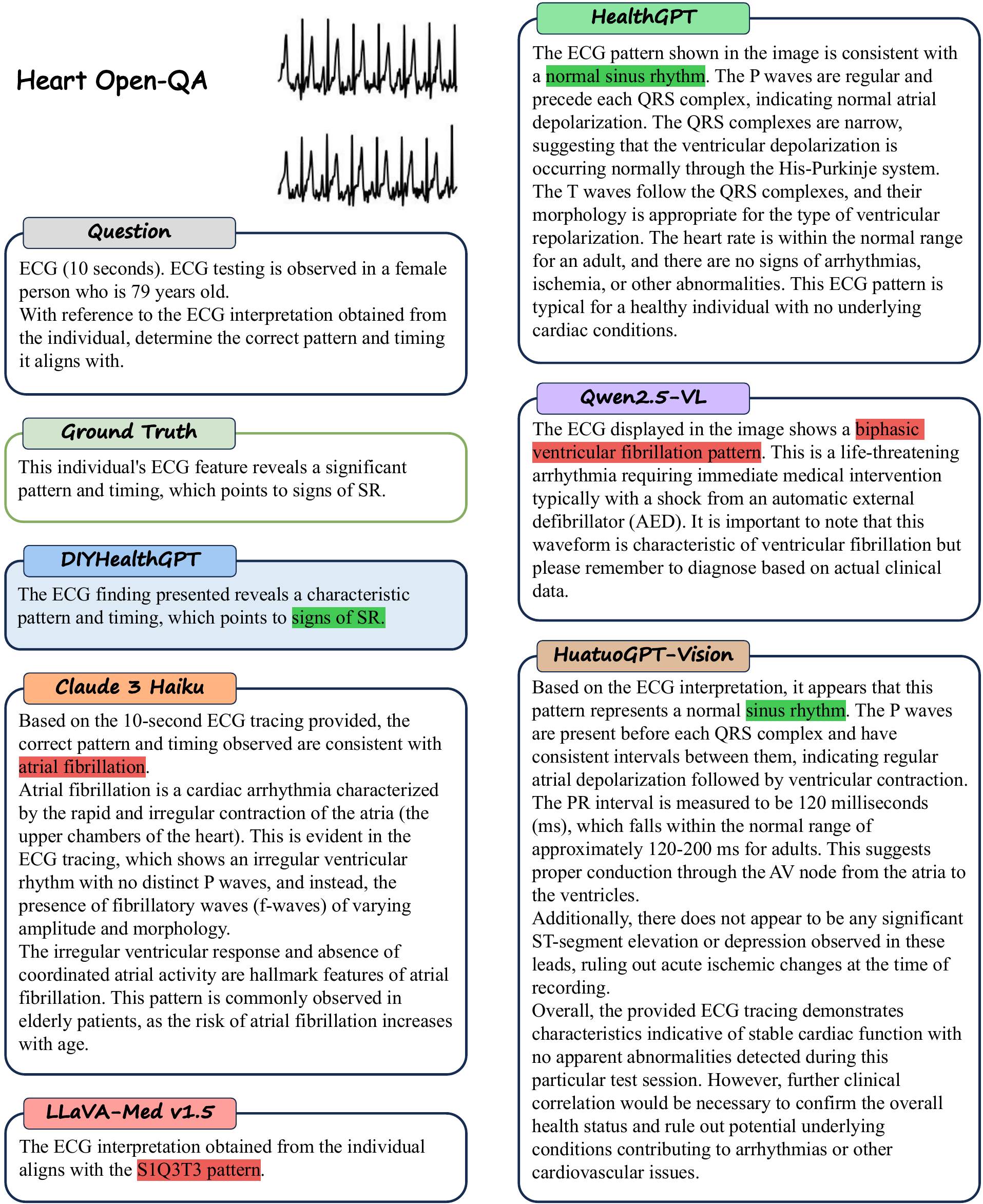}
\caption{A case study of the Heart task under \textit{open-QA} settings.}
\label{fig:case_heart_open}
\end{figure}

\begin{figure}[!ht]
\centering
\includegraphics[width=\textwidth]{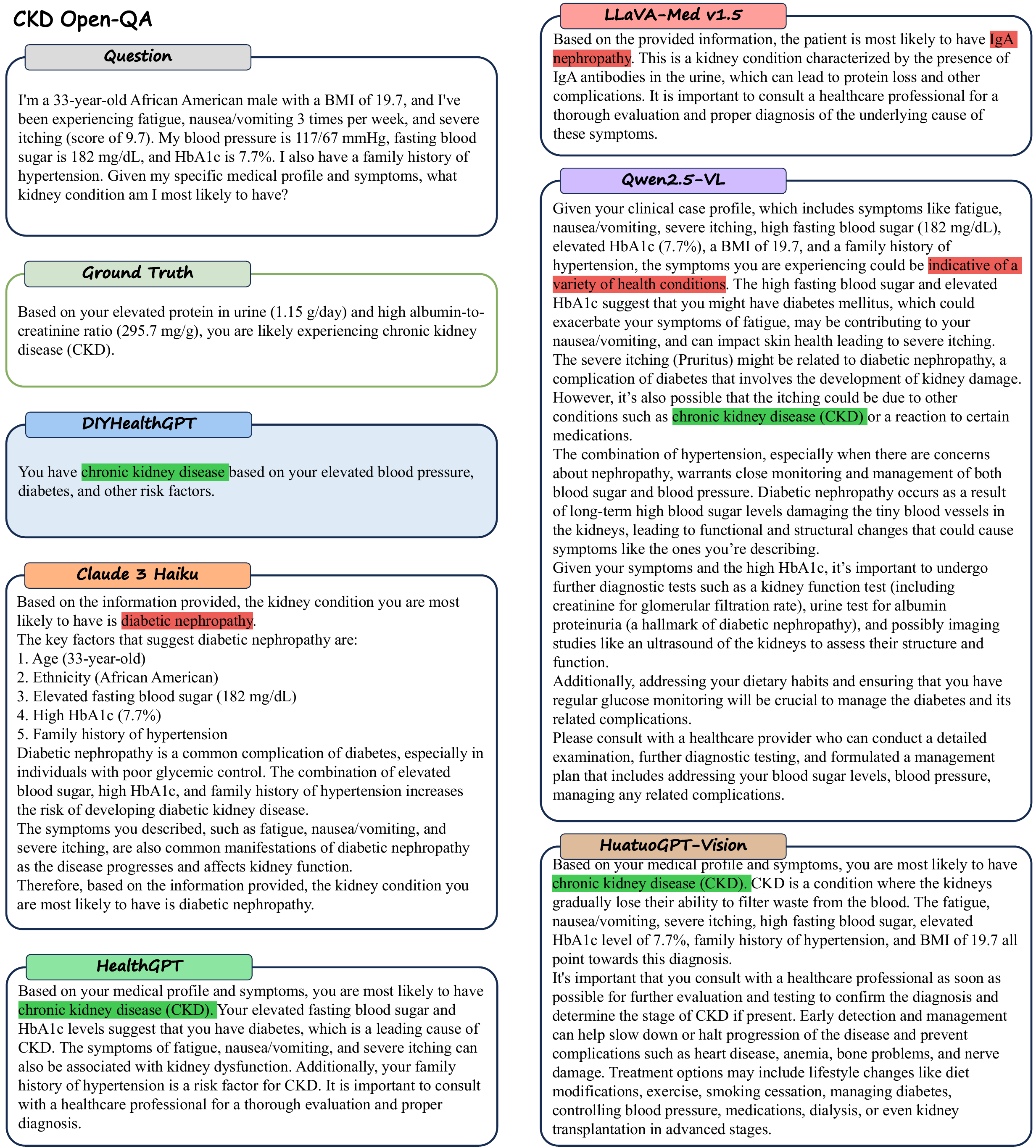}
\caption{A case study of the CKD task under \textit{open-QA} settings.}
\label{fig:case_ckd_open}
\end{figure}

\begin{figure}[!ht]
\centering
\includegraphics[width=\textwidth]{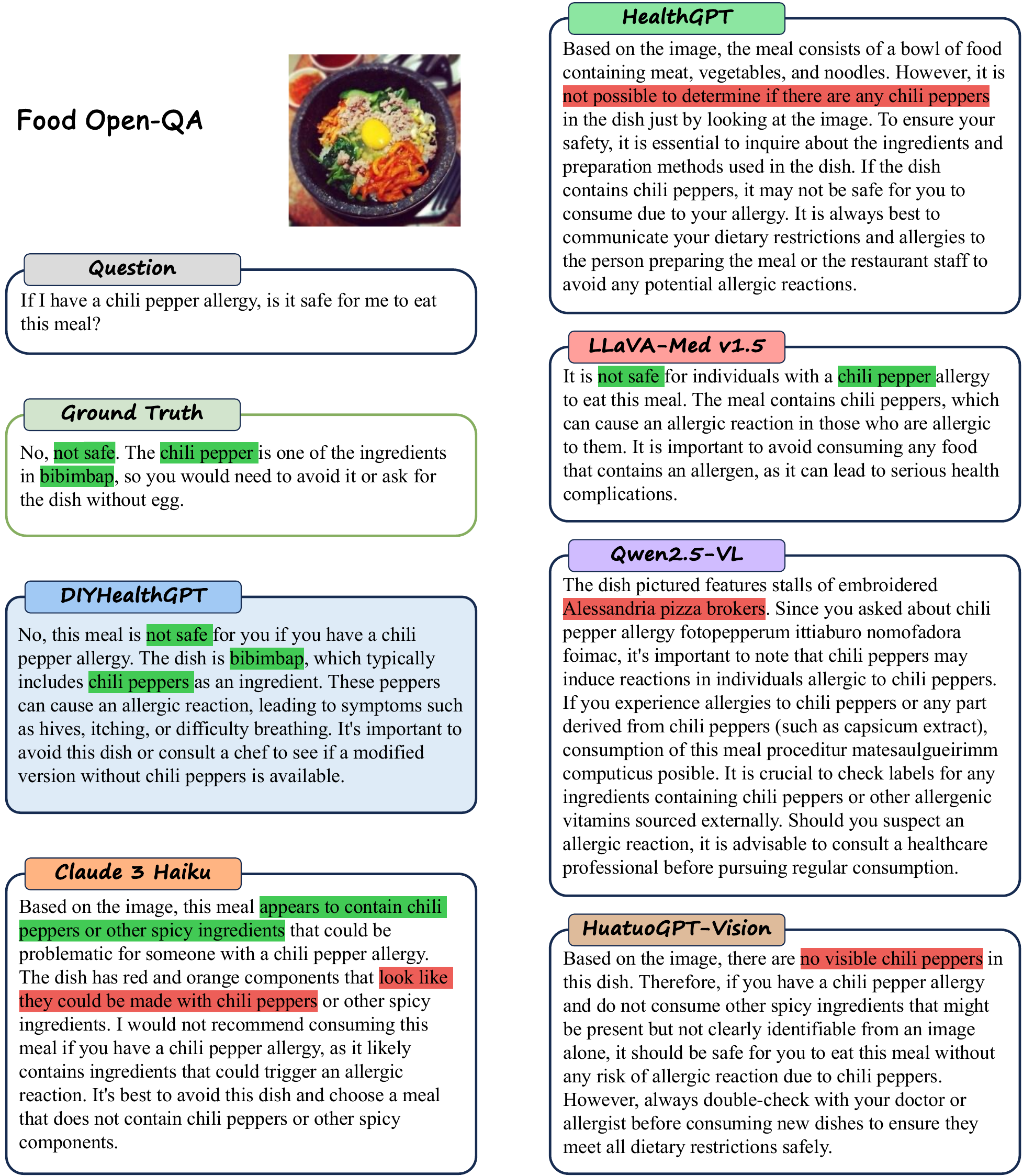}
\caption{A case study of the Food task under \textit{open-QA} settings.}
\label{fig:case_food_open}
\end{figure}

\begin{figure}[!ht]
\centering
\includegraphics[width=\textwidth]{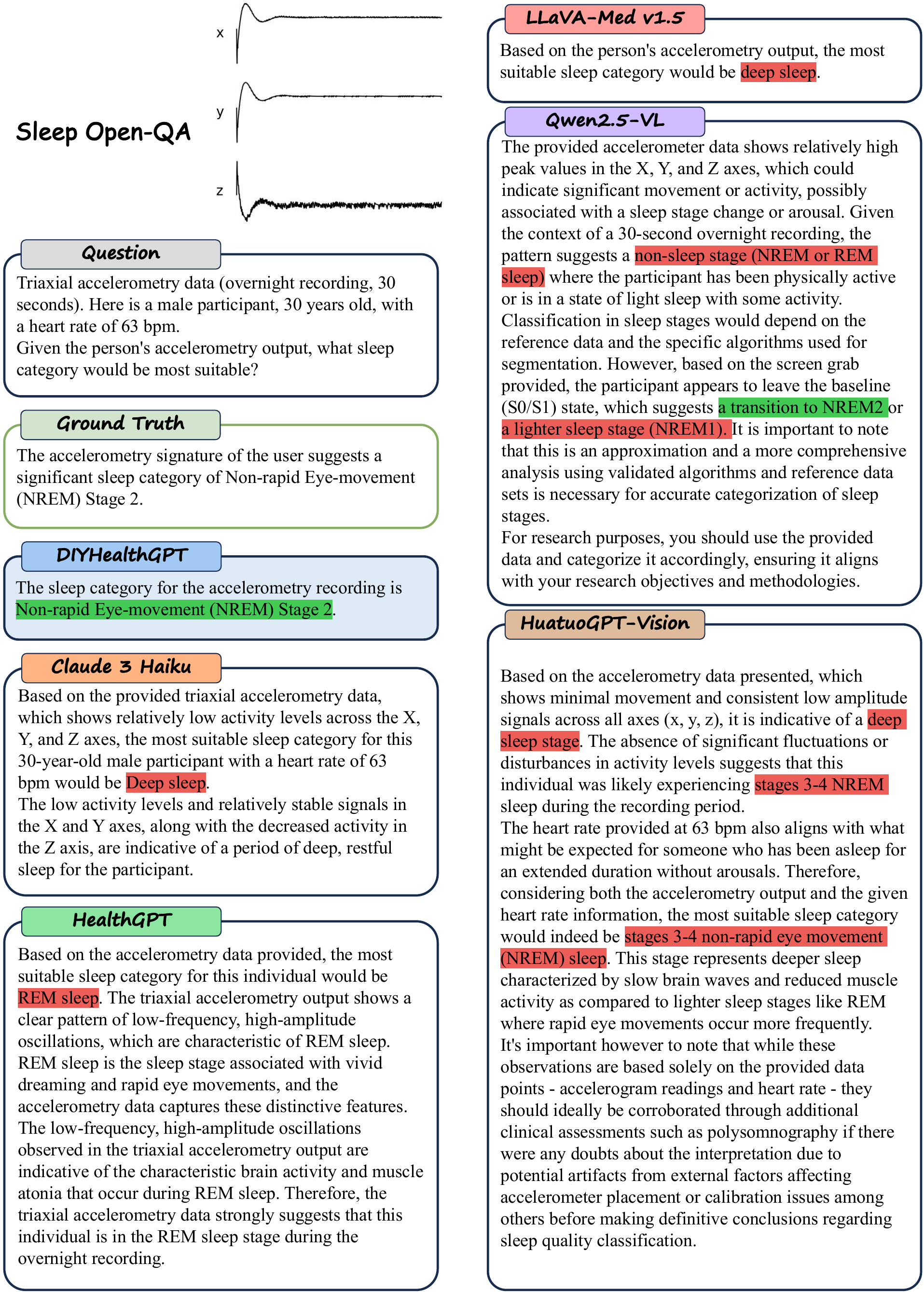}
\caption{A case study of the Sleep task under \textit{open-QA} settings.}
\label{fig:case_sleep_open}
\end{figure}

\begin{figure}[!ht]
\centering
\includegraphics[width=\textwidth]{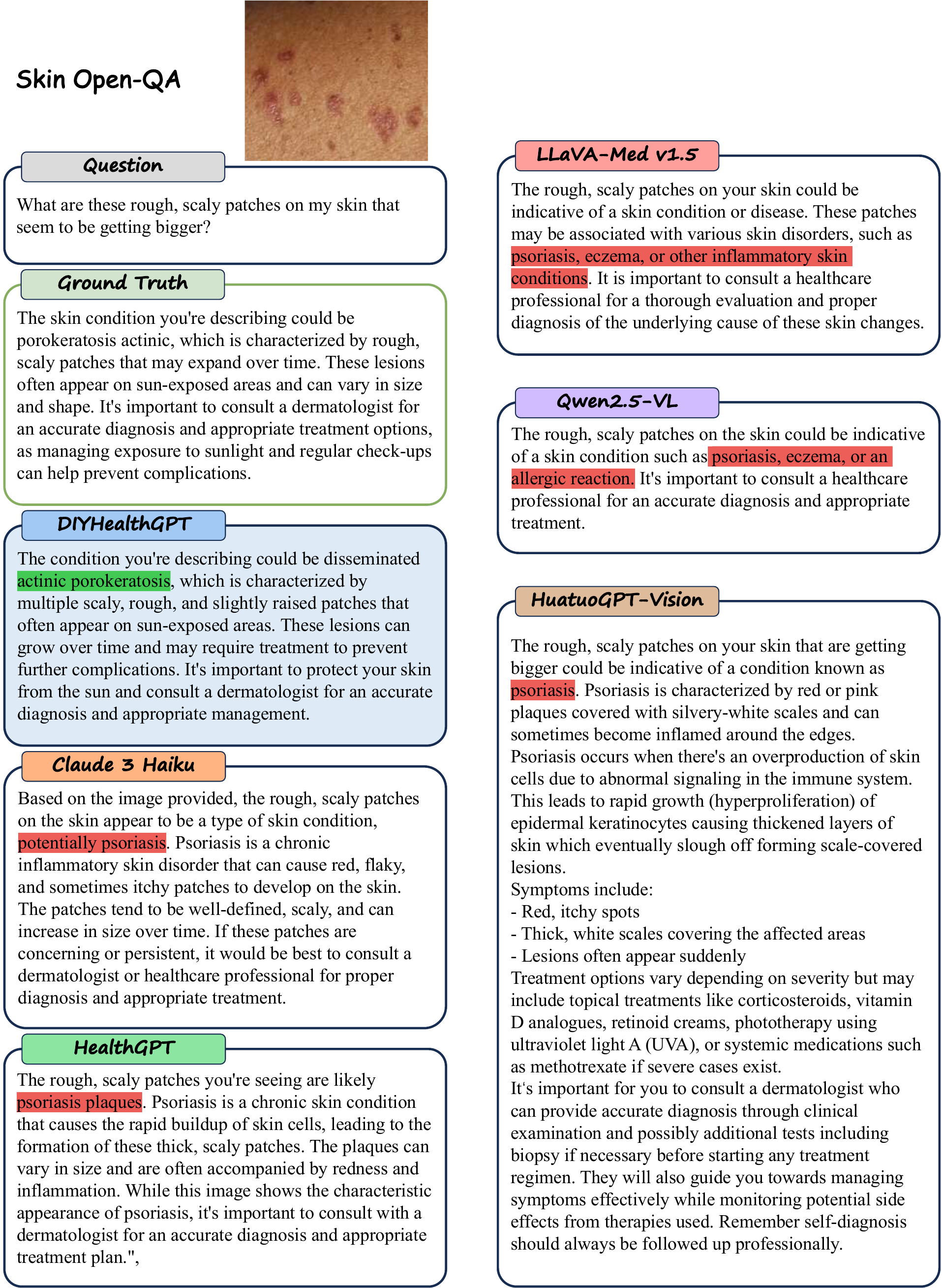}
\caption{A case study of the Skin task under \textit{open-QA} settings.}
\label{fig:case_skin_open}
\end{figure}

\begin{figure}[!ht]
\centering
\includegraphics[width=\textwidth]{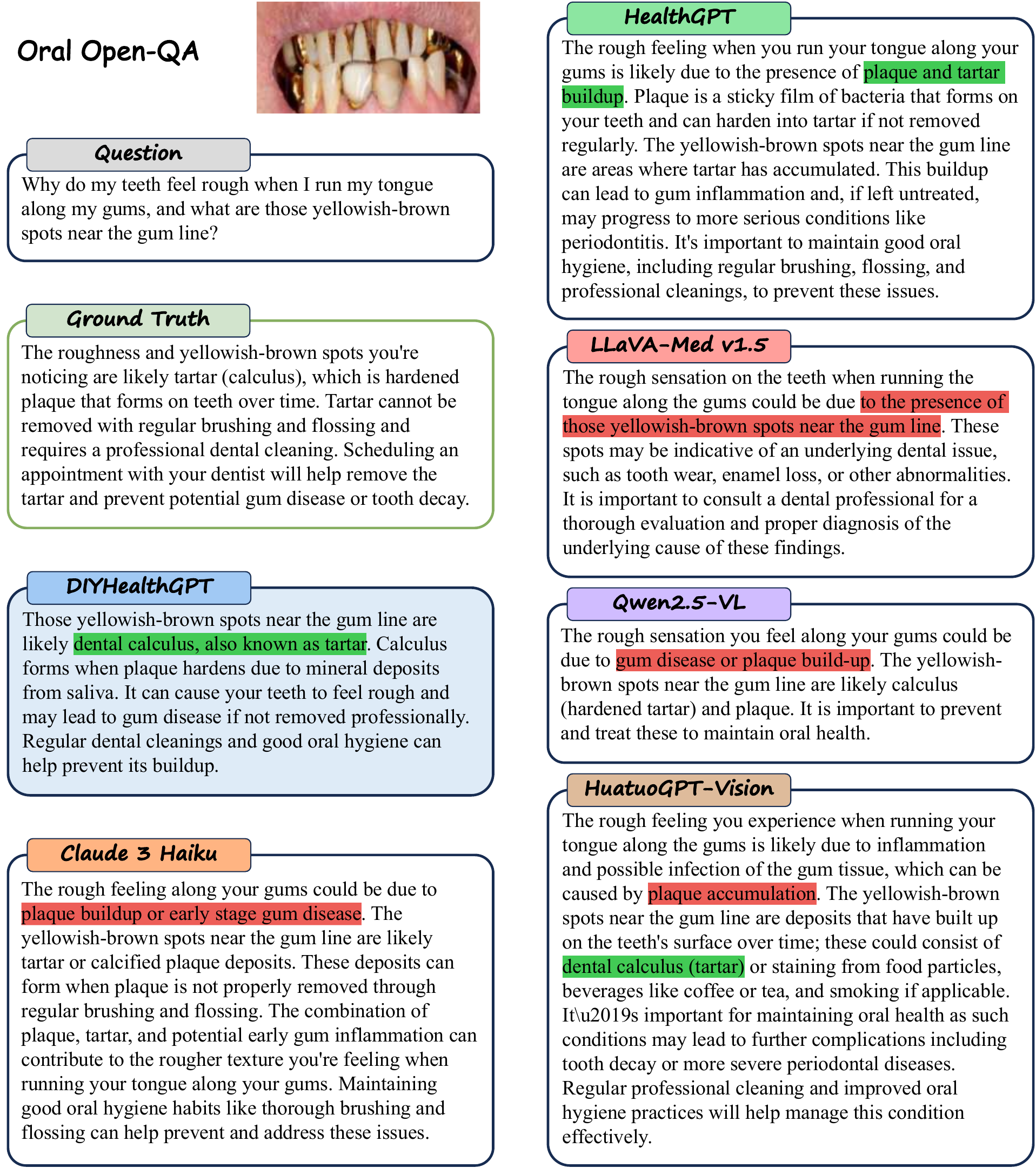}
\caption{A case study of the Oral task under \textit{open-QA} settings.}
\label{fig:case_oral_open}
\end{figure}

In this section, we present case studies comparing the responses generated by our proposed \Name with three open-source medical foundation models (LLaVa-Med v1.5, HuatuoGPT-Vision, HealthGPT), one open-source general-purpose model (Qwen2.5-VL), and one closed-source general-purpose model (Claude 3 Hiku), under both closed-QA and open-QA settings.
Figures~\ref{fig:case_s2d_close} to~\ref{fig:case_oral_close} illustrate the results for all closed-QA tasks (excluding DR, which does not support closed-QA). Across these cases, \Name consistently produces responses aligned with the ground truth answers, demonstrating strong task-specific understanding in home care scenarios.
Figures~\ref{fig:case_s2d_open} to~\ref{fig:case_oral_open} show the results for eleven tasks in open-QA settings. For example, in Figure~\ref{fig:case_s2d_open}, \Name generates answers closer to the ground truth than other models, indicating its ability to produce accurate and contextually appropriate responses to open-ended queries.

%%%%%%%%%%%%%%%%%%%%%%%%%%%%%%%%%%%%%%%%%%%%%%%%%%%%%%%%%%%%%%%%%%%%%%%%%%%%%%%
%%%%%%%%%%%%%%%%%%%%%%%%%%%%%%%%%%%%%%%%%%%%%%%%%%%%%%%%%%%%%%%%%%%%%%%%%%%%%%%

\end{document}